\pdfoutput=1
\documentclass{pasj02}
\usepackage{soul}
\usepackage{color}

\jyear{2024}
\Received{}
\Accepted{}

\usepackage{subcaption}
\usepackage{lscape}
\usepackage{graphicx}
\usepackage{natbib}
\usepackage{multicol}
\usepackage{threeparttable}
\bibpunct[:]{(}{)}{,}{a}{}{,}

\begin{document}
\title{Search for Exoplanetary Ring Systems with TESS}

\author{
Tsubasa \textsc{umetani},\altaffilmark{1}\email{umetani-tsubasa@ed.tmu.ac.jp, tobachya@gmail.com} 
Masataka \textsc{aizawa},\altaffilmark{2,3,4}
Yuichiro \textsc{ezoe},\altaffilmark{1}
and
Yoshitaka \textsc{ishisaki}\altaffilmark{1}
}%
\altaffiltext{1}{Department of Physics, Tokyo Metropolitan University, 1-1 Minami-Osawa, Hachioji-shi, Tokyo 192-0397, Japan }
\altaffiltext{2}{Tsung-Dao Lee Institute, Shanghai Jiao Tong University, Shengrong Road 520, 201210 Shanghai, P. R. China}
\altaffiltext{3} {Cluster for Pioneering Research, RIKEN, 2-1 Hirosawa, Wako, Saitama 351-0198, Japan} 
\altaffiltext{4}{College of Science, Ibaraki University, 2-1-1 Bunkyo, Mito, Ibaraki 310-8512, Japan}




\maketitle

\begin{abstract}
Photometric surveys for exoplanetary ring systems have not confirmed any object with Saturn-sized ring.
We systematically analyse 308 TESS planet candidates, mainly comprised of giant short-period planets orbiting nearby bright stars.
These targets are selected based on the optimistic detectability of rings, assuming a favourable ring orientation.
We develop a pipeline with a two-step noise reduction and compare the fitting results of both ringless and ringed transit models to the resulting phase-folded light curves.
Although we identify six systems where ringed models are statistically favoured, visual inspection of the signals suggests that none of them is conclusively attributed to the presence of rings.
Assuming the ring orientation favourable for detection, we determine the 3$\sigma$ upper limits on ring sizes for 125 objects.
Using these ring size limits, we derive upper limits on the ring occurrence rate, such as rings with an outer radius larger than 1.8 times the planetary radii occurring at rates lower than 2\%. 
However, these limits can be relaxed if tidal alignment between the spin and orbital axes holds.
We explore an alternative detection method using transit depth variations by ring precession and estimate that 10 and 13 systems are likely detectable in TESS and Kepler data, respectively.
 
\end{abstract}

\section{Introduction}\label{Introduction}
In our Solar System, planetary rings have been confirmed not only around the Jovian planets--Jupiter, Saturn, Uranus, and Neptune--but also around the asteroids: Chariklo \citep{2014Natur.508...72B}, Chiron \citep{2015A&A...576A..18O}, Haumea \citep{2017Natur.550..219O} and Quaoar \citep{2023Natur.614..239M}. 
These discoveries naturally suggest the universality of ringed planets in the Universe, as in exoplanets. 
However, to date, there has not been any conclusive detection of ringed planets similar to Saturn. 

Methods so far proposed to detect exoplanetary rings include photometric and spectroscopic approaches. 
In the photometric cases, ring features can be studied as follows: distortions of transit light curves especially observable during ingress and egress \citep{1999CRASB.327..621S}, reflection effects on the phase curve \citep{2004A&A...420.1153A, 2005ApJ...618..973D}, increase of flux during transit by forward scattering \citep{2004ApJ...616.1193B} and transit depth variations due to spin precession \citep{2010ApJ...716..850C}.
In the spectroscopic cases, the following features are useful for detection: additional velocity anomaly by the Rossiter effect \citep{2009ApJ...690....1O}, reflected light spectra for short-period planets \citep{2015A&A...583A..50S}, variation of line profiles around fast rotating stars \citep{2017MNRAS.472.2713D} and transmission spectra \citep{2022ApJ...930...50O}.

Low-density planets, with densities below 0.3 g cm$^{-3}$ and referred to as ``super-puffs'', can be indicative of ringed planets from both photometric and spectroscopic perspectives.
An increase in transit duration and depth caused by rings can lead to underestimations of planetary densities \citep{2015ApJ...803L..14Z}. 
\citet{2020AJ....159..131P} suggested that the featureless spectra of super-puffs can be explained by planetary rings. 
In this context, HIP 41378 f is an interesting target for exploring the possibility of the ring because of its extremely low-density \citet{} and featureless transmission spectra obtained by the Hubble Space Telescope (HST) \citep[]{2019arXiv191107355S, 2022ApJ...927L...5A}.  
While the evidence for the ring hypothesis remains inconclusive based on light curves and transmission spectra \citep[]{2020A&A...635L...8A, 2022ApJ...927L...5A}, the high-precision transmission data from the James Webb Space Telescope (JWST) \citep{2006SSRv..123..485G} offers the potential to distinguish between different scenarios \citep{2022ApJ...927L...5A}.

Studies on ring size and orientation demonstrate that analysing observational data is essential to verify ring detectability in photometric cases.
Rings of Saturn-like size and obliquity angle are detectable by Kepler \citep{2004ApJ...616.1193B}, the predecessor of TESS.
Detecting rings around close-in planets is challenging because of the primarily rocky composition of these rings, which makes them more compact than ice rings \citep{2011ApJ...734..117S}.
Furthermore, tidal forces from the host star dampen the obliquity of rings, making detection even more difficult \citep{2004ApJ...616.1193B}.
However, if the initial axial tilt of the planet was large, it can maintain a non-zero obliquity \citep{2005NewAR..49..478B, 2010Sci...327..977B}, making rings easier to detect.

There have been several attempts to search for rings using high-precision photometric data obtained by space satellites. 
\citet{2001ApJ...552..699B} studied the ring hypothesis for HD 209458 b using HST transit data and obtained upper limits on the ring size as 1.8 times the planetary radii, assuming the tidal alignment. 
Heising, Marcy and Schlichting (\citeyear{2015ApJ...814...81H}) fitted ringed models to the light curve of 21 Kepler objects, mainly consisting of hot Jupiters, assuming eight combinations of ring orientation and size.
They found that in at least 12 planetary systems, certain orientations of Saturn-sized rings could be excluded.
\citet{2017AJ....153..193A} searched for any distortion in transit light curves provided by Kepler for 89 long-period planets and selected KIC 10403228 as a ringed planet candidate. 
However, they estimated this planetary orbital period to be 450 years, and it is challenging to confirm the object by follow-up observations.
Also, in \citet{2018AJ....155..206A}, the ringless and ringed transit models were fitted to light curves of 168 objects with high signal-to-noise ratios in the short-cadence data of Kepler observations. 
They found no ringed planet candidates, although they identified some other physical phenomena that could be mistaken for ring signals. 
They also showed that, assuming tidal alignment, the occurrence rate of rings with an outer diameter greater than twice the planetary radius was less than 15 \%.
Beyond Saturn-sized rings, \citet{2015ApJ...800..126K} interpreted the unusual transit as planetary rings around J1407 b extending to the radius beyond 0.6 au, based on SuperWASP photometric data.

Since Kepler data were already extensively studied by our previous studies \citep[]{2017AJ....153..193A,2018AJ....155..206A}, we focus on a search for new ringed planet candidates using photometric data from TESS \citep{2015JATIS...1a4003R}. 
Given that TESS targets bright nearby stars, it allows for easier and more effective follow-up observations if any candidates are detected. 

Section \ref{Transit modelling and optimisation to detect ring signals} outlines the transit model utilized for the detection of ringed planets. 
Section \ref{Target Selection} details the selection criteria for our observational targets from TESS objects of interest \citep[TOI;][]{2021ApJS..254...39G}.
In section \ref{Data Analysis}, we describe the preprocessing procedure applied to the TESS data and the configuration for model fitting in detecting ringed planets.
Section \ref{Result} presents our findings, followed by section \ref{Discussion}, where we explore the implications of our TESS-derived results in comparison with those from previous Kepler analyses and consider alternative strategies for finding ringed planets.
Finally, section \ref{Conclusion} gives our concluding thoughts and prospects for future research.

\section{Transit modelling and optimisation to detect ring signals} \label{Transit modelling and optimisation to detect ring signals}
\subsection{Transit models for ringless and ringed planets} \label{Transit models for ringless and ringed planets}
Here, we introduce two models used in this study to compare the goodness of fit: the transit without rings, referred to as ``the ringless model'', and the one with rings as ``the ringed model''.

\begin{figure}
  \centering
  \includegraphics[width=80mm]{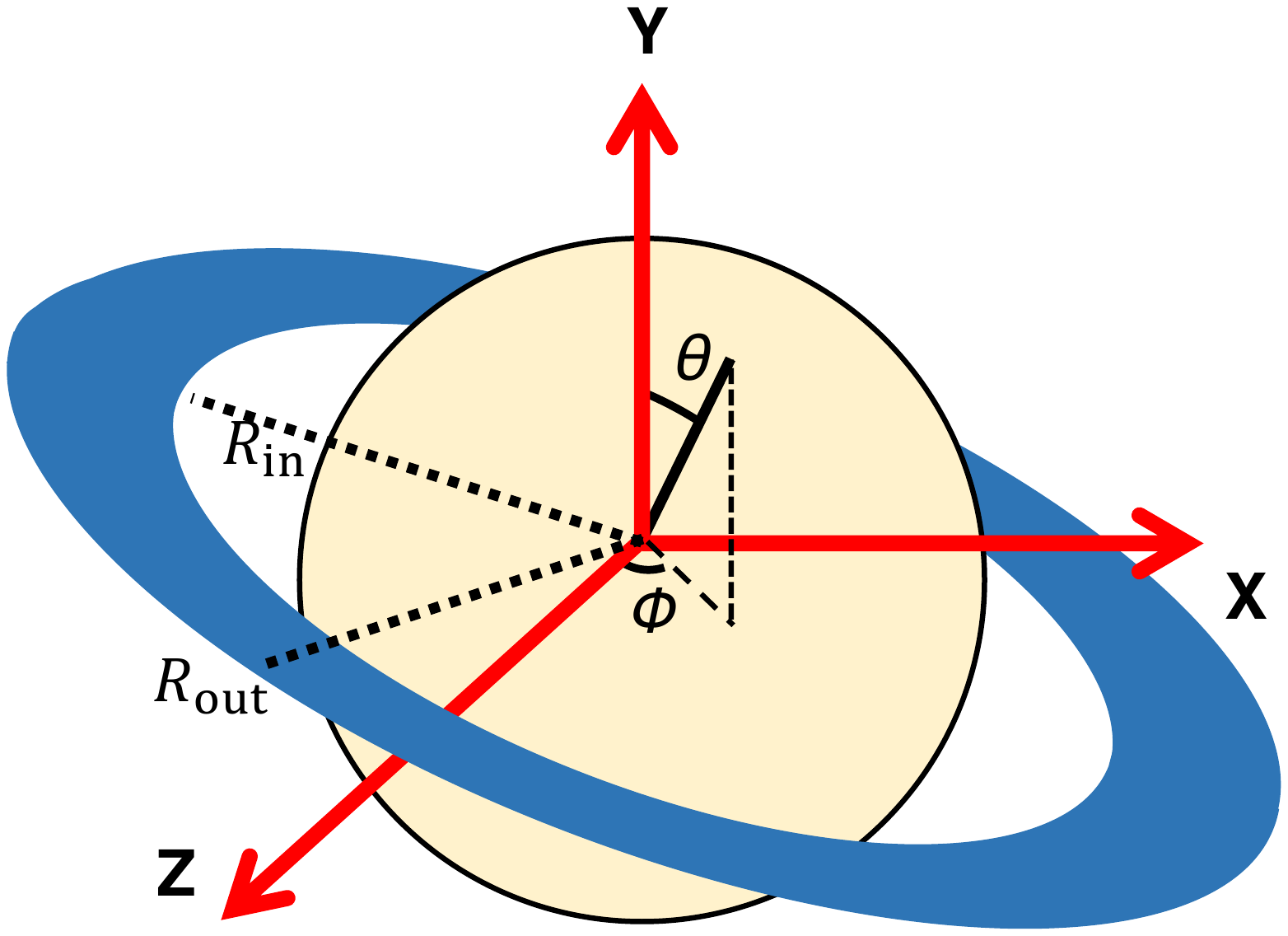}
  \caption{Schematic illustration of our ringed planet model. $X$-axis aligns with the direction of the line of nodes of the planetary orbit projected onto the stellar disk, $Y$-axis is normal to this plane, and $Z$-axis points to the observer. $\theta$ is the angle between $Y$-axis and the ring axis, $\phi$ is the azimuthal angle, $R_{\rm in}$ and $R_{\rm out}$ denote the inner and outer radius of the ring, respectively.}
  \label{fig:ring_model_explain}
\end{figure}

In both models, we define $(x,y)$ as coordinates centred on the stellar and approximate $I(x, y)$, the intensity for a given position $(x,y)$ in the stellar disk, as

\begin{equation} \label{eq: approximate stellar intensity}
I(x, y) \propto \left[1-2 q_{2} \sqrt{q_{1}}(1-\mu)-\sqrt{q_{1}}\left(1-2 q_{2}\right)(1-\mu)^{2}\right]. \\
\end{equation}

Here, both $q_{1}$ and $q_{2}$ are limb-darkening parameters, $\mu \equiv \sqrt{1-(x^2+y^2)/R_{\star}^2}$ and $R_{\star}$ is the stellar radius. 
For simplicity, we consider circular orbits for all of the planets. 

We employ the ringless model developed by \citet{2002ApJ...580L.171M} and implemented in \texttt{batman} \citep{2015PASP..127.1161K}. 
This model has six parameters: $q_1, q_2, t_0, R_{\rm p}/R_{\star}, a/R_{\star}$ and $b$. 
Here, $t_0$ is the phase shift from each transit midpoint estimated from the orbital period with known transit midpoint $t_{\rm mid}$, $R_{\rm p}/R_{\star}$ is the ratio of planetary to stellar radii, $a/R_{\star}$ is the semi-major axis normalised by the stellar radius and $b$ is the impact parameter, respectively.

For the ringed model, we employ the same transit model in \citet{2017AJ....153..193A}. 
The detailed algorithm is given in appendix A in the paper for interested readers.
Configuration of the planet and ring during transit is illustrated in figure \ref{fig:ring_model_explain}.
We define $X$-axis as the direction of the line of nodes of the planetary orbit projected onto the stellar disk, $Y$-axis as the normal to the projected planetary orbit, and $Z$-axis as the direction towards the observer.
We assume a two-dimensional, single ring with the optical depth $\tau$, inner radius $R_{\rm{in}}$, and outer radius $R_{\rm{out}}$.
$\theta$ is the angle between the normal of the planetary orbital plane and the ring axis, and $\phi$ is the azimuthal angle. 
We ignore the forward scattering by the ring and assume fully opaque rings $(\tau= \infty)$ for simplicity. 
Our ringed model has 10 parameters: $q_1, q_2, t_0, R_{\rm p}/R_{\star}, a/R_{\star}, b, \theta, \phi,  R_{\rm{in}}$ and $R_{\rm{out}}$.
We transform $R_{\rm{in}}$ and $R_{\rm{out}}$ into dimensionless parameters using planetary radius, $R_{\rm p}$, as $r_{\rm{in} / \rm{p}} \equiv R_{\rm{in}}/R_{\rm p}$ and $r_{\rm out/in} \equiv R_{\rm{out}}/R_{\rm{in}}$.

\subsection{Optimisation of model fitting parameters for detecting ring signals}
Ring signals appear as a residual when fitting the ringless model to the ringed transit light curve (in detail, \citet{2004ApJ...616.1193B}).
To quantitatively evaluate the significance of ring signals, we fit the ringless and ringed models to the light curve and calculate $\chi^2$ value defined by equation (\ref{eq: chisqaure}),

\begin{equation}\label{eq: chisqaure}
\chi^2=\sum_{i}\left(\left[d\left(t_{i}\right)-m\left(t_{i}\right)\right] / \Delta d\left(t_{i}\right)\right)^2.
\end{equation}

Here, $t_{i}$ is the time for each bin, $d\left(t_{i}\right)$ is the relative flux, $m\left(t_{i}\right)$ is the model flux and $\Delta d\left(t_{i}\right)$ is uncertainty of each bin, respectively.
We optimise fitting parameters across all models using the Levenberg-Marquardt algorithm, implemented by \texttt{lmfit} \citep{newville2014}.
We also use \texttt{lmfit} for the fourth-order polynomial model fitting (see section \ref{first step: estimation of transit parameters}).
For optimising fitting parameters of the ringless and ringed model, we set the initial, minimum, and maximum values as indicated in table \ref{tab: ringless model parameters setting} and \ref{tab: ringed model parameters setting}, respectively. 
In table \ref{tab: ringed model parameters setting}, the initial values of $R_{\rm p}/R_{\star}$ and $a/R_{\star}$ are varied as fitting parameters around the best-fit values derived from the ringless model.
We fix the planetary orbital period in day $P_{\rm{orb}}$, for both models. 
In the case of the ringless model fitting, we adjust six free parameters: $q_1$, $q_2$, $t_0$, $R_{\rm p}/R_{\star}$, $a/R_{\star}$ and $b$.
For the ringed model, we have nine free parameters: $q_1$, $q_2$, $R_{\rm p}/R_{\star}$, $a/R_{\star}$, $b$, $\theta$, $\phi$ and $r_{\rm out/in}$.
We set the initial values of $R_{\rm p}/R_{\star}$ and $a/R_{\star}$ by varying the fitting parameters around the best fit derived from the ringless model.
To simplify the model, we should assume $r_{\rm{in}/\rm{p}}=1.0$.
However, since our code generates an error when $r_{\rm{in}/\rm{p}} = 1.0$, we adopt $r_{\rm{in}/\rm{p}} = 1.01$. 
This slight adjustment does not qualitatively alter the interpretation of our results.

\begin{table}
\tbl{Ringless model parameters setting.}{%
    \begin{tabular}{cccc}
\hline
Parameter & Initial value & Min & Max \\    
\hline
$q_1$ & 0.0 -- 1.0 & 0.0 & 1.0 \\
$q_2$ & 0.0 -- 1.0 & 0.0 & 1.0 \\
$t_0$ & $-0.05$ -- 0.05 & $-0.5$ & 0.5 \\
$R_{\rm p}/R_{\star}$ & 0.01 -- 0.5 & 0.01 & 0.7 \\
$a/R_{\star}$ & 1 -- 100 & 1 & 30000 \\
$b$ & 0 -- 1.0 & 0.0 & 1.0 \\
\hline
    \end{tabular}}
    \label{tab: ringless model parameters setting}
\end{table}

\begin{table}
\tbl{Ringed model parameters setting.}{%
    \begin{tabular}{cccc}
\hline
Parameter & Initial value & Min & Max \\
\hline
$q_1$ & 0.0 -- 1.0 & 0.0 & 1.0 \\
$q_2$ & 0.0 -- 1.0 & 0.0 & 1.0 \\
$t_0$ & $-0.05$ -- 0.05 & $-0.5$ & 0.5 \\
$R_{\rm p}/R_{\star}$ & $0.5 \times R_{\rm p}/R_{\star}$ -- $ 0.7 \times R_{\rm p}/R_{\star}$ & 0.001 & 0.5 \\
$a/R_{\star}$ & $a/R_{\star} $ -- $ 1.5\times a/R_{\star}$ & 1 & 200 \\
$b$ & 0.0 -- 1.0 & 0.0 & 1.0 \\
$\theta$ & $0$ -- $\pi$ & ... & $\pi$ \\
$\phi$ & $0$ -- $2\pi$ & ... & $2\pi$ \\
$r_{\rm out/in}$ & 1.00 -- 1.80 & 1.00 & 1.80 \\
\hline
\end{tabular}}\label{tab: ringed model parameters setting}
\end{table}

The ringed model requires higher computational resources than the ringless model, and it is more likely to converge to local minima during optimisation.
Our simplification of the model ($\tau=\infty$ and $r_{\rm{in}/\rm{p}} = 1.01$) allows for efficient optimisation by limiting the parameter space.
This approximated model can detect ring signals reasonably well, though not perfectly if complicated signals are present in the data.

To test whether ring signals are statistically significant, we perform a statistical test using $F$ statistics and calculate the $p$-value defined by equations (4) and (5) following \citet{2018AJ....155..206A},

\begin{equation}\label{eq: F_obs}
F_{\rm obs}=\frac{\left(\chi_{\rm{ringless, min}}^2-\chi_{\rm{ring, min}}^2\right) /\left(N_{\rm ring}-N_{\rm ringless}\right)}{\chi_{\rm {ring, min }}^2 /\left(N_d-N_{\rm {ring}}-1\right)} ,
\end{equation}
\begin{equation}\label{eq: p_value}
p=1-\int_0^{F_{\rm obs}} F\left(f \mid N_d-N_{\rm ring}-1, N_{\rm ring}-N_{\rm ringless}\right) df .
\end{equation}

Here, $\chi_{\rm{ringless, min}}^2$ and $\chi_{\rm{ring, min}}^2$ are the minimum $\chi^2$ values obtained from the ringless and ringed model fittings, $N_{\rm ringless}$ and $N_{\rm ring}$ are the numbers of free parameters in the ringless and ringed models, and $N_d$ is the number of bins, respectively. 
We define the significance of ring signals as cases where the $p$-value $< 0.0027$, representing the $3\sigma$ limit.

\section{Target Selection}\label{Target Selection}
Since ring signals are generally tiny, with amplitudes less than 1\% of transit depth when the ring size is comparable to the planetary radius, we need to pick up targets which have good statistical quality for the following intensive analysis, resulting in the selection of 308 TOIs.

\subsection{Optimistic estimation on ring detectability} \label{Optimistic estimation on ring detectability}
To avoid overlooking ringed planet candidates, we optimistically select our targets based on ring detectability.
Ring signals are roughly proportional to the signal-to-noise $(S/N)$ ratio of the planetary transit, which is primarily determined by the transit depth $\delta$, transit duration $T_{\rm dur}$, uncertainty of transit depth, and the cadence. 
If we rebin the data to have $N_d$ points with a fixed time range covering the transit, $(S/N)$ is solely determined by the revised uncertainty $\sigma (N_d)$, the noise amplitude for the rebinned data with $N_d$ points, and $\delta$.

Even with fixed ($\delta, \sigma (N_d))$, the ring signal also depends on the ring parameters. 
Thus, to determine the optimistic detectability of the ring, we investigate ring parameters that maximize the signal in this section.

\subsubsection{Optimising ring orientation and impact parameter that maximize ring detectability} \label{Optimising ring orientation and impact parameter that maximize ring detectability}
It is known that ring signals can be high for ringed planets with large impact parameters $b$ and moderate values for $\theta$ and $\phi$ \citep[e.g.,][]{2004ApJ...616.1193B, 2009ApJ...690....1O, 2015ApJ...814...81H}. 
To verify parameter values that maximise ring signals in our ringed model, we set
$\delta=0.01$ and $\sigma (N_d=500)=0.0001$,  to be a reference parameter setting and find the values of $\theta$, $\phi$, and $b$ that maximise $\Delta \chi^2$, as

\begin{equation}\label{eq: delta chisquare}
\Delta \chi^2 \equiv \chi_{\rm{ringless,min}}^2 - \chi_{\rm{ring,min}}^2
\end{equation}

We also set $r_{\rm{out}/\rm{in}} = 1.80$ assuming the ring density is 2 g $\rm{cm}^{-3}$, based on iron-nickel composition, and the planetary density is 0.9 g $\rm{cm}^{-3}$, typical for Jovian planets in the Solar System, with densities from 0.7 to 1.6 g $\rm{cm}^{-3}$ \citep{2005AREPS..33..493G}. 
Next, we vary $\theta$ and $\phi$ from 0$^\circ$ to 90$^\circ$ with an interval of 15$^\circ$, and vary $b$ from 0.0 to 1.0 with an interval of 0.1, simulating 539 light curves and fitting both ringless and ringed models.
In this simulation, we fix the following transit parameters, which are typical values for close-in planets and identical to those of TOI-495.01, as given by \citet{2020A&A...635A.205B}: $q_1=0.26$, $q_2=0.36$, $t_0=0$, $P_{\rm{orb}}=1.27$ and $a/R_{\star}=3.81$ and adjust $R_{\rm p}/R_{\star}$ to achieve the same $\delta$ value against the given $\theta, \phi$ and $b$ values.
To cover the transit, the length of the binning range is chosen to be $1.4$  $T_{\rm dur}$. 

\begin{figure}
  \centering
  \includegraphics[width=80mm]{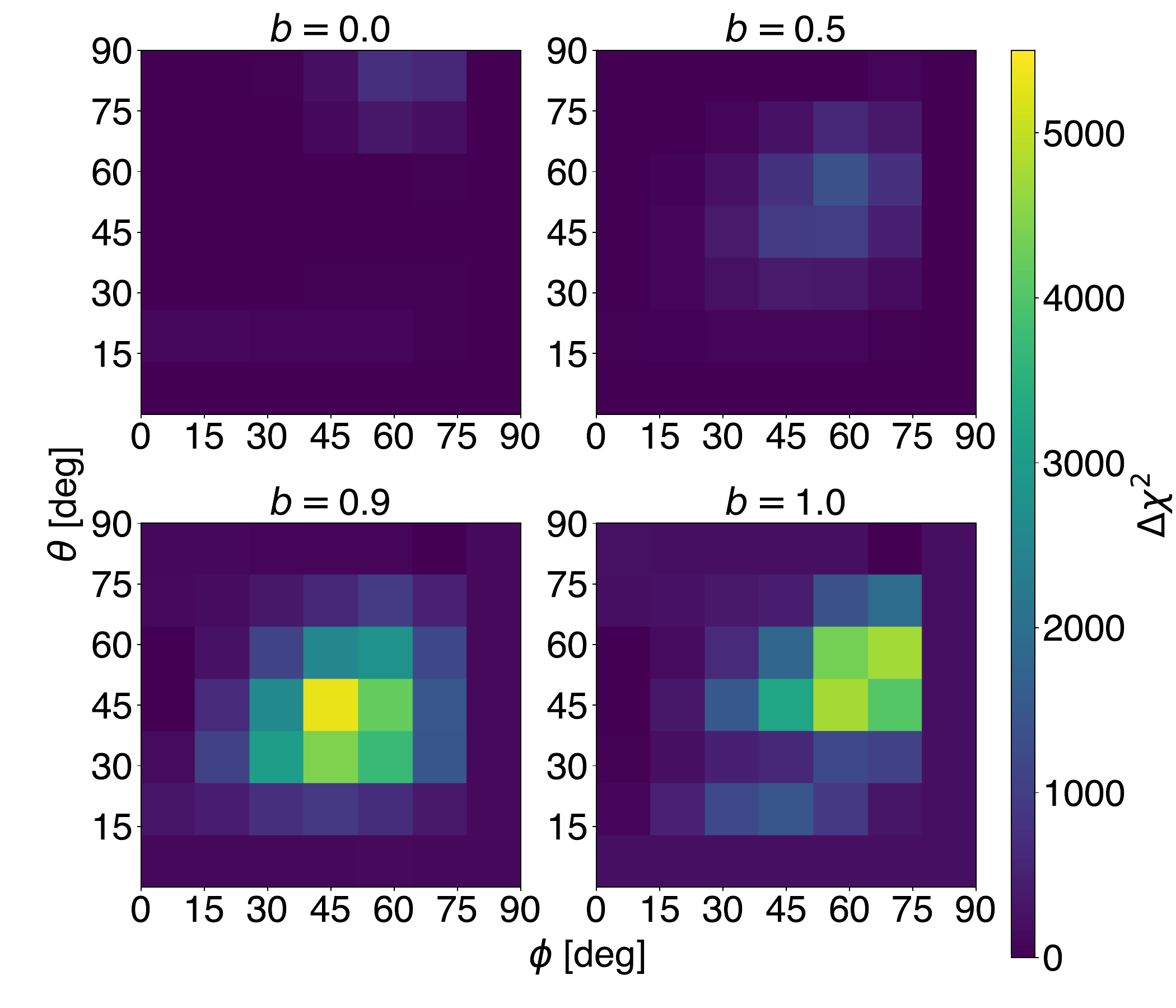}
  \caption{Variations of $\Delta \chi^{2}$, equation (\ref{eq: delta chisquare}), at each corresponding value of $\theta$, $\phi$ and $b$. The larger $\Delta \chi^{2}$ indicates that ring signals can be detected more significantly.}
  \label{fig: most detectable angle}
\end{figure}

Finally, as shown in figure \ref{fig: most detectable angle}, we find the combination of $\theta=45^\circ, \phi = 45^\circ$, and $b=0.9$ is the most suitable for ring detection in our ringed model, consistent with previous studies.
We note that this is an optimistic estimate of detectability based solely on $\delta$, $T_{\rm dur}$, and $\sigma (N_d)$. 
Ideally, prior information for $b$ can be used to refine detectability and narrow down the targets. 
However, we adopt a rougher approach to avoid missing potential targets.

\subsubsection{Ring detectability with F statistics under given transit depth and duration}\label{Ring detectability with F statistics under given transit depth and duration}
Given the parameters $\theta=45^\circ, \phi = 45^\circ$, and $b=0.9$, we consider a threshold for the detectable ring signal assuming $(\delta, \sigma (N_d))$ under the given $T_{\rm dur}$, which is directly obtained from the observation as shown in figure \ref{fig: detectable area}. 
We describe the selection of TOIs in section \ref{Selection of TESS planet candidates}.

To achieve the $5\sigma$ detection, $\Delta \chi^2$ needs to exceed $\Delta \chi_{\rm{5\sigma}}^2$ that is the value of $\Delta \chi^2$ at $5\sigma$ derived from equation (\ref{eq: p_value}).
Assuming $\chi_{\rm{ring, min}}^2=500$, the same as $N_d$, we obtain $\Delta \chi_{5\sigma}^2\simeq34$ using equation (\ref{eq: F_obs}) and (\ref{eq: p_value}).
We fit the ringless transit model to simulate light curves with $\sigma (N_d=500)=0.0001$ and $\delta$ set at 0.001, 0.010, 0.020, 0.030, 0.040, 0.050, 0.060, 0.070, 0.080, 0.090, 0.100 and 0.200, and obtain each $\Delta \chi^2$. 
We obtain uncertainty at $5\sigma$ limit $\sigma_{\rm{5\sigma}}$ for each $\delta$ and interpolated these values to set criteria in the $\delta$-$\sigma$ plane using the scaling relation given as 

\begin{equation}\label{eq: scaling sigma}
    \sigma_{\rm{5\sigma}} = \sigma (N_d=500) \left(\frac{\Delta \chi^2}{\Delta \chi_{\rm{5\sigma}}^2}\right)^\frac{1}{2}.
\end{equation}

For values of $\delta$ less than 0.001 or greater than 0.200, $\sigma_{\rm{5\sigma}}$ is extrapolated from the slope of the preceding two data points.
In these ranges of $\delta$, there are few TOIs that can serve as targets, so this extrapolation does not affect the target selection.

\subsection{Selection of TESS planet candidates}\label{Selection of TESS planet candidates}
We select TOIs released by the Science Processing Operations Center (SPOC, \citet{2016SPIE.9913E..3EJ}) from the ExoFOP-TESS TOI list on 2022 September 13  (hereafter, ETL22) \footnote{https://exofop.ipac.caltech.edu/tess}, excluding those with TESS Disposition of ``EB'' (Eclipsing Binary), TESS Follow-up Observing Program Working Group Disposition of ``FP'' (False Positive) or ``FA'' (False Alarm).
We choose the 2 min cadence data among the 20 s, 2 min, and 30 min options, as it provides sufficient time resolution to detect ring signals and more observation data of objects compared to the 20 s one.

In selecting the target TOIs which satisfy our criteria described in section \ref{Optimistic estimation on ring detectability}, we retrieve $\delta$ from ETL22 and estimate $\sigma (N_d=500)$ with the following equation (\ref{eq: sigma from cdpp}) for each TOI,

\begin{equation}\label{eq: sigma from cdpp}
    \sigma (N_d=500) = \rm{CDPP}\,(1\,\rm{hr}) \sqrt{\frac{1 hr}{1.4 \it{T}_{\rm{dur}}/\rm{500}}} \frac{1}{\sqrt{\it{N}_{\rm transit}}}.
\end{equation}

Here, $\rm{CDPP}\,(1\,\rm{hr})$ is the root mean square-combined differential photometric precision for 1 hr, indicating the estimated average $1\sigma$ uncertainty.
$N_{\rm transit}$ is the number of observed transits for each TOI. 
We scale values of $\rm{CDPP}\,(1\,\rm{hr})$ to obtain $\sigma (N_d=500)$ for the time range per bin when dividing 1.4 $T_{\rm dur}$ of each TOI into 500 bins (see section \ref{Optimising ring orientation and impact parameter that maximize ring detectability} for detail). 
Then, we exclude TOIs for which $\rm{CDPP}\,(1\,\rm{hr})$ cannot be obtained from target selection.

\begin{figure}
  \centering
  \includegraphics[width=80mm]{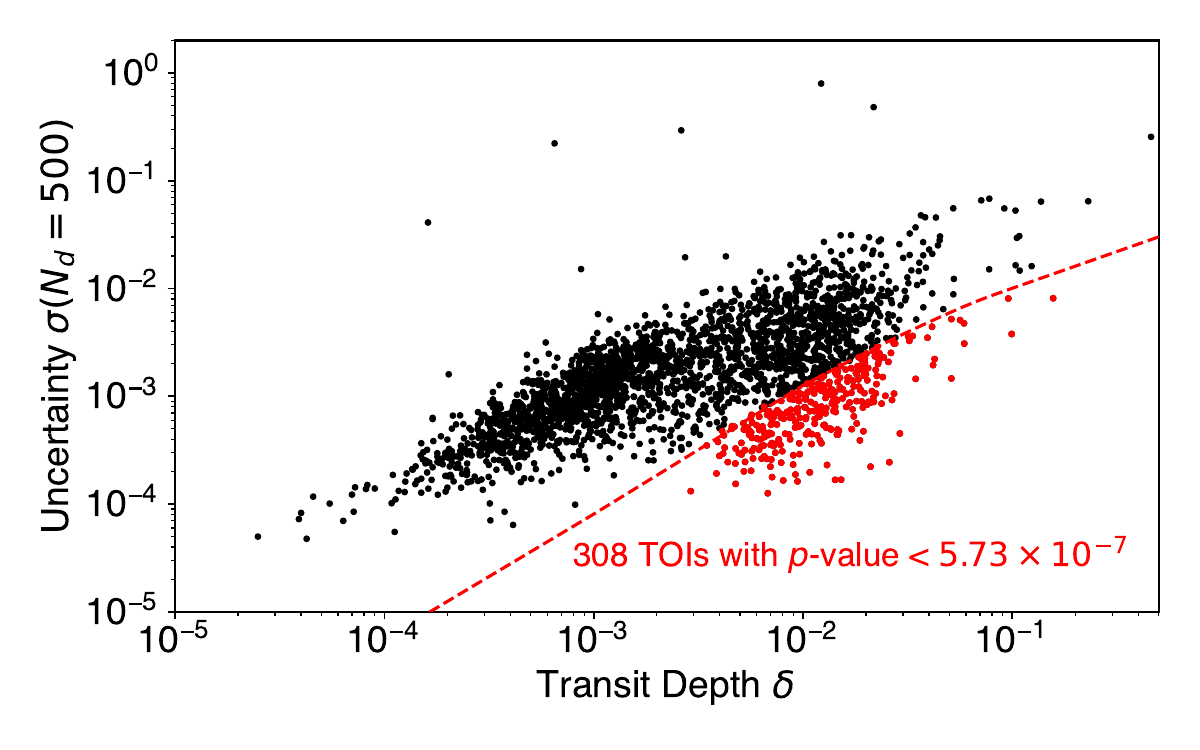}
  \caption{Criteria for detecting ring signals, roughly determined by assessing the transit depth and uncertainty in each bin for 2691 TOIs. Red dots under the dashed red curve indicate objects whose data quality satisfies $p$-value $<5.73\times {10^{-7}}$.}
  \label{fig: detectable area}
\end{figure}

In figure \ref{fig: detectable area}, the dashed red curve indicates $p$-value$=5.73 \times 10^{-7}$, representing $5\sigma$ limits. 
The red points located below the dashed red line are TOIs for which rings can be detected under the most optimistic ring parameters.
Applying our $5\sigma$ detection criteria, 308 TOIs satisfy these conditions out of a total of 2691 TOIs. 

\section{Data Analysis} \label{Data Analysis}
\begin{figure*}
  \centering
  \includegraphics[width=160mm]{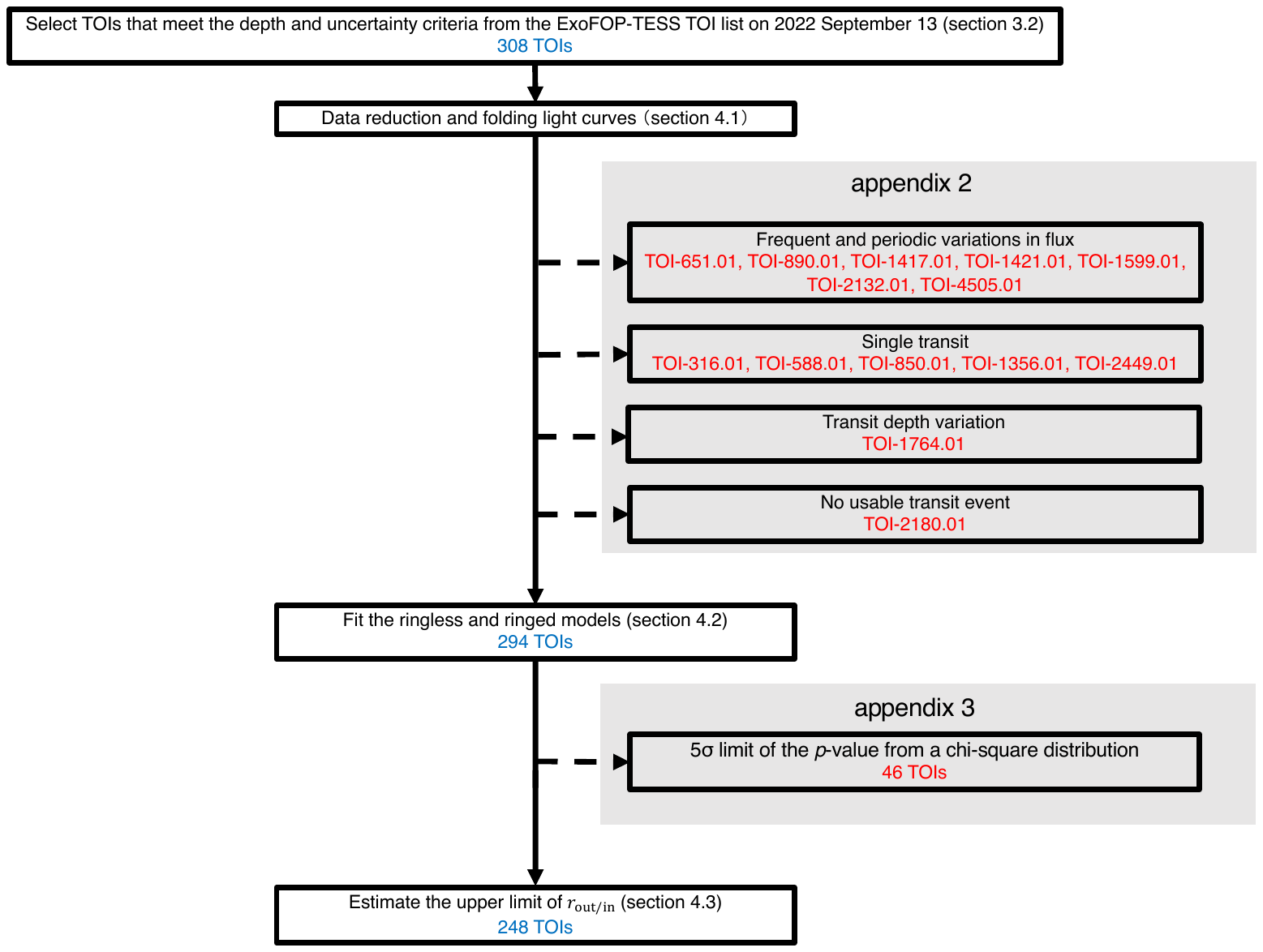}
  \caption{Flowchart of our entire analysis. In each step, the analysis procedure and the target objects are presented. In appendix section, the objects excluded from our analysis and the reasons for their exclusion are presented. In appendix \ref{Exceptional cases}, we describe each case in detail, while in appendix \ref{Objects with p-value<0.0027 in excepted TOIs}, we discuss the objects among the 50 that exhibit interesting signals.
  }
  \label{fig: analysis_flow}
\end{figure*}

In this section, we describe the data processing and model comparison used to detect ring signals from the light curve of 308 TOIs selected in section \ref{Target Selection}.
Figure \ref{fig: analysis_flow} shows the flowchart of our entire analysis. 
We reduce noises and folded light curves of each TOI in section \ref{Data reduction and making phase-folded light curve}, compare the ringless and ringed model against the data in section \ref{Comparison of fitting results for models with and without rings} and derive the upper limits of $r_{\rm out/in}$ in section \ref{Obtain upper limit of r_out}.

\begin{figure*}
  \centering
  \includegraphics[width=155mm]{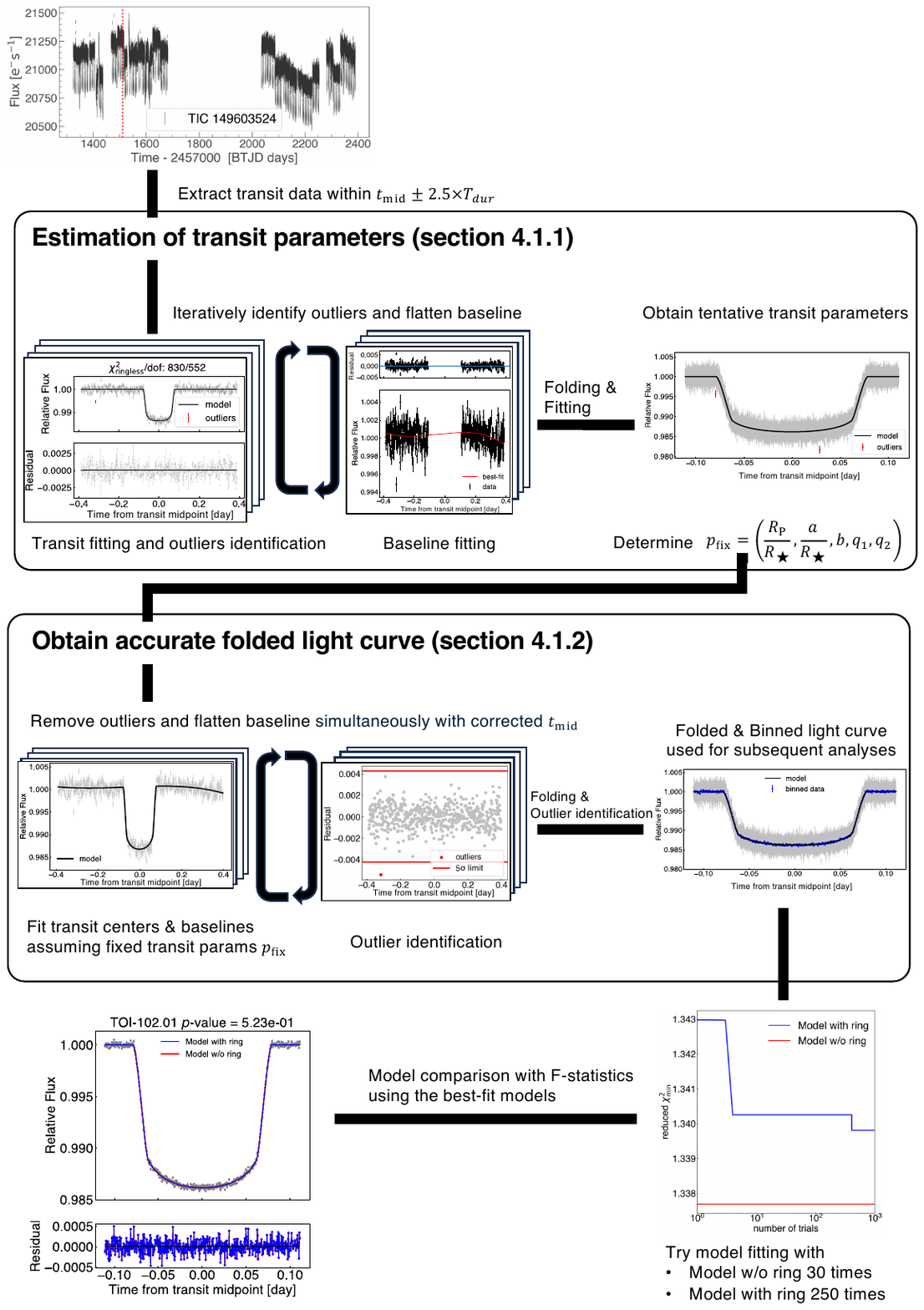}
  \caption{Detailed procedure from initial processing of each transit data to the derivation of the significance of ring signals in the case of TOI-102.01.}
  \label{fig: preprocess_flow}
\end{figure*}

\subsection{Data reduction and phase-folded light curves}\label{Data reduction and making phase-folded light curve}
This section describes the preprocessing step for 308 TOIs prior to fitting with the ringless and ringed models in section \ref{Comparison of fitting results for models with and without rings}. 
To accurately correct instrumental or stellar noises, we divide analyses into two steps as detailed in section \ref{first step: estimation of transit parameters} and section \ref{second step: correct light curve using transit parameters}, as illustrated in figure \ref{fig: preprocess_flow}.

In this study, we analyse the Pre-search Data Conditioning Simple Aperture Photometry (PDC-SAP) flux. 
Additionally, we also employ SAP flux for the comparison and confirm that the results are consistent (see appendix \ref{SAP flux analysis}). 

\subsubsection{First step: estimation of transit parameters} \label{first step: estimation of transit parameters}
For light curves of each TOI, we extract the time range of all transit events sufficient for fitting the baseline, $t_{\rm mid} \pm \, 2.5\, T_{\rm dur}$, by referring to known $t_{\rm mid}$, $P_{\rm{orb}}$ and $T_{\rm dur}$ given by ETL22.
If more than 10 \% of the data is missing due to data gaps in any extracted transit, we do not analyse the data to avoid inaccurate determination of baselines.
For multi-planetary systems, we exclude events where the transits of other planets overlap.

For the retrieved data, we employ the following preprocessing procedures.

\begin{enumerate}
\item Baseline flattening:
We fit a fourth-order polynomial model to the out-of-transit light curve with $|t-t_{\rm mid}|>0.7\; T_{\rm dur}$. 
We then normalise the light curve, including the transit, by dividing it by the derived polynomial. 

\item  Outlier removal by ringless model fitting: 
we fit the ringless model to the normalised light curve, and exclude $5\sigma$ outliers in the residuals, the difference between the data and the optimised model. 
In this ringless model fitting, we fix $P_{\rm{orb}}$ and optimise six free parameters, $q_1$, $q_2$, $t_0$, $R_{\rm p}/R_{\star}$, $a/R_{\star}$ and $b$. 
In fitting, we optimise parameters starting from 30 randomly generated initial values within the ranges shown in table \ref{tab: ringless model parameters setting} to pick up the best parameters. 
We confirm that the adopted number 30 is enough to obtain a reasonable solution, avoiding trapping in a local minimum.  
\item We repeat the above baseline flattening and outlier removal process until no outliers are detected. 
\end{enumerate}

For $i$-th transit event, we define $t_{\rm{mid, new}, \it{i}} = t_{\rm{mid}, \it{i}} - t_{0, \it{i}}$, where $t_{\rm{mid}, \it{i}}$ is the transit midpoint of the $i$-th event, and $t_{0, \it{i}}$ is the phase shift from $t_{\rm{mid}, \it{i}}$, and we use these values in section \ref{second step: correct light curve using transit parameters}.
Using $t_{ {\rm mid, new}, i}$ and derived baselines, we stack these light curves and fit them with the ringless model to update the planetary transit parameters and remove outliers. 
We refine $P_{\rm{orb}}$ and $T_{\rm dur}$, based on the series of $t_{\rm mid} - t_0$ and transit parameters, respectively.
In case only two transit events are usable in the light curves, we adopt $P_{\rm{orb}}$ value listed in ETL22.

In this step, we eliminate 14 TOIs from our targets because they cannot be handled by this systematic preprocessing (in detail, appendix \ref{Exceptional cases}).

\subsubsection{Second step: correct light curve using transit parameters} \label{second step: correct light curve using transit parameters}
Using the series of $t_{ {\rm mid, new}, i}$ and $T_{\rm dur}$ derived in section \ref{first step: estimation of transit parameters}, we re-extract each transit event and perform simultaneous optimisation for the transit model and the polynomial function while fixing the derived transit parameters except for $t_0$.
After determining baselines and transit centres, we stack light curves and fit the ringless model to the data. 
We remove $5\sigma$ outliers again and finally obtain the phase-folded light curve, which is then analysed by ringless and ringed models in section \ref{Comparison of fitting results for models with and without rings}.

\subsection{Comparison of fitting results for models with and without rings}\label{Comparison of fitting results for models with and without rings}
Using the phase-folded data produced in section \ref{Data reduction and making phase-folded light curve}, we fit ringless and ringed models for 294 TOIs to search for possible ring signals. 

To find ring signals in the light curve clearly, we extract a part of the light curve within transit centre $\pm \, 0.7 \, T_{\rm dur}$ and bin it into 500 bins by calculating the mean of the flux and the weighted mean of the errors within each bin at equal time intervals. 
This procedure results in a binning resolution of approximately 8--180 s for our targets, and if $N_{\rm transit}$ is small or $T_{\rm dur}$ is short, there may be fewer than 500 bins in the data.

To compare the goodness of fit between the ringless and ringed models, we optimise parameters for both models. 
The free parameters and the optimisation process of the ringless model are the same as those in section \ref{first step: estimation of transit parameters}. 
In the optimisation with the ringed model, we have nine free parameters, $q_1, q_2, t_0, R_{\rm p}/R_{\star}, a/R_{\star}, b, \theta, \phi$ and $r_{\rm out/in}$, as shown in table \ref{tab: ringed model parameters setting}.
We employ the parameter values obtained from the ringless model fitting as initial values for these six free parameters, $R_{\rm p}/R_{\star}$ and $a/R_{\star}$. 
We repeat the fitting process 250 times and confirm that this number is sufficient to avoid possible trapping in a local minimum and obtain a reasonable solution.

As described in section \ref{Optimistic estimation on ring detectability}, the comparison of the two models is performed using $F$ statistics and the $p$-value, defined in equation (\ref{eq: F_obs}) and equation (\ref{eq: p_value}) against the null hypothesis as in \citet{2018AJ....155..206A}.
In this paper, $N_{\rm ringless}=6$, $N_{\rm ring}=9$, and $N_d=500$.
We consider TOIs with $p$-value $< 0.0027$ as candidates for ringed planets.

For 46 TOIs in our targets, $\chi_{\rm{ringless, min}}^2$ and $\chi_{\rm{ring, min}}^2$ are similar values, both larger than 1.34 or smaller than 0.72, which is the $5\sigma$ limit of the chi-squared distribution. 
This suggests the under(over)estimation of uncertainties, which we consider observational errors associated with TESS light curves in our analysis. 
We suspect that the underestimation is due to systematic errors, but instead of investigating the cause, we exclude these systems from the primary analysis.
The result of the model fitting and the upper limits of $r_{\rm out/in}$ for these excluded TOIs are listed in table \ref{ringed model and ringless model fitting result in excepted TOIs} in appendix \ref{Objects with p-value<0.0027 in excepted TOIs}.

To verify that our analysis flow can detect ring signals, we simulate the 
light curve using the transit parameters of TOI-495.01, assuming $b=0.9$, and the ring orientation as discussed in section \ref{Optimising ring orientation and impact parameter that maximize ring detectability}.
We adjust $R_{\rm p}/R_{\star}$ to be consistent with the $\delta$ value of TOI-495.01 in ETL22 and generate light curves with an equal number of transit events for this planet. 
We add white noise with an amplitude of 0.1\% of the stellar flux and simulate stellar noise represented by a sinusoidal variation with an amplitude of 0.5\%. 
The period of the sinusoidal function is set to be twice the transit duration, with a randomly assigned phase for each transit.
Subsequently, we apply the procedures outlined in section \ref{Data reduction and making phase-folded light curve} and this section to the simulated data.
Our verification confirm that the pipeline effectively removes noise, enabling the significant detection of the assumed ring signals.

\subsection{Upper limit of $r_{\rm out/in}$}\label{Obtain upper limit of r_out}
We estimate an upper limit of $r_{\rm out/in}$, $(r_{\rm out/in})_{\rm{upp}}$, for 248 TOIs.
We adopt the similar approach to \citet{2018AJ....155..206A} in estimating $(r_{\rm out/in})_{\rm{upp}}$.
We define $(r_{\rm out/in})_{\rm{upp}}$ as the value of $(r_{\rm out/in})$ at the $3\sigma$ upper limit, where $\Delta \chi^2_{\rm limit} = 9$, calculated using equation (\ref{eq: delta chisquare limit}). 

\begin{equation}\label{eq: delta chisquare limit}
\Delta \chi^2_{\rm limit} \equiv \frac{\left(\chi_{\rm{ring,min}}^2 - \chi_{\rm{ringless,min}}^2\right)}{\left(\chi_{\rm{ringless,min}}^2 / \rm{dof}\right)}
\end{equation}

To find $(r_{\rm out/in})_{\rm{upp}}$ that satisfies this criterion, we fit the ringed model to the light curve while fixing $r_{\rm out/in}$ at various values.
The value of $r_{\rm out/in}$ is varied among nine values from 1.00 to 2.00 at an interval of 0.05 during the procedure.
In fitting, we fix the following ring parameters: $r_{\rm{in} / \rm{p}}=1.01$, $\theta=45^\circ$ and $\phi=45^\circ$. 
On the other hand, we adjust five remaining free parameters: $q_1, q_2, R_{\rm p}/R_{\star}, a/R_{\star}$ and $b$.
The $\Delta \chi^2_{\rm limit}$ values are interpolated between the ones obtained from the model fitting by fixing $r_{\rm out/in}$ at various values.

\begin{figure}
  \centering
  \includegraphics[width=80mm]{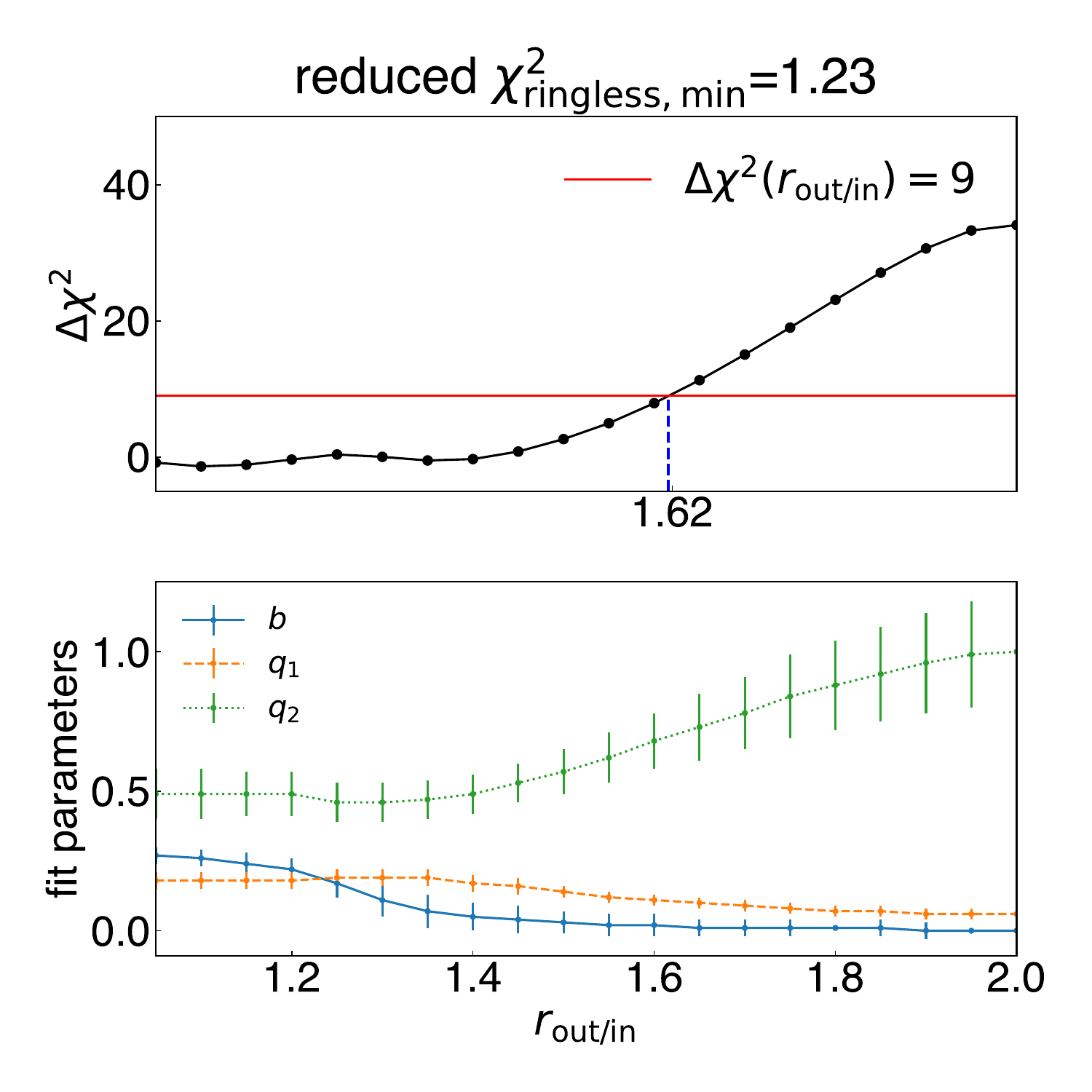}
  \caption{(Upper panel) Estimation of $(r_{\rm out/in})_{\rm upp}$ for TOI-398.01. Black curve indicates $\Delta \chi^2$ against $r_{\rm out/in}$. In this case, $(r_{\rm out/in})_{\rm upp}=1.62$. (Lower panel) Variation of $b$, $q_1$, and $q_2$ with $r_{\rm out/in}$ among the fitting parameters. A significant increase in $\Delta \chi^2$ is observed as $b$ approaches zero.}
  \label{fig: upper limit}
\end{figure}

For instance, we show the increment of $\Delta \chi^2_{\rm limit}$ against $(r_{\rm out/in})_{\rm{upp}}$ for TOI-398.01 and the ringless model fitting to the light curve in figure \ref{fig: upper limit}.  
In the upper panel, red line represents the threshold which determines the $3\sigma$ upper limit. 
 $r_{\rm out/in}$ at the intersection point with the red line is taken as $(r_{\rm out/in})_{\rm{upp}}$ for the object.
In this case, $(r_{\rm out/in})_{\rm{upp}}$ is found to be 1.62.
The lower panel illustrates $b$, $q_1$, and $q_2$, showing significant variation with the change of $r_{\rm out/in}$. 

As shown in \citet{2004ApJ...616.1193B}, as $b$ decreases, ring signals on the whole transit shape would diminish because ingress and egress become relatively minor parts of the light curve. 
In figure \ref{fig: upper limit}, the increase in $\Delta \chi^2$ associated with the increase in $r_{\rm out/in}$ is indeed suppressed by the decrease in $b$.
Consequently, when we fit the ringed model to the transit light curve without rings to determine an upper limit for $r_{\rm out/in}$, the fitting process prefer to decrease the $b$ value as the assumed $r_{\rm out/in}$ increases. 

\section{Result} \label{Result}
\subsection{No definite candidates for ringed planets}
The transit light curves of the 294 TOIs are fitted with both the ringless and ringed models to evaluate the significance of ring signals as outlined in section \ref{Data Analysis}. 
As a result, there are six TOIs with $p$-value $< 0.0027$ among our targets, except for the 46 TOIs where the noise is underestimated. 
Figure \ref{fig: Light curves of 5 systems} shows the comparison of the best-fit results for the ringless and ringed models in these six TOIs.
Here, we divide the light curves into 100 bins and investigate the time variation of the residuals in detail.
However, no significant ring signals are present in the residuals for any of the systems.
We also perform a visual inspection of all fitting results, such as the bottom-left plot in figure \ref{fig: preprocess_flow}, and cannot find any TOI with signals resembling rings.
This indicates that rings are not detected in the high $(S/N)$ planets targeted in this study.
We summarise the fitting results of the systems with $p$-value $< 0.0027$ in table \ref{tab: Fitting results for the systems with 3 sigma p-value}, the others in table \ref{tab: ringed model and ringless model fitting result}.

\begin{table*}
\rotatebox{90}{%
\begin{minipage}{\textheight} 
\begin{threeparttable}
\caption{Fitting results for the systems with $p$-value $< 0.0027$ using both ringless and ringed models.}
\label{tab: Fitting results for the systems with 3 sigma p-value}
\begin{tabular}{crrrcrrrr}
\hline
\multicolumn{1}{c}{TOI} & \multicolumn{1}{c}{$P_{\rm{orb}}$ [day]} & \multicolumn{1}{c}{$b^{*}$} & \multicolumn{1}{c}{$(R_{\rm p} / R_{\star})_{\rm{ringless}}^{*}$} & \multicolumn{1}{c}{$(r_{\rm out/in })_{\rm{upp}}$} & \multicolumn{1}{c}{$\delta$ [ppm]} & \multicolumn{1}{c}{$\sigma (N_d=500)$ [ppm]} & \multicolumn{1}{c}{$(\chi^{2}_{\rm{ringless,min}},\chi^{2}_{\rm{ring,min}}, N_d)$} & \multicolumn{1}{c}{$p$-value} \\

\hline
182.01 & 1.769180 $\pm$ 1e-06 & ... & ... & ... & 9570 $\pm$ 71 & 523 & (519.24, 503.45, 500) & 1.69e-03 \\
189.01 & 2.194064 $\pm$ 5e-06 & ... & ... & ... & 6440 $\pm$ 72 & 488 & (659.99, 636.76, 500) & 5.38E-04 \\
418.01 & 2.235985 $\pm$ 1e-06 & 0.7 $\pm$ 0.1 & 0.13 $\pm$ 0.03 & 1.45 & 19100 $\pm$ 132 & 896 & (412.16, 396.73, 446) & 8.33e-04 \\
1461.01 & 3.5687 $\pm$ 4e-04 & 0.85 $\pm$ 0.02 & 0.143 $\pm$ 0.007 & ... & 20100 $\pm$ 283 & 2493 & (327.22, 315.66, 408) & 2.49e-03 \\
2126.01 & 1.3061860 $\pm$ 4e-07 & ... & ... & 1.52 & 24700 $\pm$ 154 & 1009 & (490.66, 472.75, 500) & 3.94e-04 \\
2154.01 & 3.82408 $\pm$ 1e-05 & 0.7 $\pm$ 0.1 & 0.10 $\pm$ 0.02 & ... & 10900 $\pm$ 107 & 884 & (484.92, 467.50, 500) & 4.52e-04 \\
\hline
\end{tabular}
\begin{tablenotes}
\item[*] Due to the high uncertainties, the parameter value cannot be determined accurately and is therefore denoted as ``...''.
\end{tablenotes}
\end{threeparttable}
\end{minipage}
}

\end{table*}

\begin{figure*}[width=160mm]
  \centering
  \begin{minipage}{0.49\textwidth}
    \centering
    \includegraphics[width=\textwidth]{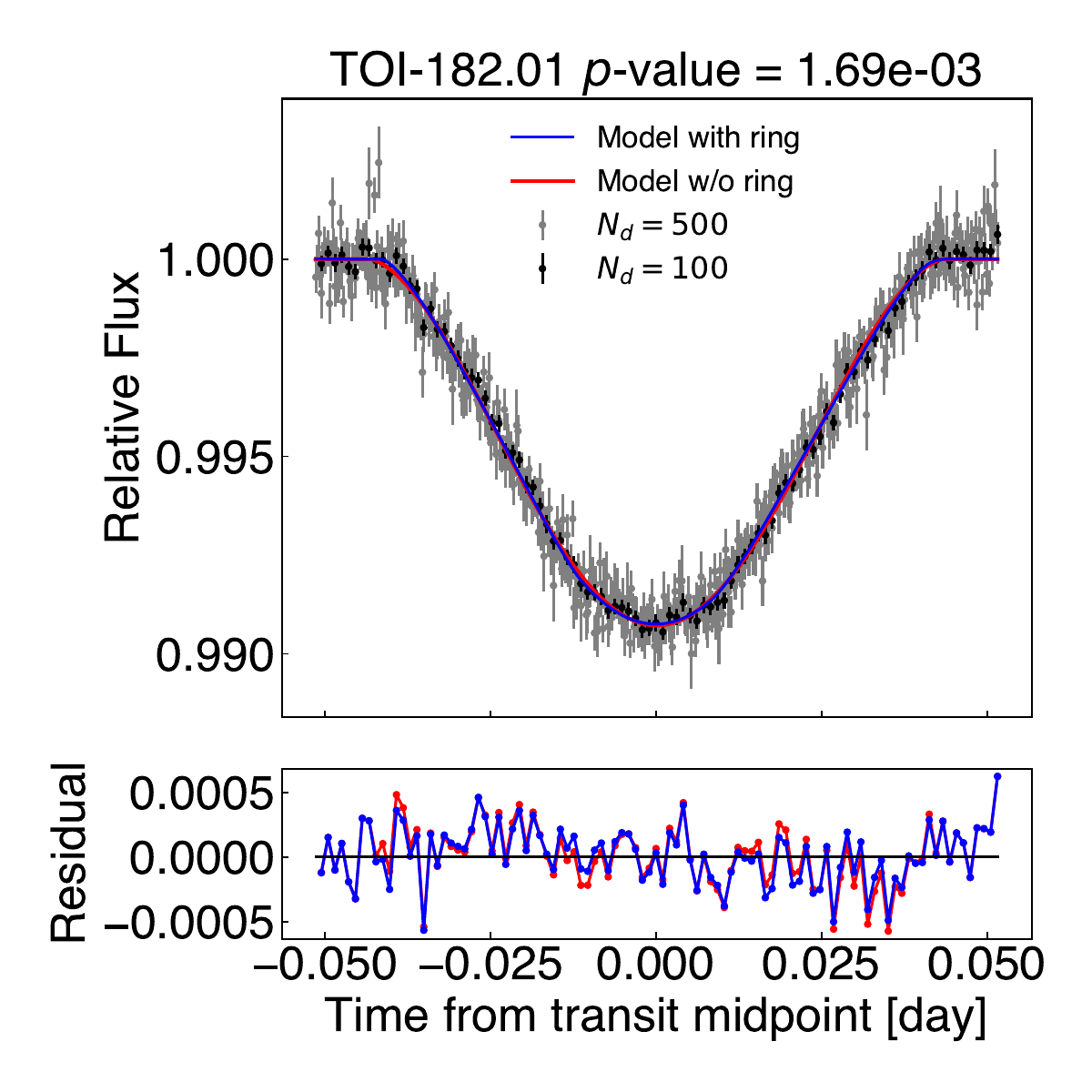}
  \end{minipage}
  \hfill
  \begin{minipage}{0.49\textwidth}
    \centering
    \includegraphics[width=\textwidth]{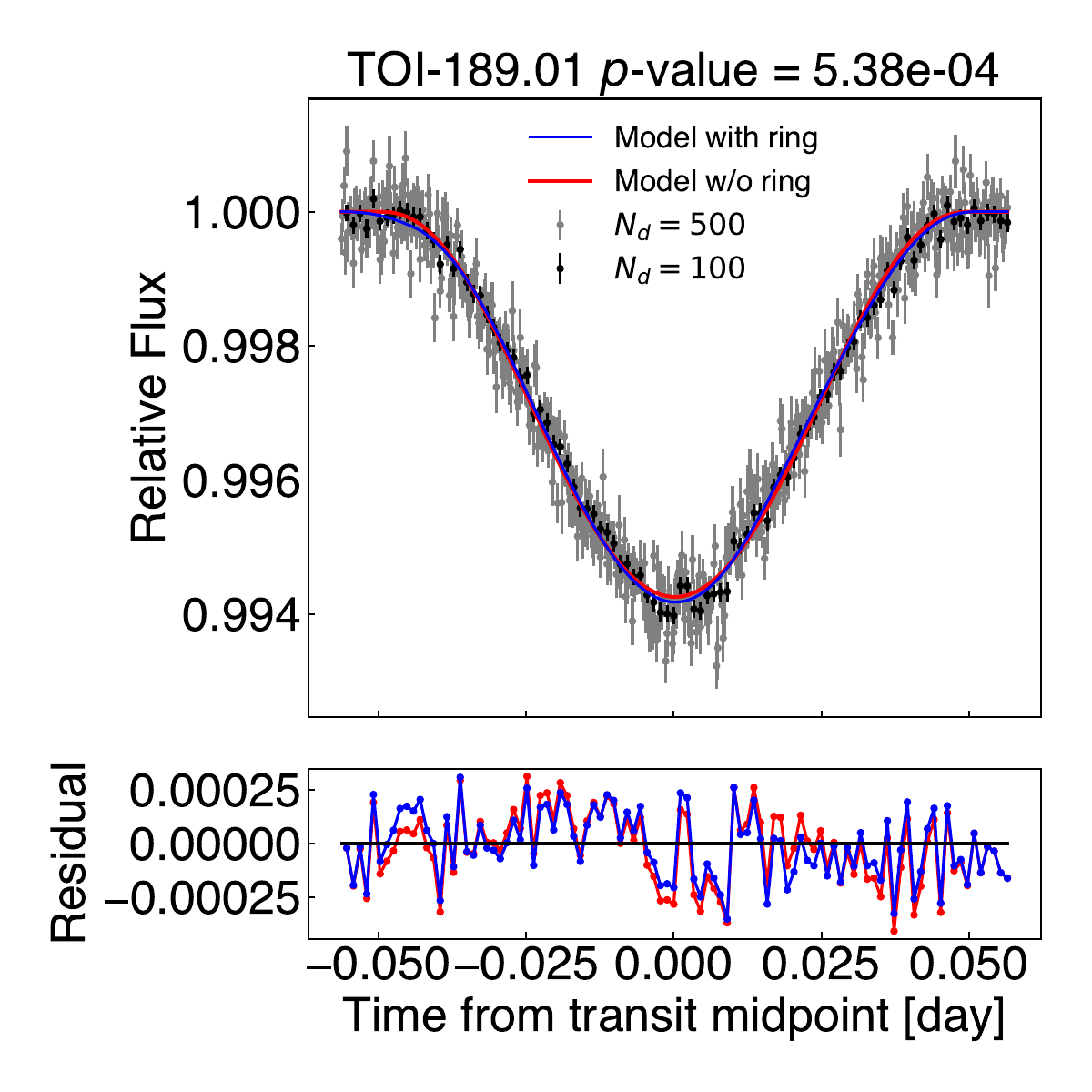}
  \end{minipage}
  \vspace{0.5cm}
  \begin{minipage}{0.49\textwidth}
    \centering
    \includegraphics[width=\textwidth]{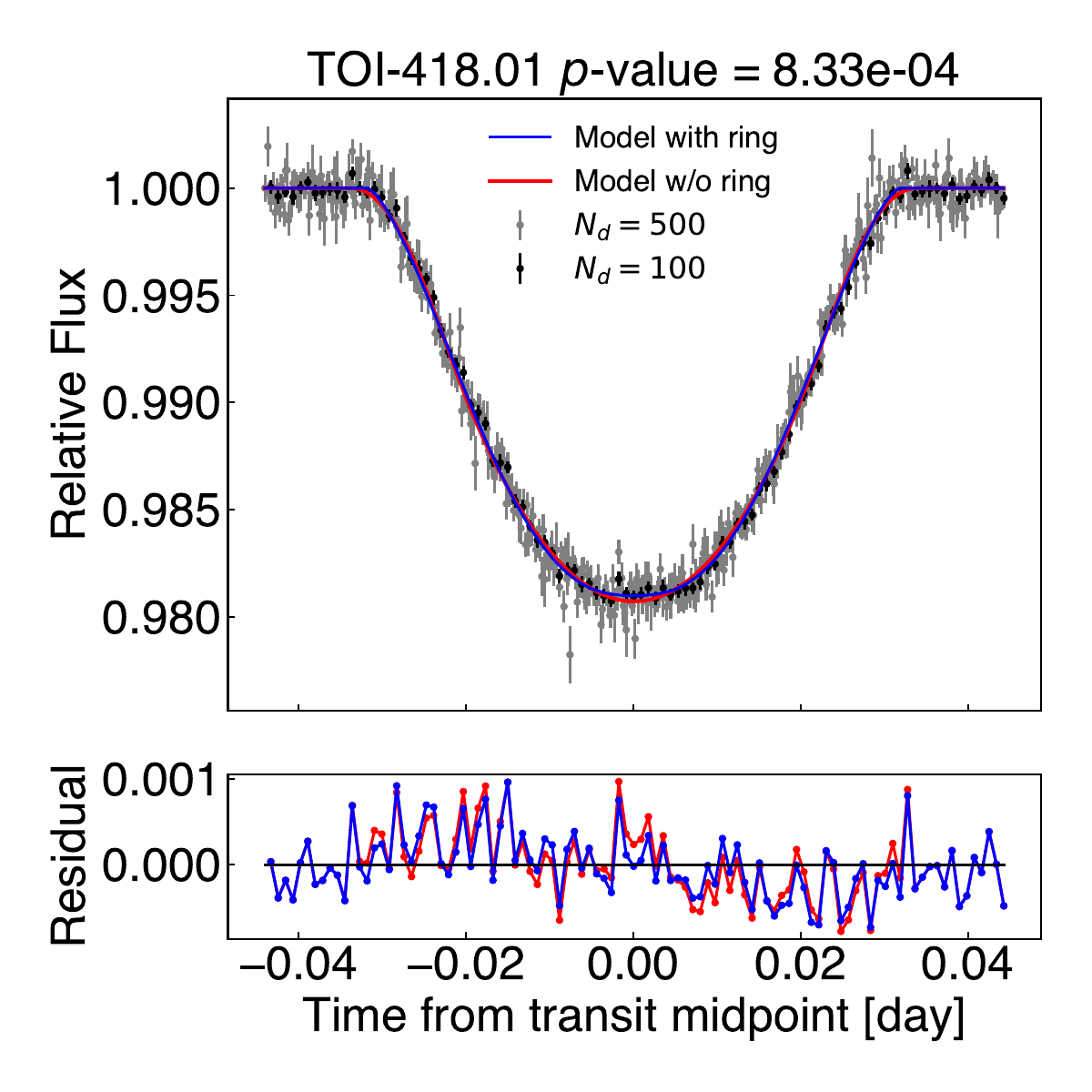}
  \end{minipage}
  \hfill
  \begin{minipage}{0.49\textwidth}
    \centering
    \includegraphics[width=\textwidth]{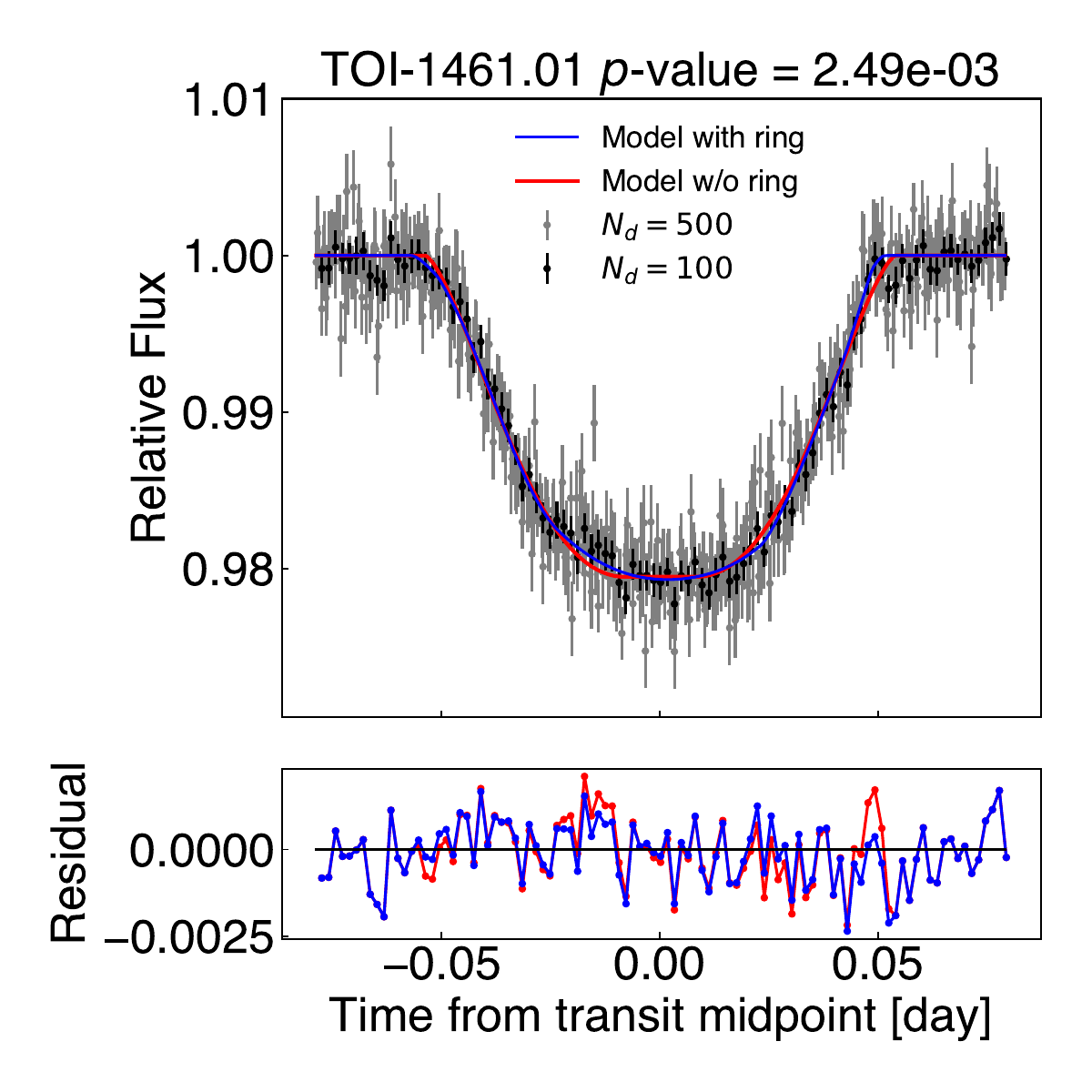}
  \end{minipage}
  
  \caption{Light curves of six systems that satisfy $3\sigma$ detection threshold. The residuals shown in the lower panels are calculated for $N_d = 100$.}
  \label{fig: Light curves of 5 systems}
\end{figure*}

\begin{figure*}[width=160mm]
\ContinuedFloat
  \centering
  \begin{minipage}{0.49\textwidth}
    \centering
    \includegraphics[width=\textwidth]{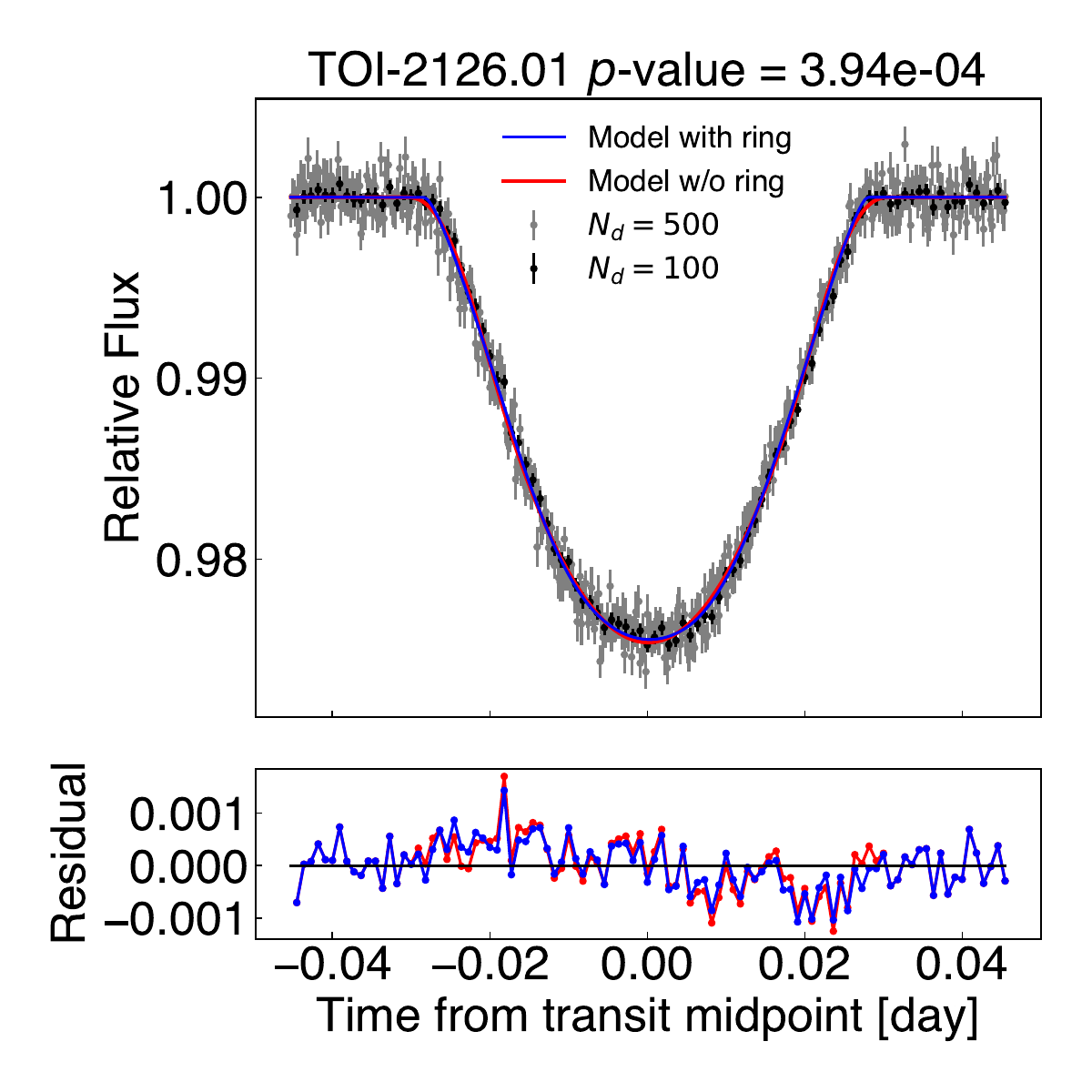}
  \end{minipage}
  \hfill
  \begin{minipage}{0.49\textwidth}
    \centering
    \includegraphics[width=\textwidth]{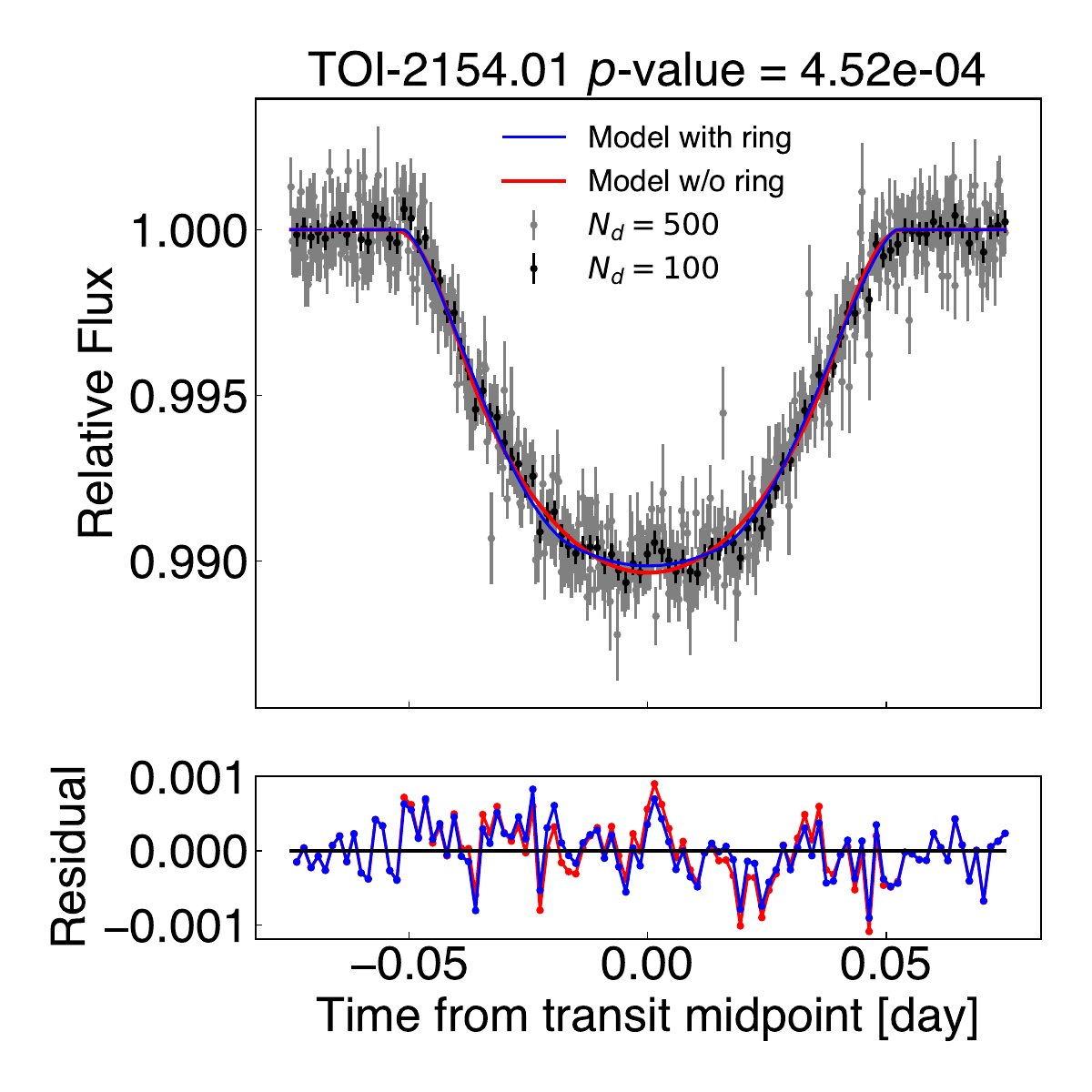}
  \end{minipage}
  
  \caption{(Continued)}
\end{figure*}

\subsection{Upper limit of $r_{\rm out/in}$}\label{result: Upper limits of ring parameters}
\begin{figure}
  \centering
  \includegraphics[width=80mm]{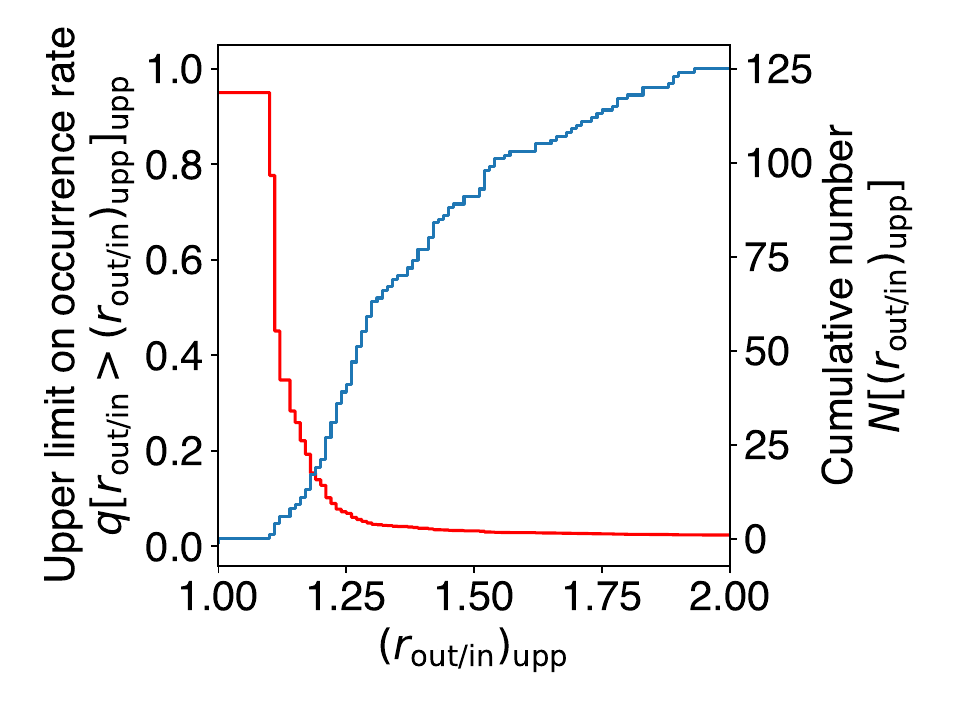}
  \caption{(Blue line) Cumulative distribution of the number of TOIs with upper limits on the ring outer radius smaller than $(r_{\rm{out/in}})_{\rm upp}$, denoted as $N[(r_{\rm out/in})_{\rm upp}]$. (red line) upper limits on the occurrence rate of rings larger than a given $(r_{\rm{out/in}})_{\rm upp}$  around a planet, denoted as $q[r_{\rm out/in}>(r_{\rm out/in})_{\rm upp}]_{\rm upp}$.}
  \label{fig: uplimit_cdf}
\end{figure}

Following the method in section \ref{Obtain upper limit of r_out}, we derive upper limits on the ring size in 125 TOIs, as presented in table \ref{tab: ringed model and ringless model fitting result}.
For TOIs where $r_{\rm out/in} = 2.00$ and $\Delta \chi^2_{\rm limit} < 9$, we cannot obtain upper limit and leave the $(r_{\rm out/in})_{\rm{upp}}$ column in table \ref{tab: ringed model and ringless model fitting result} blank.
The cumulative distribution of the number of TOIs against upper limits on the ring size, $N[(r_{\rm out/in})_{\rm{upp}}]$, and derived upper limits on ring occurrence rate for the ring size larger than $(r_{\rm out/in})_{\rm{upp}}$, $q[r_{\rm out/in}>(r_{\rm out/in})_{\rm upp}]_{\rm upp}$, are shown in figure \ref{fig: uplimit_cdf}, following \citet{2018AJ....155..206A}. 
To improve clarity, the previous notations $q[>(r_{\rm out/in})]_{\rm upp}  $  and $ N[(r_{\rm out/in})_{\rm{upp}}]$ in \citet{2018AJ....155..206A} have been modified to the current ones while maintaining the same meanings.


\section{Discussion} \label{Discussion}
\subsection{Comparison of ring survey results between Kepler and TESS observations}
\begin{figure}
  \centering
  \includegraphics[width=80mm]{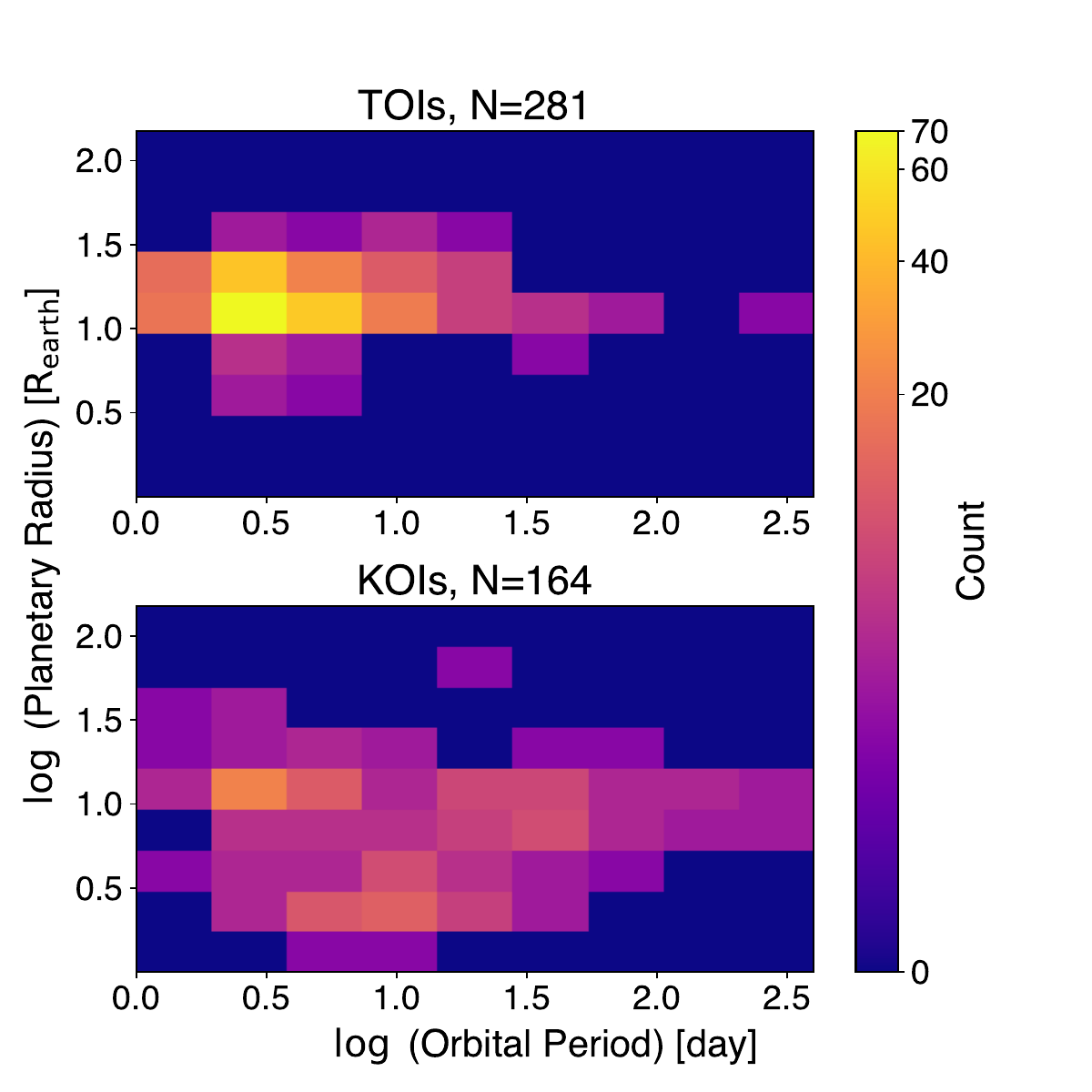}
  \caption{Comparison of planetary radius and orbital period distributions between our TESS targets (upper panel) and the Kepler targets from \citet{2018AJ....155..206A} (lower panel).}
  \label{fig: rp_period_comp}
\end{figure}

In our analysis, there are no definite ringed planet candidates among the 294 TOIs, which primarily consist of close-in planets, consistent with the null results for Kepler Object of Interest (KOI) reported by \citet{2018AJ....155..206A}. 
We investigate the differences in the distributions of radii and orbital periods for the planets analysed by this study and \citet{2018AJ....155..206A}, as shown in figure \ref{fig: rp_period_comp}.
Our criteria and those adopted in \citet{2018AJ....155..206A} both choose planets with high $(S/N)$. 
As shown in figure \ref{fig: rp_period_comp}, while the selected Kepler targets are distributed in a broader region of $(P_{\rm orb}, R_{\rm p})$, our TESS targets are concentrated toward close-in giant planets, $P_{\rm orb}$ less than 10 d and radii up to 10 Earth radii, with a sample size approximately five times larger. 

The scarcity of rings around close-in planets is consistent with the results of previous surveys (e.g., Heising, Marcy and Schlichting (\citeyear{2015ApJ...814...81H}), \citet{2018AJ....155..206A}) and the expected physical characteristics of rings, which are unfavourable for the detection \citep{2011ApJ...734..117S}.
In the case of the optically thick rings assumed in this study, the orbital decay time of ring particles due to the Poynting-Robertson drag depends on the mass surface density of the ring. 
For instance, in the case of B-ring of Saturn, this timescale ranges from 1 Myr to 1 Gyr for the semi-major axis of known close-in planets \citep{2011ApJ...734..117S}.
In total, 294 TOIs are analysed in this study, and no ringed planet candidates are found. 
From the upper limit of $(r_{\rm out/in})_{\rm{upp}}$ for 125 TOIs at $\theta = 45^\circ$ and $\phi = 45^\circ$, we determine that $q[ r_{\rm out/in}>1.8]$ is less than 2\% according to figure \ref{fig: uplimit_cdf}.  
The strict limits on the ring occurrence rate can be partly due to the ignorance of possible alignment between the ring and planetary spin planes, which can decrease the ring detectability.

\subsection{Detectability of ringed planets using precession of planetary rotation}
Kepler and TESS observe numerous planets and achieve the photometric precision necessary to detect Saturn-sized rings \citep{2004ApJ...616.1193B}, making their data ideal for ringed planet surveys.
Nevertheless, the surveys using those data have not yielded any high-confidence ringed planet candidates. 
Thus, we need an alternative approach to detect them.
Among the methods introduced in section \ref{Introduction}, using spin axis precession is favourable, as it allows the use of the extensive light curve data of Kepler and TESS, and flux variation due to ring axis precession can be as high as 1\%, greater than those obtained through model comparison.
In this section, we briefly discuss the potential number of candidates for ring detection through this method.

When the planetary spin axis aligns with the ring axis and is inclined to the orbital axis, the ring plane undergoes precession, causing variations in the azimuthal angle $\phi(t)$ and transit depth over time \citep{2010ApJ...716..850C}.
To consider cases with non-zero obliquities and potential precession, we select planets where the timescale for the spin axis aligns with the normal of the orbital plane $t_{\rm{damp}}$, given by equation (\ref{eq: damping timescale})--the same as equation (1) of \citet{2011ApJ...734..117S}-- is greater than 1 Gyr.

\begin{equation}\label{eq: damping timescale}
   t_{\rm{damp}} \simeq \frac{2 C Q_{\rm{p}}}{3 k_{\rm{p}}}\left(\frac{M_{\rm{p}}}{M_{\star}}\right)\left(\frac{a}{R_{\rm p}}\right)^3\left(\frac{P_{\rm{orb}}}{2 \pi}\right).
\end{equation}

Here, $C$ is the dimensionless planetary moment of inertia, $Q_{\rm{p}}$ is the tidal function of the planet, $k_{\rm{p}}$ is the Love number,  $M_{\rm{p}}$ is the planetary mass, $M_{\star}$ is the mass of the host star, $a$ is the semi-major axis and $R_{\rm p}$ is the planetary radius, respectively.
For our calculations, we adopt typical parameter values for Jovian planets: $C=0.25$, $Q_{\rm{p}}=10^{6.5}$ and $k_{\rm{p}}=1.5$.
For planets whose masses are not directly measured but observed via the transit method, we estimate their masses based on their radii using \texttt{forecaster} \citep{2017ApJ...834...17C}.
For $R_{\rm p} > 10 R_{\rm{earth}}$, this calculation applies $R \simeq M^{-0.04}$, potentially leading to mass estimation errors up to two orders of magnitude.
In this case, we cap $M_{\rm{p}}$ at $2.47 \times 10^{28}$ kg, the upper mass limit distinguishing planets from brown dwarfs.

For planets with $t_{\rm{damp}}>1$ Gyr, we calculate the precession period of the planetary spin, $P_{\rm{prec}}$. 
When the planet revolves on a circular orbit, this is given by \citet{1975Sci...189..377W},

\begin{equation}\label{eq: precision period}
P_{\rm{prec}}=\frac{13.3~\rm{yr}}{\cos \theta}\left(\frac{C / J_2}{13.5}\right)\left(\frac{P_{\rm{orb}}}{15 \rm{~d}}\right)^2\left(\frac{10~\rm{hr}}{P_{\rm {rot}}}\right).
\end{equation}

Here, $\theta$ is the obliquity of the ring axis from the normal of the planetary orbit, $J_2$ is the planetary zonal quadrupole moment and $P_{\rm{rot}}$ is the planetary spin period.
We set $\theta$ as $45^\circ$ and refer to the value $C / J_2=13.5$ taken from Saturn \citep{2004AJ....128.2501W}, and $P_{\rm {rot}}=10$ hr taken from Jupiter \citep{2009P&SS...57.1467H}. 

Using $P_{\rm{prec}}$, we consider $\phi(t)$ as the phase of precession to estimate the transit depth at each event, defined by equation (\ref{eq:  phi change}), which is the same as that given by equation (3) of \citet{2010ApJ...716..850C}, 

\begin{equation}\label{eq: phi change}
\phi(t)=\frac{2 \pi t}{P_{\rm{prec}}} + \phi_0.
\end{equation}
To consider an optimistic case when large transit depth variations to be observed, we set $\phi_0$ to 45$^\circ$.
To discuss the detectability, we simulate transit depth variations assuming transit depths, transit depth errors, and $P_{\rm orb}$. 
Specifically, for each target, we derive $R_{\rm p} / R_{\star}$ assuming ring size to match $\delta$ at $\phi_0$, and calculate the evolution of $(\phi(t), \delta(t))$ during $P_{\rm orb}$.
The transit depth error is derived from $\delta$, $T_{\rm dur}$, and observational noise using equation (23) of \citet{2008ApJ...689..499C},

\begin{equation}\label{eq: transit depth error}
    \sigma_\delta \simeq Q^{-1} \delta, \\
    \left(Q \equiv \sqrt{N_d} \frac{\delta}{\sigma (N_d)}\right).
\end{equation}

The transit midpoint of each event is taken from the actual observations: we retrieve observed sectors/quarters and obtain $t_{\rm mid}$ referring to known $t_{\rm mid}$, $P_{\rm{orb}}$.
We use only events with over 50\% of data that exist around $t_{\rm mid}$ to ensure accurate transit depth in actual data analysis.
We only consider the short-cadence data, 2 min cadence for TESS and 1 min cadence for Kepler to ensure sufficient data points for robust transit depth determination.
Assuming the calculated transit depth variations and transit depth errors, we simulate the time series of transit depths for all of the targets.

\begin{figure}
    \centering
    \includegraphics[width=80mm]{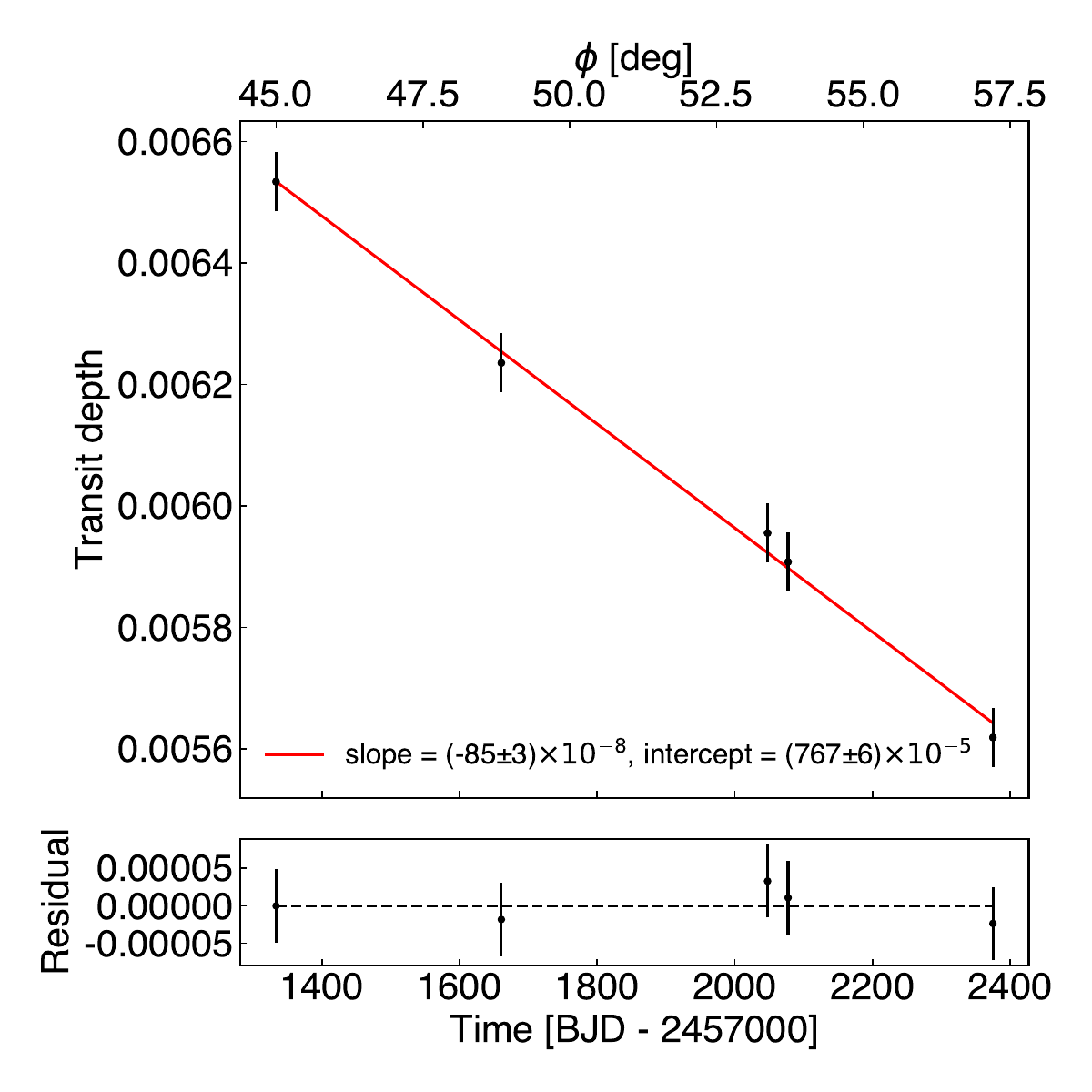}
    \caption{Transit depth variations derived from simulation data of TOI-173.01, assuming presence and precession of rings. Red line in the upper panel represents the best-fit linear function. The horizontal axis represents the actual times at which the transit events are observed, assuming that the first event occurred at a precession phase of $\phi=45^\circ$.}
    \label{fig: an example of varifying transit depth variation}
\end{figure}

We fit a linear model to the time variation of transit depth for each object and derive the linear slope of the variation.
We select targets by considering transit depth variations significant if the deviation of the slope from zero $\sigma_{\rm slope}$, exceeds the $3\sigma$ significance level.
Figure \ref{fig: an example of varifying transit depth variation} shows a simulated example of transit depth variations for TOI-173.01. 
The depth variation is clearly shown, and the slope is estimated to be  ($-85 \pm 3 )\times 10^{-8}$, which is deviated from zero by 26$\sigma$. 
As shown in tables \ref{tab: precession targets in TESS} and \ref{tab: precession targets in kepler}, we identify 10 detectable targets for TESS and 13 detectable targets for Kepler.
Most TESS candidates are about the size of Jupiter with $P_{\rm orb}$ of several tens of days, whereas Kepler candidates often have longer $P_{\rm orb}$ and radii several times that of Earth.

In reality, transit depth variations are likely influenced by systematic and stellar noises, making the analysis challenging and the above estimate optimistic. 
Nevertheless, analysing these transit depth variations provides a unique path for identifying possible ring planet candidates, which is complementary to the current approach. 
We leave the actual analysis for future studies.

\section{Conclusion}\label{Conclusion}
To search for ringed planet candidates, we systematically analyse 308 TESS planet candidates, mainly consisting of close-in planets which are five times more numerous than in \citet{2018AJ....155..206A}.
These candidates are selected based on expected ring signals significantly larger than uncertainty.
To detect tiny ring signals, we develop a pipeline that removes non-planetary variations and outliers from light curves using a two-step process, then compared the fitting results of ringless and ringed transit models to the phase-folded light curves.
As a result, we identify six systems where ringed models are statistically favoured over the ringless model.
However, visual inspection of the signals provides no conclusive evidence for the presence of rings in light curves.

Given the null detection, we determine the 3$\sigma$ upper limits of the outer radius of the rings in 125 TOIs, assuming the most favourable ring orientation for detection. 
The distribution of these upper limits suggests that the occurrence rate of rings with an outer radius exceeding 1.8 times the planetary radii is less than 2\%.
Non-detection of ringed planets can be partly attributed to the damping of ring obliquity due to tidal alignment, expected for short-period planets.

Transit depth variations caused by the precession of rings allow the detection of larger ring signals in longer-period planets compared to the transit model comparison.
We optimistically pick up 10 (table \ref{tab: precession targets in TESS}) and 13 (table \ref{tab: precession targets in kepler}) targets for TESS and Kepler targets under favourable ring orientation.
We expect this approach to be effective in future ringed planet searches.

TESS observations are scheduled to continue until 2028 September at the latest, providing additional opportunities to observe transits of existing TOIs.
With the upcoming launch of PLATO in 2026 \citep{2014ExA....38..249R}, continuous efforts to detect ringed planets, similar to Saturn, remain essential.

\begin{ack}
This paper incorporates data acquired through the TESS mission, retrieved from the MAST data archive at the Space Telescope Science Institute (STScI). The funding for the TESS mission is generously supplied by the NASA Explorer Program. STScI operates under NASA contract NAS 5-26555, managed by the Association of Universities for Research in Astronomy, Inc. Our research involved the utilisation of (1) Lightkurve, a Python package designed for the analysis of Kepler and TESS data (Lightkurve Collaboration et al. 2018), and (2) data retrieved from, as well as tools provided by ETL22. We thank Takaya Ohashi for an enlightening discussion regarding this study. 
\end{ack}

\begin{table*}
\rotatebox{90}{%
\begin{minipage}{\textheight} 
\caption{Fitting results for the systems with $p$-value $> 0.0027$ using both ringless and ringed models.}
\label{tab: ringed model and ringless model fitting result}
\begin{tabular}{crrrcrrrr}
\hline
\multicolumn{1}{c}{TOI} & \multicolumn{1}{c}{$P_{\rm{orb}}$ [day]} & \multicolumn{1}{c}{$b^{*}$} & \multicolumn{1}{c}{$(R_{\rm p} / R_{\star})_{\rm{ringless}}^{*}$} & \multicolumn{1}{c}{$(r_{\rm out/in })_{\rm{upp}}$} & \multicolumn{1}{c}{$\delta$ [ppm]} & \multicolumn{1}{c}{$\sigma (N_d=500)$ [ppm]} & \multicolumn{1}{c}{$(\chi^{2}_{\rm{ringless,min}},\chi^{2}_{\rm{ring,min}}, N_d)$} & \multicolumn{1}{c}{$p$-value} \\

\hline
101.01 & 1.430372 $\pm$ 1e-06 & 0.73 $\pm$ 0.02 & 0.143 $\pm$ 0.007 & 1.23 & 19200 $\pm$ 317 & 2232 & (412.18, 409.73, 500) & 4.04e-01 \\
102.01 & 4.4119390 $\pm$ 7e-07 & 0.24 $\pm$ 0.02 & 0.1109 $\pm$ 0.0002 & 1.42 & 15220 $\pm$ 15 & 168 & (660.82, 657.80, 500) & 5.23e-01 \\
105.01 & 2.1846660 $\pm$ 8e-07 & 0.47 $\pm$ 0.02 & 0.1029 $\pm$ 0.0005 & ... & 11710 $\pm$ 46 & 433 & (443.65, 443.03, 500) & 8.78e-01 \\
112.01 & 2.4998040 $\pm$ 7e-07 & ... & 0.1128 $\pm$ 0.0006 & 1.45 & 14620 $\pm$ 53 & 491 & (520.01, 514.90, 500) & 1.84e-01 \\
114.01 & 3.288787 $\pm$ 1e-06 & ... & 0.0774 $\pm$ 0.0003 & 1.51 & 6990 $\pm$ 21 & 222 & (493.45, 489.68, 500) & 2.88e-01 \\
116.01 & 2.798579 $\pm$ 2e-06 & 0.45 $\pm$ 0.04 & 0.121 $\pm$ 0.001 & ... & 16700 $\pm$ 104 & 875 & (420.49, 418.20, 500) & 4.44e-01 \\
121.01 & 14.7637 $\pm$ 3e-04 $^{\dagger}$ & ... & 0.111 $\pm$ 0.001 & ... & 14530 $\pm$ 66 & 909 & (270.46, 268.87, 275) & 6.69e-01 \\
129.01 & 0.9809719 $\pm$ 8e-07 & ... & ... & 1.39 & 6200 $\pm$ 160 & 646 & (474.03, 466.04, 500) & 3.97e-02 \\
143.01 & 2.310970 $\pm$ 1e-06 & 0.70 $\pm$ 0.02 & 0.081 $\pm$ 0.001 & 1.30 & 6830 $\pm$ 43 & 437 & (501.97, 499.80, 500) & 5.47e-01 \\
150.01 & 5.857418 $\pm$ 6e-06 & 0.48 $\pm$ 0.03 & 0.0780 $\pm$ 0.0004 & ... & 6490 $\pm$ 26 & 348 & (458.32, 456.58, 500) & 6.01e-01 \\
157.01 & 2.084539 $\pm$ 1e-06 & 0.766 $\pm$ 0.008 & 0.116 $\pm$ 0.003 & 1.21 & 12900 $\pm$ 108 & 778 & (494.54, 492.59, 500) & 5.87e-01 \\
159.01 & 3.762836 $\pm$ 3e-06 & 0.35 $\pm$ 0.04 & 0.1019 $\pm$ 0.0005 & ... & 11360 $\pm$ 51 & 556 & (497.22, 495.49, 500) & 6.37e-01 \\
163.01 & 4.231115 $\pm$ 3e-06 & 0.27 $\pm$ 0.07 & 0.0806 $\pm$ 0.0004 & ... & 7240 $\pm$ 30 & 347 & (407.57, 404.62, 500) & 3.13e-01 \\
165.01 & 7.75766 $\pm$ 3e-05 & 0.72 $\pm$ 0.01 & 0.079 $\pm$ 0.001 & 1.22 & 7640 $\pm$ 38 & 382 & (466.12, 463.48, 500) & 4.25e-01 \\
170.01 & 3.711516 $\pm$ 6e-06 & 0.60 $\pm$ 0.04 & 0.105 $\pm$ 0.002 & ... & 12300 $\pm$ 168 & 1510 & (440.99, 439.23, 500) & 5.80e-01 \\
173.01 & 29.7536 $\pm$ 5e-04 & 0.3 $\pm$ 0.1 & 0.0752 $\pm$ 0.0005 & ... & 6530 $\pm$ 29 & 459 & (554.03, 553.48, 500) & 9.22e-01 \\
185.01 & 0.94145250 $\pm$ 9e-08 & 0.37 $\pm$ 0.01 & 0.0976 $\pm$ 0.0002 & 1.20 & 10770 $\pm$ 25 & 196 & (596.90, 592.65, 500) & 3.20e-01 \\
190.01 & 10.020600 $\pm$ 9e-06 & 0.44 $\pm$ 0.04 & 0.0868 $\pm$ 0.0006 & ... & 8220 $\pm$ 34 & 449 & (476.20, 473.23, 500) & 3.81e-01 \\
192.01 & 3.922712 $\pm$ 3e-06 & ... & 0.097 $\pm$ 0.001 & ... & 11410 $\pm$ 95 & 844 & (402.43, 401.03, 498) & 6.36e-01 \\
194.01 & 4.899644 $\pm$ 5e-06 & 0.70 $\pm$ 0.02 & 0.101 $\pm$ 0.002 & 1.30 & 10260 $\pm$ 78 & 784 & (413.96, 411.61, 495) & 4.31e-01 \\
195.01 & 2.0727600 $\pm$ 7e-07 & 0.44 $\pm$ 0.03 & 0.1104 $\pm$ 0.0006 & ... & 13510 $\pm$ 56 & 500 & (566.48, 563.79, 481) & 5.23e-01 \\
199.01 & 104.87 $\pm$ 1e-02 & 0.60 $\pm$ 0.03 & 0.104 $\pm$ 0.002 & ... & 12300 $\pm$ 51 & 712 & (509.43, 508.28, 499) & 7.75e-01 \\
201.01 & 52.97818 $\pm$ 4e-05 & 0.73 $\pm$ 0.01 & 0.0780 $\pm$ 0.0009 & 1.14 & 6720 $\pm$ 20 & 262 & (534.87, 527.65, 500) & 8.32e-02 \\
231.01 & 3.361002 $\pm$ 2e-06 & 0.36 $\pm$ 0.05 & 0.146 $\pm$ 0.001 & ... & 24000 $\pm$ 171 & 1507 & (471.49, 469.44, 499) & 5.44e-01 \\
232.01 & 1.3382310 $\pm$ 4e-07 & ... & 0.150 $\pm$ 0.001 & 1.46 & 27200 $\pm$ 131 & 1057 & (467.00, 465.66, 500) & 7.03e-01 \\
241.01 & 1.3866530 $\pm$ 9e-07 & 0.25 $\pm$ 0.08 & 0.1125 $\pm$ 0.0007 & 1.88 & 14200 $\pm$ 132 & 1096 & (481.09, 479.99, 500) & 7.70e-01 \\
243.01 & 3.437306 $\pm$ 5e-06 & 0.70 $\pm$ 0.02 & 0.123 $\pm$ 0.003 & 1.30 & 15000 $\pm$ 130 & 1481 & (461.43, 459.00, 500) & 4.60e-01 \\
246.01 & 7.872890 $\pm$ 6e-06 & ... & 0.113 $\pm$ 0.002 & ... & 14600 $\pm$ 127 & 1355 & (464.98, 463.24, 481) & 6.23e-01 \\
250.01 & 1.628430 $\pm$ 1e-06 & 0.32 $\pm$ 0.09 & 0.111 $\pm$ 0.001 & ... & 13800 $\pm$ 136 & 1153 & (483.15, 480.80, 500) & 4.95e-01 \\
264.01 & 2.216744 $\pm$ 2e-06 & 0.68 $\pm$ 0.03 & 0.065 $\pm$ 0.001 & 1.53 & 4580 $\pm$ 50 & 528 & (493.42, 491.24, 500) & 5.37e-01 \\
267.01 & 5.752583 $\pm$ 7e-06 & ... & 0.0676 $\pm$ 0.0003 & 1.77 & 5270 $\pm$ 24 & 316 & (575.37, 574.56, 500) & 8.75e-01 \\
272.01 & 3.309844 $\pm$ 2e-06 & 0.63 $\pm$ 0.04 & 0.133 $\pm$ 0.004 & 1.28 & 19200 $\pm$ 226 & 1867 & (397.99, 395.52, 433) & 4.51e-01 \\
275.01 & 0.919563 $\pm$ 4e-06 & ... & ... & 1.23 & 9180 $\pm$ 87 & 512 & (496.99, 491.95, 500) & 1.72e-01 \\
388.01 & 2.903684 $\pm$ 3e-06 & 0.3 $\pm$ 0.2 & 0.0645 $\pm$ 0.0007 & ... & 4560 $\pm$ 40 & 498 & (499.46, 496.56, 500) & 4.13e-01 \\
391.01 & 1.955093 $\pm$ 1e-06 & 0.70 $\pm$ 0.02 & 0.136 $\pm$ 0.004 & 1.33 & 19600 $\pm$ 145 & 1068 & (413.73, 411.17, 430) & 4.54e-01 \\
398.01 & 1.3600290 $\pm$ 3e-07 & 0.28 $\pm$ 0.03 & 0.1305 $\pm$ 0.0005 & 1.62 & 19430 $\pm$ 58 & 473 & (606.62, 603.10, 500) & 4.15e-01 \\
404.01 & 2.962642 $\pm$ 3e-06 & 0.69 $\pm$ 0.04 & 0.161 $\pm$ 0.009 & ... & 27100 $\pm$ 409 & 3097 & (395.41, 386.00, 444) & 1.49e-02 \\
413.01 & 3.662392 $\pm$ 1e-06 & 0.56 $\pm$ 0.01 & 0.1076 $\pm$ 0.0004 & 1.17 & 12250 $\pm$ 34 & 366 & (464.84, 463.41, 500) & 6.80e-01 \\
415.01 & 3.611267 $\pm$ 5e-06 & 0.69 $\pm$ 0.03 & 0.076 $\pm$ 0.001 & 1.57 & 5680 $\pm$ 65 & 673 & (487.61, 484.65, 500) & 3.94e-01 \\
423.01 & 3.161569 $\pm$ 1e-06 & 0.28 $\pm$ 0.06 & 0.123 $\pm$ 0.001 & ... & 17480 $\pm$ 67 & 650 & (530.65, 526.81, 500) & 3.12e-01 \\
450.01 & 10.71477 $\pm$ 2e-05 & ... & ... & 1.39 & 56300 $\pm$ 639 & 5093 & (425.19, 424.32, 482) & 8.09e-01 \\
453.01 & 2.7058 $\pm$ 2e-04 & ... & 0.0768 $\pm$ 0.0005 & ... & 7500 $\pm$ 44 & 547 & (477.87, 474.50, 499) & 3.26e-01 \\
471.01 & 3.585708 $\pm$ 2e-06 & 0.75 $\pm$ 0.01 & 0.108 $\pm$ 0.003 & 1.31 & 11810 $\pm$ 71 & 645 & (533.32, 524.90, 500) & 5.01e-02 \\

\hline
\end{tabular}
\end{minipage}
}

\end{table*}

\begin{table*}
\rotatebox{90}{%
\begin{minipage}{\textheight} 
\flushleft{Table \ref{tab: ringed model and ringless model fitting result}. (Continued)}
\begin{tabular}{crrrcrrrr}
\hline
\multicolumn{1}{c}{TOI} & \multicolumn{1}{c}{$P_{\rm{orb}}$ [day]} & \multicolumn{1}{c}{$b^{*}$} & \multicolumn{1}{c}{$(R_{\rm p} / R_{\star})_{\rm{ringless}}^{*}$} & \multicolumn{1}{c}{$(r_{\rm out/in })_{\rm{upp}}$} & \multicolumn{1}{c}{$\delta$ [ppm]} & \multicolumn{1}{c}{$\sigma (N_d=500)$ [ppm]} & \multicolumn{1}{c}{$(\chi^{2}_{\rm{ringless,min}},\chi^{2}_{\rm{ring,min}}, N_d)$} & \multicolumn{1}{c}{$p$-value} \\
\hline
472.01 & 0.568261 $\pm$ 7e-06 & ... & ... & 1.14 & 42500 $\pm$ 465 & 2213 & (469.28, 463.01, 500) & 8.57e-02 \\
477.01 & 2.944429 $\pm$ 3e-06 & 0.45 $\pm$ 0.06 & 0.134 $\pm$ 0.002 & ... & 20500 $\pm$ 213 & 1826 & (434.57, 433.52, 500) & 7.56e-01 \\
479.01 & 2.781736 $\pm$ 2e-06 & 0.77 $\pm$ 0.02 & 0.113 $\pm$ 0.006 & 1.29 & 12800 $\pm$ 120 & 964 & (426.40, 423.08, 449) & 3.30e-01 \\
481.01 & 10.331159 $\pm$ 8e-06 & ... & 0.0621 $\pm$ 0.0005 & ... & 4750 $\pm$ 18 & 237 & (501.85, 501.64, 500) & 9.77e-01 \\
483.01 & 4.378085 $\pm$ 6e-06 & ... & 0.0785 $\pm$ 0.0007 & ... & 7070 $\pm$ 35 & 448 & (489.64, 488.81, 500) & 8.42e-01 \\
490.01 & 2.810601 $\pm$ 2e-06 & 0.86 $\pm$ 0.03 & 0.11 $\pm$ 0.02 & 1.18 & 11980 $\pm$ 84 & 647 & (476.11, 474.07, 469) & 5.79e-01 \\
501.01 & 1.7100533 $\pm$ 8e-07 & 0.85 $\pm$ 0.02 & 0.11 $\pm$ 0.01 & 1.11 & 12680 $\pm$ 74 & 543 & (480.11, 477.66, 500) & 4.73e-01 \\
503.01 & 3.677355 $\pm$ 4e-06 & 0.91 $\pm$ 0.03 & 0.08 $\pm$ 0.02 & ... & 3920 $\pm$ 56 & 376 & (543.17, 540.39, 500) & 4.72e-01 \\
505.01 & 3.329466 $\pm$ 2e-06 & 0.28 $\pm$ 0.06 & 0.0980 $\pm$ 0.0005 & 1.78 & 10600 $\pm$ 42 & 470 & (432.92, 432.70, 500) & 9.70e-01 \\
507.01 & 0.8994659 $\pm$ 4e-07 & 0.68 $\pm$ 0.08 & 0.24 $\pm$ 0.03 & 1.30 & 58800 $\pm$ 789 & 4750 & (419.71, 413.33, 500) & 5.72e-02 \\
508.01 & 4.6119 $\pm$ 4e-04 & 0.42 $\pm$ 0.06 & 0.0972 $\pm$ 0.0008 & ... & 11890 $\pm$ 71 & 788 & (485.49, 476.74, 489) & 3.33e-02 \\
511.01 & 3.355244 $\pm$ 3e-06 & 0.37 $\pm$ 0.06 & 0.1005 $\pm$ 0.0008 & ... & 12310 $\pm$ 69 & 720 & (423.96, 422.96, 500) & 7.62e-01 \\
527.01 & 18.08921 $\pm$ 3e-05 & 0.91 $\pm$ 0.09 & 0.12 $\pm$ 0.02 & 1.29 & 5530 $\pm$ 50 & 440 & (426.01, 421.15, 408) & 2.06e-01 \\
567.01 & 1.537366 $\pm$ 1e-06 & 0.68 $\pm$ 0.03 & 0.131 $\pm$ 0.004 & 1.43 & 17900 $\pm$ 220 & 1648 & (435.08, 434.98, 500) & 9.90e-01 \\
573.01 & 13.57735 $\pm$ 6e-05 & ... & ... & 1.18 & 157000 $\pm$ 1134 & 8113 & (222.69, 219.94, 233) & 4.29e-01 \\
585.01 & 5.5472 $\pm$ 4e-04 & 0.25 $\pm$ 0.05 & 0.1175 $\pm$ 0.0006 & 1.71 & 11570 $\pm$ 60 & 732 & (535.20, 533.66, 440) & 7.43e-01 \\
587.01 & 8.0442 $\pm$ 5e-04 & 0.55 $\pm$ 0.02 & 0.0689 $\pm$ 0.0003 & 1.39 & 4940 $\pm$ 23 & 296 & (545.09, 539.68, 500) & 1.79e-01 \\
615.01 & 4.6616 $\pm$ 9e-04 & 0.53 $\pm$ 0.05 & 0.102 $\pm$ 0.001 & ... & 12000 $\pm$ 107 & 1177 & (391.85, 391.15, 412) & 8.70e-01 \\
621.01 & 3.11233 $\pm$ 3e-05 & 0.91 $\pm$ 0.02 & 0.08 $\pm$ 0.02 & 1.16 & 6340 $\pm$ 30 & 263 & (503.90, 501.10, 500) & 4.34e-01 \\
622.01 & 6.402525 $\pm$ 7e-06 & 0.63 $\pm$ 0.03 & 0.0606 $\pm$ 0.0006 & 1.42 & 4150 $\pm$ 28 & 293 & (557.68, 551.12, 500) & 1.21e-01 \\
625.01 & 4.7869 $\pm$ 5e-04 & 0.33 $\pm$ 0.09 & 0.0853 $\pm$ 0.0007 & ... & 8390 $\pm$ 58 & 715 & (338.90, 336.29, 327) & 4.83e-01 \\
626.01 & 4.4010 $\pm$ 2e-04 & 0.42 $\pm$ 0.07 & 0.0766 $\pm$ 0.0007 & ... & 6900 $\pm$ 38 & 457 & (529.44, 529.30, 500) & 9.87e-01 \\
640.01 & 5.0038 $\pm$ 3e-04 & 0.89 $\pm$ 0.07 & 0.09 $\pm$ 0.06 & 1.66 & 6900 $\pm$ 69 & 725 & (432.18, 424.56, 424) & 6.10e-02 \\
655.01 & 0.7888387 $\pm$ 7e-07 & 0.66 $\pm$ 0.02 & 0.146 $\pm$ 0.003 & 1.16 & 22400 $\pm$ 187 & 1296 & (480.50, 479.33, 500) & 7.55e-01 \\
656.01 & 0.813473 $\pm$ 1e-06 & 0.684 $\pm$ 0.009 & 0.161 $\pm$ 0.002 & 1.10 & 26600 $\pm$ 152 & 923 & (456.14, 452.22, 493) & 2.43e-01 \\
671.01 & 10.197030 $\pm$ 3e-06 $^{\dagger}$ & 0.70 $\pm$ 0.02 & 0.122 $\pm$ 0.002 & 1.19 & 16110 $\pm$ 88 & 1054 & (514.87, 506.61, 407) & 9.23e-02 \\
677.01 & 11.23661 $\pm$ 1e-05 & 0.72 $\pm$ 0.02 & 0.095 $\pm$ 0.002 & 1.21 & 9410 $\pm$ 85 & 707 & (486.04, 483.44, 399) & 5.54e-01 \\
679.01 & 11.17554 $\pm$ 7e-05 & 0.49 $\pm$ 0.07 & 0.0823 $\pm$ 0.0009 & ... & 7870 $\pm$ 70 & 904 & (375.81, 369.31, 400) & 7.80e-02 \\
683.01 & 3.405886 $\pm$ 5e-06 & 0.77 $\pm$ 0.02 & 0.127 $\pm$ 0.008 & 1.29 & 16600 $\pm$ 175 & 1481 & (409.68, 403.75, 413) & 1.18e-01 \\
744.01 & 4.317684 $\pm$ 5e-06 & 0.92 $\pm$ 0.04 & 0.113 $\pm$ 0.005 & 1.75 & 11280 $\pm$ 98 & 728 & (498.12, 486.98, 484) & 1.33e-02 \\
747.01 & 1.682794 $\pm$ 2e-06 & 0.59 $\pm$ 0.03 & 0.0868 $\pm$ 0.0008 & 1.54 & 8080 $\pm$ 65 & 619 & (463.16, 460.15, 500) & 3.63e-01 \\
748.01 & 2.021959 $\pm$ 3e-06 & 0.77 $\pm$ 0.01 & 0.099 $\pm$ 0.003 & 1.24 & 9440 $\pm$ 81 & 772 & (627.91, 627.63, 500) & 9.75e-01 \\
749.01 & 0.6355280 $\pm$ 3e-07 & ... & ... & ... & 51200 $\pm$ 846 & 5200 & (451.79, 447.47, 500) & 1.94e-01 \\
758.01 & 13.273550 $\pm$ 8e-06 $^{\dagger}$ & 0.50 $\pm$ 0.03 & 0.139 $\pm$ 0.001 & ... & 21060 $\pm$ 61 & 872 & (479.49, 478.03, 400) & 7.54e-01 \\
760.01 & 12.340 $\pm$ 2e-03 & 0.73 $\pm$ 0.02 & 0.118 $\pm$ 0.004 & 1.26 & 13390 $\pm$ 75 & 1037 & (473.39, 471.68, 489) & 6.29e-01 \\
764.01 & 5.631666 $\pm$ 9e-06 & ... & 0.1105 $\pm$ 0.0008 & 1.56 & 13880 $\pm$ 64 & 902 & (455.72, 450.99, 500) & 1.63e-01 \\
766.01 & 3.446453 $\pm$ 5e-06 & 0.62 $\pm$ 0.06 & 0.121 $\pm$ 0.005 & 1.27 & 16400 $\pm$ 174 & 1492 & (379.55, 374.95, 429) & 1.64e-01 \\
767.01 & 3.7648 $\pm$ 3e-04 & 0.42 $\pm$ 0.06 & 0.141 $\pm$ 0.002 & ... & 22300 $\pm$ 193 & 1759 & (258.28, 257.41, 270) & 8.32e-01 \\
769.01 & 4.981680 $\pm$ 6e-06 & 0.38 $\pm$ 0.08 & 0.126 $\pm$ 0.002 & ... & 18100 $\pm$ 184 & 1669 & (431.85, 429.01, 500) & 3.58e-01 \\
774.01 & 4.465634 $\pm$ 5e-06 & 0.3 $\pm$ 0.1 & 0.124 $\pm$ 0.002 & ... & 17300 $\pm$ 136 & 1396 & (460.71, 459.55, 500) & 7.45e-01 \\
778.01 & 4.633610 $\pm$ 4e-06 & 0.63 $\pm$ 0.02 & 0.0832 $\pm$ 0.0006 & 1.22 & 7660 $\pm$ 38 & 398 & (518.56, 515.63, 500) & 4.27e-01 \\
780.01 & 3.052402 $\pm$ 2e-06 & ... & 0.1342 $\pm$ 0.0007 & 1.52 & 22300 $\pm$ 123 & 1063 & (549.56, 546.34, 499) & 4.11e-01 \\
818.01 & 3.1185 $\pm$ 2e-04 & 0.82 $\pm$ 0.04 & 0.11 $\pm$ 0.02 & 1.83 & 12200 $\pm$ 188 & 1477 & (284.98, 281.44, 275) & 3.45e-01 \\
\hline
\end{tabular}%
\end{minipage}
}
\end{table*}

\begin{table*}
\rotatebox{90}{%
\begin{minipage}{\textheight} 
\flushleft{Table \ref{tab: ringed model and ringless model fitting result}. (Continued)}
\begin{tabular}{crrrcrrrr}
\hline
\multicolumn{1}{c}{TOI} & \multicolumn{1}{c}{$P_{\rm{orb}}$ [day]} & \multicolumn{1}{c}{$b^{*}$} & \multicolumn{1}{c}{$(R_{\rm p} / R_{\star})_{\rm{ringless}}^{*}$} & \multicolumn{1}{c}{$(r_{\rm out/in })_{\rm{upp}}$} & \multicolumn{1}{c}{$\delta$ [ppm]} & \multicolumn{1}{c}{$\sigma (N_d=500)$ [ppm]} & \multicolumn{1}{c}{$(\chi^{2}_{\rm{ringless,min}},\chi^{2}_{\rm{ring,min}}, N_d)$} & \multicolumn{1}{c}{$p$-value} \\
\hline
820.01 & 3.7522 $\pm$ 3e-04 & 0.53 $\pm$ 0.07 & 0.093 $\pm$ 0.001 & ... & 9500 $\pm$ 103 & 1087 & (261.89, 258.49, 292) & 2.97e-01 \\
822.01 & 7.133514 $\pm$ 7e-06 & ... & 0.123 $\pm$ 0.002 & ... & 18200 $\pm$ 199 & 1896 & (370.11, 366.65, 372) & 3.33e-01 \\
827.01 & 1.228236 $\pm$ 1e-06 & ... & ... & 1.78 & 7500 $\pm$ 100 & 720 & (459.77, 449.05, 500) & 9.02e-03 \\
828.01 & 5.322 $\pm$ 1e-03 & 0.74 $\pm$ 0.02 & 0.082 $\pm$ 0.002 & 1.51 & 6560 $\pm$ 70 & 762 & (262.90, 259.26, 290) & 2.71e-01 \\
830.01 & 15.04052 $\pm$ 2e-05 & ... & ... & ... & 18000 $\pm$ 194 & 1606 & (330.21, 325.50, 358) & 1.71e-01 \\
834.01 & 2.675550 $\pm$ 1e-06 & 0.32 $\pm$ 0.05 & 0.1135 $\pm$ 0.0006 & ... & 13990 $\pm$ 76 & 729 & (435.08, 432.76, 500) & 4.55e-01 \\
842.01 & 2.786385 $\pm$ 3e-06 & 0.83 $\pm$ 0.02 & 0.11 $\pm$ 0.02 & 1.22 & 11900 $\pm$ 188 & 1301 & (416.24, 414.40, 500) & 5.39e-01 \\
862.01 & 3.31022 $\pm$ 1e-05 & 0.836 $\pm$ 0.006 & 0.074 $\pm$ 0.004 & 1.27 & 5080 $\pm$ 41 & 369 & (565.18, 560.76, 500) & 2.78e-01 \\
899.01 & 12.84619 $\pm$ 3e-05 & 0.72 $\pm$ 0.05 & 0.090 $\pm$ 0.005 & ... & 8900 $\pm$ 106 & 990 & (485.73, 485.31, 500) & 9.36e-01 \\
905.01 & 3.739567 $\pm$ 3e-06 & 0.84 $\pm$ 0.03 & 0.13 $\pm$ 0.02 & 1.11 & 15500 $\pm$ 148 & 1156 & (397.10, 396.45, 490) & 8.52e-01 \\
919.01 & 20.02723 $\pm$ 6e-05 & ... & ... & 1.32 & 17500 $\pm$ 140 & 1233 & (487.03, 484.98, 466) & 5.89e-01 \\
924.01 & 12.12737 $\pm$ 3e-05 & ... & ... & 1.62 & 18800 $\pm$ 154 & 1156 & (477.09, 474.85, 500) & 5.11e-01 \\
934.01 & 3.781740 $\pm$ 5e-06 & 0.55 $\pm$ 0.04 & 0.131 $\pm$ 0.002 & ... & 19000 $\pm$ 218 & 1772 & (440.19, 438.85, 500) & 6.82e-01 \\
966.01 & 3.4092 $\pm$ 2e-04 & 0.50 $\pm$ 0.04 & 0.124 $\pm$ 0.001 & ... & 16400 $\pm$ 151 & 1481 & (421.85, 419.90, 366) & 6.49e-01 \\
1019.01 & 5.23414 $\pm$ 8e-05 & 0.70 $\pm$ 0.01 & 0.145 $\pm$ 0.002 & 1.24 & 20540 $\pm$ 84 & 887 & (579.63, 574.33, 431) & 2.76e-01 \\
1050.01 & 3.735485 $\pm$ 3e-06 & 0.33 $\pm$ 0.06 & 0.123 $\pm$ 0.001 & ... & 17030 $\pm$ 96 & 1077 & (474.68, 473.48, 500) & 7.44e-01 \\
1069.01 & 2.977642 $\pm$ 4e-06 & 0.53 $\pm$ 0.05 & 0.105 $\pm$ 0.001 & ... & 12240 $\pm$ 82 & 802 & (437.19, 434.81, 474) & 4.70e-01 \\
1076.01 & 2.549561 $\pm$ 6e-06 & 0.3 $\pm$ 0.1 & 0.113 $\pm$ 0.002 & ... & 15000 $\pm$ 138 & 1217 & (455.07, 454.64, 500) & 9.27e-01 \\
1085.01 & 3.254119 $\pm$ 3e-06 & 0.62 $\pm$ 0.01 & 0.1105 $\pm$ 0.0008 & 1.26 & 13620 $\pm$ 43 & 491 & (580.15, 576.83, 500) & 4.20e-01 \\
1107.01 & 4.07819 $\pm$ 9e-05 & ... & 0.0733 $\pm$ 0.0003 & ... & 6000 $\pm$ 44 & 538 & (497.40, 493.71, 500) & 3.02e-01 \\
1135.01 & 8.027730 $\pm$ 5e-06 & 0.37 $\pm$ 0.05 & 0.0713 $\pm$ 0.0003 & ... & 5830 $\pm$ 27 & 329 & (602.06, 600.51, 500) & 7.37e-01 \\
1165.01 & 2.2552880 $\pm$ 2e-07 & 0.529 $\pm$ 0.006 & 0.1329 $\pm$ 0.0003 & 1.11 & 21020 $\pm$ 28 & 222 & (511.94, 508.78, 500) & 3.85e-01 \\
1173.01 & 7.064397 $\pm$ 5e-06 & 0.75 $\pm$ 0.02 & 0.090 $\pm$ 0.002 & 1.22 & 8670 $\pm$ 74 & 631 & (420.77, 418.37, 500) & 4.23e-01 \\
1176.01 & 14.00775 $\pm$ 1e-05 & ... & 0.0730 $\pm$ 0.0003 & 1.68 & 6170 $\pm$ 24 & 318 & (412.36, 405.68, 500) & 4.58e-02 \\
1181.01 & 2.1031940 $\pm$ 8e-07 & 0.32 $\pm$ 0.03 & 0.0769 $\pm$ 0.0002 & 1.80 & 7120 $\pm$ 16 & 177 & (525.71, 524.37, 500) & 7.42e-01 \\
1186.01 & 11.24631 $\pm$ 3e-05 & 0.41 $\pm$ 0.03 & 0.0893 $\pm$ 0.0004 & ... & 8990 $\pm$ 18 & 283 & (525.23, 524.22, 500) & 8.14e-01 \\
1190.01 & 3.695011 $\pm$ 1e-06 & ... & 0.0821 $\pm$ 0.0002 & 1.42 & 7870 $\pm$ 21 & 242 & (640.22, 637.09, 500) & 4.93e-01 \\
1194.01 & 2.310645 $\pm$ 1e-06 & 0.82 $\pm$ 0.01 & 0.082 $\pm$ 0.005 & 1.23 & 7030 $\pm$ 86 & 544 & (381.42, 379.53, 500) & 4.88e-01 \\
1198.01 & 3.612771 $\pm$ 8e-06 & 0.35 $\pm$ 0.06 & 0.0999 $\pm$ 0.0007 & ... & 10780 $\pm$ 65 & 726 & (501.12, 499.75, 500) & 7.18e-01 \\
1248.01 & 4.36013 $\pm$ 9e-05 & 0.59 $\pm$ 0.04 & 0.0716 $\pm$ 0.0009 & ... & 5570 $\pm$ 57 & 475 & (405.55, 405.06, 500) & 8.97e-01 \\
1251.01 & 5.963056 $\pm$ 4e-06 & 0.699 $\pm$ 0.009 & 0.0983 $\pm$ 0.0009 & 1.25 & 11000 $\pm$ 47 & 445 & (444.24, 440.56, 500) & 2.53e-01 \\
1257.01 & 5.452649 $\pm$ 5e-06 & ... & 0.0810 $\pm$ 0.0007 & 1.83 & 7750 $\pm$ 68 & 693 & (442.29, 440.02, 471) & 4.99e-01 \\
1259.01 & 3.4779790 $\pm$ 6e-07 & 0.11 $\pm$ 0.07 & 0.1493 $\pm$ 0.0007 & 1.27 & 29030 $\pm$ 54 & 451 & (472.05, 467.61, 500) & 2.01e-01 \\
1264.01 & 2.74416 $\pm$ 2e-05 & 0.79 $\pm$ 0.02 & 0.071 $\pm$ 0.003 & 1.41 & 5650 $\pm$ 70 & 462 & (433.37, 432.48, 500) & 7.99e-01 \\
1265.01 & 2.2047380 $\pm$ 8e-07 & 0.49 $\pm$ 0.02 & 0.0777 $\pm$ 0.0003 & ... & 6530 $\pm$ 25 & 277 & (482.22, 480.80, 500) & 6.95e-01 \\
1268.01 & 8.157733 $\pm$ 9e-06 & ... & 0.089 $\pm$ 0.001 & ... & 10340 $\pm$ 59 & 637 & (479.61, 478.72, 500) & 8.22e-01 \\
1282.01 & 0.9689924 $\pm$ 5e-07 & 0.40 $\pm$ 0.05 & 0.1100 $\pm$ 0.0008 & ... & 13500 $\pm$ 131 & 1151 & (374.10, 373.17, 500) & 7.46e-01 \\
1283.01 & 10.33853 $\pm$ 6e-05 & 0.2 $\pm$ 0.1 & 0.121 $\pm$ 0.001 & ... & 18010 $\pm$ 83 & 900 & (358.84, 354.27, 429) & 1.47e-01 \\
1295.01 & 3.196884 $\pm$ 2e-06 & 0.53 $\pm$ 0.02 & 0.0859 $\pm$ 0.0004 & 1.32 & 7960 $\pm$ 29 & 308 & (459.22, 457.92, 500) & 7.09e-01 \\
1296.01 & 3.944374 $\pm$ 4e-06 & 0.31 $\pm$ 0.07 & 0.0765 $\pm$ 0.0005 & ... & 7110 $\pm$ 30 & 367 & (394.00, 390.79, 500) & 2.60e-01 \\
1298.01 & 4.53708 $\pm$ 5e-05 & 0.4 $\pm$ 0.1 & 0.059 $\pm$ 0.001 & ... & 4190 $\pm$ 40 & 434 & (520.19, 518.40, 500) & 6.39e-01 \\
1300.01 & 2.871698 $\pm$ 3e-06 & ... & 0.0846 $\pm$ 0.0002 & 1.41 & 8680 $\pm$ 34 & 399 & (531.56, 528.82, 500) & 4.68e-01 \\
1302.01 & 5.66662 $\pm$ 1e-05 & 0.63 $\pm$ 0.02 & 0.0930 $\pm$ 0.0009 & 1.24 & 9780 $\pm$ 51 & 537 & (479.50, 477.57, 500) & 5.76e-01 \\

\hline
\end{tabular}%
\end{minipage}
}
\end{table*}

\begin{table*}
\rotatebox{90}{%
\begin{minipage}{\textheight} 
\flushleft{Table \ref{tab: ringed model and ringless model fitting result}. (Continued)}
\begin{tabular}{crrrcrrrr}
\hline
\multicolumn{1}{c}{TOI} & \multicolumn{1}{c}{$P_{\rm{orb}}$ [day]} & \multicolumn{1}{c}{$b^{*}$} & \multicolumn{1}{c}{$(R_{\rm p} / R_{\star})_{\rm{ringless}}^{*}$} & \multicolumn{1}{c}{$(r_{\rm out/in })_{\rm{upp}}$} & \multicolumn{1}{c}{$\delta$ [ppm]} & \multicolumn{1}{c}{$\sigma (N_d=500)$ [ppm]} & \multicolumn{1}{c}{$(\chi^{2}_{\rm{ringless,min}},\chi^{2}_{\rm{ring,min}}, N_d)$} & \multicolumn{1}{c}{$p$-value} \\
\hline
1307.01 & 2.534615 $\pm$ 8e-06 & 0.88 $\pm$ 0.02 & 0.08 $\pm$ 0.01 & 1.23 & 5190 $\pm$ 72 & 560 & (467.79, 457.19, 500) & 1.05e-02 \\
1311.01 & 6.14072 $\pm$ 2e-05 & 0.8 $\pm$ 0.1 & 0.12 $\pm$ 0.02 & 1.65 & 9400 $\pm$ 113 & 953 & (462.26, 460.94, 500) & 7.05e-01 \\
1366.01 & 14.8622 $\pm$ 5e-04 $^{\dagger}$ & ... & ... & 1.48 & 11100 $\pm$ 110 & 1162 & (317.52, 308.35, 305) & 3.42e-02 \\
1419.01 & 2.899739 $\pm$ 2e-06 & 0.62 $\pm$ 0.02 & 0.112 $\pm$ 0.002 & 1.52 & 13800 $\pm$ 115 & 856 & (409.09, 406.09, 500) & 3.08e-01 \\
1425.01 & 1.031836 $\pm$ 2e-06 & ... & ... & 1.19 & 5280 $\pm$ 46 & 316 & (413.11, 407.97, 500) & 1.05e-01 \\
1431.01 & 2.6502310 $\pm$ 9e-07 & 0.88 $\pm$ 0.01 & 0.078 $\pm$ 0.008 & 1.12 & 5220 $\pm$ 25 & 200 & (625.91, 619.44, 500) & 1.65e-01 \\
1455.01 & 3.623144 $\pm$ 2e-06 & 0.773 $\pm$ 0.007 & 0.129 $\pm$ 0.004 & 1.21 & 18050 $\pm$ 55 & 559 & (421.42, 415.27, 500) & 6.56e-02 \\
1456.01 & 18.7118 $\pm$ 9e-04 & ... & 0.0699 $\pm$ 0.0003 & 1.42 & 6290 $\pm$ 25 & 342 & (632.19, 622.59, 500) & 5.74e-02 \\
1458.01 & 2.7760 $\pm$ 1e-04 & 0.35 $\pm$ 0.09 & 0.107 $\pm$ 0.001 & ... & 12700 $\pm$ 114 & 1090 & (382.89, 380.07, 409) & 3.99e-01 \\
1612.01 & 2.864133 $\pm$ 2e-06 & 0.74 $\pm$ 0.01 & 0.092 $\pm$ 0.001 & 1.26 & 9600 $\pm$ 42 & 365 & (486.10, 482.72, 500) & 3.30e-01 \\
1627.01 & 5.017216 $\pm$ 4e-06 & 0.2 $\pm$ 0.1 & 0.111 $\pm$ 0.001 & ... & 14240 $\pm$ 75 & 803 & (400.83, 398.69, 500) & 4.54e-01 \\
1628.01 & 2.143638 $\pm$ 1e-06 & 0.65 $\pm$ 0.02 & 0.093 $\pm$ 0.001 & 1.35 & 9200 $\pm$ 59 & 576 & (393.13, 392.02, 500) & 7.07e-01 \\
1636.01 & 1.7731 $\pm$ 1e-04 & ... & ... & ... & 26300 $\pm$ 340 & 2525 & (325.89, 320.09, 419) & 6.13e-02 \\
1637.01 & 4.0769 $\pm$ 4e-04 & 0.43 $\pm$ 0.09 & 0.098 $\pm$ 0.002 & ... & 10500 $\pm$ 114 & 1313 & (425.86, 424.67, 430) & 7.57e-01 \\
1649.01 & 3.45577 $\pm$ 4e-05 & 0.89 $\pm$ 0.03 & 0.09 $\pm$ 0.02 & 1.15 & 6110 $\pm$ 47 & 400 & (573.18, 570.16, 489) & 4.70e-01 \\
1670.01 & 40.7500 $\pm$ 1e-04 & 0.77 $\pm$ 0.01 & 0.077 $\pm$ 0.002 & 1.37 & 6280 $\pm$ 31 & 398 & (402.43, 400.51, 426) & 5.74e-01 \\
1714.01 & 3.474475 $\pm$ 3e-06 & 0.2 $\pm$ 0.1 & 0.1028 $\pm$ 0.0006 & 1.69 & 12920 $\pm$ 45 & 510 & (482.54, 478.89, 500) & 2.94e-01 \\
1721.01 & 4.125060 $\pm$ 5e-06 & 0.29 $\pm$ 0.09 & 0.0875 $\pm$ 0.0007 & ... & 8890 $\pm$ 55 & 632 & (446.07, 445.30, 498) & 8.40e-01 \\
1725.01 & 1.0914170 $\pm$ 5e-07 & 0.36 $\pm$ 0.03 & 0.1162 $\pm$ 0.0005 & ... & 17090 $\pm$ 54 & 494 & (391.48, 389.94, 500) & 5.86e-01 \\
1766.01 & 2.703390 $\pm$ 2e-06 & 0.68 $\pm$ 0.01 & 0.0955 $\pm$ 0.0008 & 1.28 & 10340 $\pm$ 47 & 443 & (452.32, 448.62, 500) & 2.58e-01 \\
1767.01 & 4.35314 $\pm$ 7e-05 & 0.57 $\pm$ 0.04 & 0.094 $\pm$ 0.001 & ... & 10270 $\pm$ 86 & 937 & (509.13, 506.95, 478) & 5.70e-01 \\
1779.01 & 1.881723 $\pm$ 1e-06 & ... & 0.292 $\pm$ 0.004 & 1.21 & 99300 $\pm$ 621 & 3782 & (366.74, 364.10, 490) & 3.25e-01 \\
1796.01 & 2.643898 $\pm$ 1e-06 & 0.85 $\pm$ 0.01 & 0.087 $\pm$ 0.008 & 1.23 & 6400 $\pm$ 97 & 524 & (414.61, 410.50, 355) & 3.28e-01 \\
1810.01 & 1.3273470 $\pm$ 7e-07 & ... & 0.122 $\pm$ 0.001 & 1.52 & 18900 $\pm$ 143 & 1135 & (392.75, 392.19, 500) & 8.75e-01 \\
1811.01 & 3.713067 $\pm$ 5e-06 & 0.78 $\pm$ 0.01 & 0.14 $\pm$ 0.01 & 1.21 & 19600 $\pm$ 242 & 1704 & (380.62, 379.38, 421) & 7.20e-01 \\
1815.01 & 2.555331 $\pm$ 7e-06 & 0.31 $\pm$ 0.06 & 0.0755 $\pm$ 0.0004 & ... & 6550 $\pm$ 33 & 418 & (464.48, 463.68, 500) & 8.40e-01 \\
1823.01 & 38.8136 $\pm$ 4e-04 & 0.43 $\pm$ 0.08 & 0.090 $\pm$ 0.002 & ... & 9570 $\pm$ 68 & 900 & (463.05, 460.65, 500) & 4.67e-01 \\
1825.01 & 10.18242 $\pm$ 2e-05 & 0.78 $\pm$ 0.03 & 0.101 $\pm$ 0.008 & 1.38 & 10600 $\pm$ 120 & 1105 & (463.83, 462.68, 500) & 7.49e-01 \\
1826.01 & 4.141978 $\pm$ 3e-06 & 0.812 $\pm$ 0.008 & 0.106 $\pm$ 0.005 & 1.17 & 11380 $\pm$ 78 & 614 & (391.69, 389.13, 500) & 3.59e-01 \\
1829.01 & 6.2897 $\pm$ 5e-04 & 0.66 $\pm$ 0.03 & 0.144 $\pm$ 0.004 & 1.25 & 24300 $\pm$ 240 & 2294 & (505.72, 503.04, 500) & 4.55e-01 \\
1833.01 & 3.693 $\pm$ 1e-03 & 0.48 $\pm$ 0.04 & 0.0888 $\pm$ 0.0006 & ... & 8580 $\pm$ 38 & 442 & (416.90, 414.27, 460) & 4.15e-01 \\
1840.01 & 7.84561 $\pm$ 5e-05 & 0.45 $\pm$ 0.08 & 0.078 $\pm$ 0.001 & ... & 7020 $\pm$ 48 & 649 & (368.69, 367.79, 384) & 8.22e-01 \\
1841.01 & 4.301193 $\pm$ 8e-06 & 0.3 $\pm$ 0.1 & 0.135 $\pm$ 0.003 & ... & 22100 $\pm$ 248 & 2346 & (356.50, 355.26, 499) & 6.37e-01 \\
1844.01 & 3.213057 $\pm$ 2e-06 & 0.31 $\pm$ 0.09 & 0.139 $\pm$ 0.002 & ... & 22500 $\pm$ 191 & 1545 & (366.23, 362.70, 482) & 2.06e-01 \\
1845.01 & 3.661002 $\pm$ 5e-06 & 0.44 $\pm$ 0.06 & 0.129 $\pm$ 0.002 & ... & 18700 $\pm$ 216 & 1915 & (394.67, 393.55, 500) & 7.06e-01 \\
1859.01 & 63.4834 $\pm$ 2e-04 & ... & 0.066 $\pm$ 0.002 & ... & 5640 $\pm$ 52 & 635 & (497.33, 495.06, 440) & 5.79e-01 \\
1864.01 & 1.6453270 $\pm$ 9e-07 & 0.30 $\pm$ 0.05 & 0.1268 $\pm$ 0.0008 & 1.89 & 19310 $\pm$ 89 & 785 & (513.27, 511.92, 500) & 7.29e-01 \\
1870.01 & 2.76340 $\pm$ 3e-05 & 0.54 $\pm$ 0.05 & 0.106 $\pm$ 0.002 & ... & 12400 $\pm$ 123 & 1271 & (465.04, 457.36, 500) & 4.27e-02 \\
1874.01 & 17.86841 $\pm$ 8e-05 & 0.86 $\pm$ 0.05 & 0.11 $\pm$ 0.04 & 1.34 & 11200 $\pm$ 85 & 1108 & (390.83, 385.98, 500) & 1.05e-01 \\
1877.01 & 3.801591 $\pm$ 4e-06 & 0.51 $\pm$ 0.04 & 0.131 $\pm$ 0.002 & ... & 19000 $\pm$ 198 & 1802 & (440.83, 438.34, 500) & 4.26e-01 \\
1907.01 & 5.4773 $\pm$ 2e-04 & 0.46 $\pm$ 0.06 & 0.0904 $\pm$ 0.0009 & ... & 9010 $\pm$ 73 & 929 & (441.98, 439.67, 500) & 4.63e-01 \\
1927.01 & 4.11177 $\pm$ 2e-05 & 0.86 $\pm$ 0.03 & 0.15 $\pm$ 0.01 & ... & 22100 $\pm$ 336 & 2690 & (376.86, 373.99, 500) & 2.90e-01 \\
1934.01 & 4.5112 $\pm$ 3e-04 & 0.42 $\pm$ 0.08 & 0.100 $\pm$ 0.001 & ... & 11170 $\pm$ 99 & 1163 & (350.08, 346.79, 380) & 3.21e-01 \\
\hline
\end{tabular}%
\end{minipage}
}
\end{table*}

\begin{table*}
\rotatebox{90}{%
\begin{minipage}{\textheight} 
\flushleft{Table \ref{tab: ringed model and ringless model fitting result}. (Continued)}
\begin{tabular}{crrrcrrrr}
\hline
\multicolumn{1}{c}{TOI} & \multicolumn{1}{c}{$P_{\rm{orb}}$ [day]} & \multicolumn{1}{c}{$b^{*}$} & \multicolumn{1}{c}{$(R_{\rm p} / R_{\star})_{\rm{ringless}}^{*}$} & \multicolumn{1}{c}{$(r_{\rm out/in })_{\rm{upp}}$} & \multicolumn{1}{c}{$\delta$ [ppm]} & \multicolumn{1}{c}{$\sigma (N_d=500)$ [ppm]} & \multicolumn{1}{c}{$(\chi^{2}_{\rm{ringless,min}},\chi^{2}_{\rm{ring,min}}, N_d)$} & \multicolumn{1}{c}{$p$-value} \\
\hline
1963.01 & 12.651370 $\pm$ 7e-06 $^{\dagger}$ & 0.61 $\pm$ 0.03 & 0.159 $\pm$ 0.003 & 1.38 & 25600 $\pm$ 183 & 2082 & (409.11, 407.35, 340) & 7.00e-01 \\
1970.01 & 2.55999 $\pm$ 4e-05 & 0.62 $\pm$ 0.03 & 0.103 $\pm$ 0.001 & 1.48 & 11300 $\pm$ 117 & 1061 & (409.58, 407.71, 500) & 5.24e-01 \\
1976.01 & 7.70 $\pm$ 2e-02 & 0.753 $\pm$ 0.009 & 0.092 $\pm$ 0.002 & 1.21 & 7400 $\pm$ 43 & 704 & (521.27, 518.56, 500) & 4.65e-01 \\
2012.01 & 3.056527 $\pm$ 4e-06 & ... & 0.084 $\pm$ 0.001 & 1.93 & 8800 $\pm$ 58 & 652 & (452.24, 450.85, 500) & 6.80e-01 \\
2014.01 & 4.810106 $\pm$ 5e-06 & 0.44 $\pm$ 0.03 & 0.0819 $\pm$ 0.0004 & ... & 7370 $\pm$ 20 & 299 & (569.38, 567.46, 500) & 6.47e-01 \\
2020.01 & 5.633467 $\pm$ 2e-06 & 0.48 $\pm$ 0.03 & 0.0701 $\pm$ 0.0003 & ... & 5290 $\pm$ 21 & 245 & (560.37, 557.50, 500) & 4.71e-01 \\
2031.01 & 5.7156 $\pm$ 2e-04 & 0.40 $\pm$ 0.07 & 0.105 $\pm$ 0.001 & ... & 12200 $\pm$ 102 & 1126 & (381.02, 375.07, 427) & 8.69e-02 \\
2040.01 & 3.86 $\pm$ 2e-02 & ... & 0.108 $\pm$ 0.002 & ... & 14700 $\pm$ 193 & 1701 & (278.40, 276.59, 290) & 6.10e-01 \\
2109.01 & 0.67248 $\pm$ 2e-05 & 0.75 $\pm$ 0.02 & 0.079 $\pm$ 0.003 & ... & 7890 $\pm$ 92 & 681 & (487.86, 476.87, 460) & 1.65e-02 \\
2119.01 & 7.200856 $\pm$ 3e-06 & 0.626 $\pm$ 0.009 & 0.222 $\pm$ 0.003 & 1.12 & 51200 $\pm$ 182 & 1463 & (464.62, 458.47, 428) & 1.34e-01 \\
2121.01 & 2.31974 $\pm$ 3e-05 & ... & ... & ... & 19500 $\pm$ 199 & 1552 & (545.56, 544.73, 500) & 8.62e-01 \\
2124.01 & 3.553926 $\pm$ 4e-06 & 0.77 $\pm$ 0.01 & 0.098 $\pm$ 0.004 & 1.26 & 9370 $\pm$ 73 & 763 & (449.81, 449.61, 500) & 9.73e-01 \\
2125.01 & 5.031632 $\pm$ 8e-06 & 0.38 $\pm$ 0.08 & 0.0704 $\pm$ 0.0006 & ... & 5450 $\pm$ 39 & 512 & (463.45, 460.97, 500) & 4.53e-01 \\
2127.01 & 5.5081 $\pm$ 3e-04 & ... & 0.131 $\pm$ 0.005 & ... & 21600 $\pm$ 294 & 2621 & (344.96, 344.87, 476) & 9.88e-01 \\
2129.01 & 4.627660 $\pm$ 3e-06 & 0.91 $\pm$ 0.04 & 0.08 $\pm$ 0.04 & 1.26 & 6190 $\pm$ 58 & 486 & (415.90, 414.98, 500) & 7.81e-01 \\
2131.01 & 1.8468350 $\pm$ 7e-07 & 0.49 $\pm$ 0.02 & 0.1054 $\pm$ 0.0005 & ... & 12050 $\pm$ 42 & 389 & (520.36, 512.96, 500) & 7.09e-02 \\
2135.01 & 2.788473 $\pm$ 2e-06 & 0.30 $\pm$ 0.08 & 0.112 $\pm$ 0.001 & ... & 14470 $\pm$ 99 & 930 & (397.93, 397.24, 499) & 8.36e-01 \\
2137.01 & 14.15024 $\pm$ 3e-05 & 0.65 $\pm$ 0.04 & 0.20 $\pm$ 0.01 & 1.34 & 41400 $\pm$ 412 & 4414 & (312.98, 312.04, 320) & 8.18e-01 \\
2140.01 & 2.470614 $\pm$ 1e-06 & 0.85 $\pm$ 0.03 & 0.12 $\pm$ 0.02 & 1.28 & 14310 $\pm$ 97 & 714 & (506.50, 498.94, 500) & 6.07e-02 \\
2159.01 & 10.05094 $\pm$ 4e-05 & 0.4 $\pm$ 0.1 & 0.093 $\pm$ 0.001 & ... & 8660 $\pm$ 69 & 942 & (480.46, 478.20, 500) & 5.10e-01 \\
2197.01 & 4.9547 $\pm$ 2e-04 & 0.53 $\pm$ 0.05 & 0.0788 $\pm$ 0.0008 & ... & 6740 $\pm$ 62 & 672 & (358.16, 353.95, 354) & 2.54e-01 \\
2220.01 & 1.392653 $\pm$ 6e-06 & ... & ... & 1.54 & 7220 $\pm$ 81 & 636 & (441.92, 430.80, 500) & 5.86e-03 \\
2222.01 & 2.245685 $\pm$ 8e-06 & 0.2 $\pm$ 0.1 & 0.107 $\pm$ 0.001 & ... & 12600 $\pm$ 120 & 1174 & (445.65, 445.34, 500) & 9.52e-01 \\
2251.01 & 11.57434 $\pm$ 4e-05 & 0.49 $\pm$ 0.06 & 0.114 $\pm$ 0.002 & ... & 14300 $\pm$ 122 & 1645 & (393.32, 390.16, 500) & 2.65e-01 \\
2380.01 & 5.21536 $\pm$ 3e-05 & 0.38 $\pm$ 0.08 & 0.0693 $\pm$ 0.0006 & ... & 5220 $\pm$ 38 & 491 & (489.64, 487.24, 500) & 4.91e-01 \\
2547.01 & 2.311420 $\pm$ 3e-06 & 0.35 $\pm$ 0.05 & 0.1160 $\pm$ 0.0008 & ... & 16090 $\pm$ 90 & 797 & (486.53, 482.29, 500) & 2.32e-01 \\
2548.01 & 8.523 $\pm$ 7e-03 & 0.68 $\pm$ 0.02 & 0.128 $\pm$ 0.003 & 1.37 & 18900 $\pm$ 262 & 2251 & (433.81, 430.18, 351) & 4.11e-01 \\
2579.01 & 3.71466 $\pm$ 1e-05 & ... & 0.1001 $\pm$ 0.0009 & 1.74 & 12620 $\pm$ 71 & 770 & (395.80, 394.59, 469) & 7.05e-01 \\
3460.01 & 4.6289 $\pm$ 4e-04 & 0.40 $\pm$ 0.06 & 0.0694 $\pm$ 0.0005 & ... & 5380 $\pm$ 35 & 485 & (473.94, 469.09, 500) & 1.69e-01 \\
3678.01 & 4.8542 $\pm$ 2e-04 & 0.50 $\pm$ 0.05 & 0.0853 $\pm$ 0.0008 & ... & 7840 $\pm$ 48 & 633 & (475.01, 474.63, 500) & 9.41e-01 \\
3908.01 & 1.46811 $\pm$ 2e-05 & ... & ... & 1.44 & 18500 $\pm$ 156 & 1226 & (438.29, 435.66, 468) & 4.31e-01 \\
3960.01 & 3.23998 $\pm$ 4e-05 & ... & 0.124 $\pm$ 0.002 & ... & 18400 $\pm$ 253 & 2301 & (455.20, 454.64, 500) & 8.94e-01 \\
4059.01 & 2.038174 $\pm$ 8e-06 & 0.3 $\pm$ 0.1 & 0.133 $\pm$ 0.002 & ... & 23000 $\pm$ 278 & 2148 & (444.14, 443.10, 500) & 7.64e-01 \\
4087.01 & 3.177478 $\pm$ 5e-06 & 0.2 $\pm$ 0.1 & 0.1065 $\pm$ 0.0009 & ... & 14190 $\pm$ 84 & 782 & (463.57, 462.48, 500) & 7.65e-01 \\
4103.01 & 12.27408 $\pm$ 4e-05 & 0.3 $\pm$ 0.2 & 0.107 $\pm$ 0.002 & ... & 14400 $\pm$ 154 & 1588 & (252.48, 250.79, 295) & 5.90e-01 \\
4138.01 & 3.66004 $\pm$ 2e-05 & 0.4 $\pm$ 0.1 & 0.081 $\pm$ 0.002 & ... & 8600 $\pm$ 58 & 684 & (416.02, 412.24, 500) & 2.14e-01 \\
4145.01 & 4.0664 $\pm$ 1e-04 & 0.82 $\pm$ 0.05 & 0.13 $\pm$ 0.03 & 1.41 & 19700 $\pm$ 311 & 2234 & (383.46, 380.02, 419) & 2.97e-01 \\
4381.01 & 1.49742 $\pm$ 5e-05 & ... & 0.186 $\pm$ 0.002 & 1.28 & 39300 $\pm$ 427 & 3495 & (467.25, 458.69, 500) & 2.83e-02 \\
4427.01 & 4.61827 $\pm$ 1e-05 & 0.68 $\pm$ 0.02 & 0.141 $\pm$ 0.003 & 1.18 & 20300 $\pm$ 156 & 1406 & (413.11, 408.58, 500) & 1.44e-01 \\
4436.01 & 1.759059 $\pm$ 4e-06 & 0.3 $\pm$ 0.1 & 0.0973 $\pm$ 0.0007 & ... & 10500 $\pm$ 105 & 1085 & (458.53, 458.25, 500) & 9.59e-01 \\
4458.01 & 3.93938 $\pm$ 3e-05 & ... & 0.0599 $\pm$ 0.0003 & ... & 4110 $\pm$ 36 & 475 & (554.60, 554.28, 500) & 9.63e-01 \\
4468.01 & 2.77086 $\pm$ 1e-05 & 0.3 $\pm$ 0.1 & 0.137 $\pm$ 0.002 & ... & 22600 $\pm$ 355 & 2867 & (509.10, 506.85, 500) & 5.38e-01 \\
4486.01 & 2.691558 $\pm$ 5e-06 & 0.46 $\pm$ 0.05 & 0.0814 $\pm$ 0.0006 & ... & 7120 $\pm$ 51 & 582 & (638.25, 629.83, 500) & 8.90e-02 \\
\hline
\end{tabular}%
\end{minipage}
}
\end{table*}

\begin{table*}
\rotatebox{90}{%
\begin{minipage}{\textheight} 
\begin{threeparttable}
\flushleft{Table \ref{tab: ringed model and ringless model fitting result}. (Continued)}
\begin{tabular}{crrrcrrrr}
\hline
\multicolumn{1}{c}{TOI} & \multicolumn{1}{c}{$P_{\rm{orb}}$ [day]} & \multicolumn{1}{c}{$b^{*}$} & \multicolumn{1}{c}{$(R_{\rm p} / R_{\star})_{\rm{ringless}}^{*}$} & \multicolumn{1}{c}{$(r_{\rm out/in })_{\rm{upp}}$} & \multicolumn{1}{c}{$\delta$ [ppm]} & \multicolumn{1}{c}{$\sigma (N_d=500)$ [ppm]} & \multicolumn{1}{c}{$(\chi^{2}_{\rm{ringless,min}},\chi^{2}_{\rm{ring,min}}, N_d)$} & \multicolumn{1}{c}{$p$-value} \\
\hline
4516.01 & 3.7227 $\pm$ 3e-04 & ... & 0.129 $\pm$ 0.003 & ... & 20000 $\pm$ 181 & 1604 & (412.93, 410.57, 436) & 4.86e-01 \\
4518.01 & 3.409 $\pm$ 2e-03 & 0.4 $\pm$ 0.1 & 0.117 $\pm$ 0.003 & ... & 15100 $\pm$ 191 & 1910 & (227.01, 224.14, 256) & 3.72e-01 \\
4530.01 & 3.65278 $\pm$ 8e-05 & 0.48 $\pm$ 0.06 & 0.131 $\pm$ 0.002 & ... & 19100 $\pm$ 245 & 2203 & (444.49, 439.72, 499) & 1.52e-01 \\
4536.01 & 4.0461 $\pm$ 2e-04 & ... & 0.0814 $\pm$ 0.0009 & ... & 7540 $\pm$ 60 & 740 & (444.62, 443.70, 500) & 7.99e-01 \\
4606.01 & 11.169 $\pm$ 2e-03 & ... & 0.0875 $\pm$ 0.0004 & ... & 9660 $\pm$ 49 & 630 & (517.25, 515.51, 491) & 6.55e-01 \\
4618.01 & 3.79988 $\pm$ 7e-05 & 0.70 $\pm$ 0.03 & 0.159 $\pm$ 0.009 & 1.27 & 27500 $\pm$ 471 & 3087 & (419.28, 415.11, 484) & 1.92e-01 \\
4623.01 & 3.25879 $\pm$ 5e-05 & 0.61 $\pm$ 0.03 & 0.146 $\pm$ 0.003 & 1.29 & 22800 $\pm$ 244 & 1995 & (333.16, 330.86, 384) & 4.58e-01 \\
5093.01 & 2.29841 $\pm$ 3e-05 & ... & 0.159 $\pm$ 0.005 & ... & 33400 $\pm$ 475 & 3621 & (456.81, 455.97, 500) & 8.24e-01 \\
5094.01 & 3.33667 $\pm$ 5e-05 & 0.3 $\pm$ 0.2 & 0.077 $\pm$ 0.001 & ... & 6500 $\pm$ 109 & 803 & (451.98, 448.13, 473) & 2.65e-01 \\
5096.01 & 2.87533 $\pm$ 2e-05 & 0.66 $\pm$ 0.01 & 0.145 $\pm$ 0.002 & 1.26 & 21700 $\pm$ 128 & 967 & (439.09, 434.48, 479) & 1.75e-01 \\
5144.01 & 2.65568 $\pm$ 2e-05 & ... & 0.1321 $\pm$ 0.0008 & 1.52 & 20260 $\pm$ 94 & 836 & (502.11, 498.48, 500) & 3.14e-01 \\
5149.01 & 27.37150 $\pm$ 8e-05 $^{\dagger}$ & 0.44 $\pm$ 0.07 & 0.101 $\pm$ 0.001 & ... & 11050 $\pm$ 78 & 1334 & (503.03, 499.67, 500) & 3.49e-01 \\
5374.01 & 2.9164 $\pm$ 1e-04 & 0.73 $\pm$ 0.02 & 0.086 $\pm$ 0.002 & 1.70 & 7260 $\pm$ 66 & 657 & (470.47, 468.61, 466) & 6.13e-01 \\
5375.01 & 1.721552 $\pm$ 3e-06 & 0.47 $\pm$ 0.04 & 0.171 $\pm$ 0.002 & ... & 31800 $\pm$ 496 & 3591 & (552.13, 549.09, 500) & 4.39e-01 \\
5379.01 & 12.74265 $\pm$ 2e-05 $^{\dagger}$ & 0.89 $\pm$ 0.05 & 0.11 $\pm$ 0.05 & ... & 10800 $\pm$ 125 & 1314 & (334.66, 330.48, 315) & 2.79e-01 \\
5384.01 & 2.9896 $\pm$ 1e-04 & 0.64 $\pm$ 0.02 & 0.108 $\pm$ 0.001 & 1.18 & 12120 $\pm$ 67 & 680 & (437.35, 435.17, 459) & 5.24e-01 \\
5394.01 & 15.1934 $\pm$ 7e-04 & 0.9 $\pm$ 0.1 & 0.10 $\pm$ 0.04 & ... & 5720 $\pm$ 64 & 570 & (583.72, 582.14, 460) & 7.48e-01 \\
5631.01 & 2.24381 $\pm$ 3e-05 & 0.55 $\pm$ 0.04 & 0.094 $\pm$ 0.001 & ... & 10310 $\pm$ 70 & 616 & (398.94, 398.26, 394) & 8.84e-01 \\
5675.01 & 4.0549 $\pm$ 1e-04 & 0.3 $\pm$ 0.2 & 0.141 $\pm$ 0.004 & ... & 23200 $\pm$ 273 & 2509 & (288.71, 287.31, 357) & 6.40e-01 \\
5676.01 & 3.5062 $\pm$ 1e-04 & ... & ... & ... & 18400 $\pm$ 206 & 1432 & (322.76, 321.34, 370) & 6.60e-01 \\
5679.01 & 1.6624 $\pm$ 5e-04 & 0.60 $\pm$ 0.08 & 0.074 $\pm$ 0.002 & ... & 12600 $\pm$ 151 & 1535 & (604.09, 591.02, 462) & 1.95e-02 \\
5793.01 & 2.6940 $\pm$ 2e-04 & ... & 0.102 $\pm$ 0.001 & 1.89 & 11700 $\pm$ 106 & 1181 & (398.52, 394.19, 434) & 2.00e-01 \\
5796.01 & 1.7430 $\pm$ 1e-04 & ... & 0.169 $\pm$ 0.003 & 1.90 & 32200 $\pm$ 394 & 3271 & (497.38, 491.29, 446) & 1.46e-01 \\
5797.01 & 2.1523 $\pm$ 2e-04 & 0.71 $\pm$ 0.03 & 0.132 $\pm$ 0.007 & 1.73 & 17000 $\pm$ 251 & 1820 & (254.87, 253.39, 292) & 6.50e-01 \\
5801.01 & 3.06786 $\pm$ 9e-05 & 0.2 $\pm$ 0.1 & 0.171 $\pm$ 0.002 & ... & 34500 $\pm$ 182 & 1449 & (321.59, 316.61, 355) & 1.45e-01 \\
5823.01 & 3.8681 $\pm$ 3e-04 & 0.68 $\pm$ 0.02 & 0.126 $\pm$ 0.003 & 1.20 & 17260 $\pm$ 93 & 754 & (435.76, 432.75, 372) & 4.72e-01 \\
6072.01 & 3.6915 $\pm$ 1e-04 & 0.63 $\pm$ 0.05 & 0.122 $\pm$ 0.004 & ... & 16300 $\pm$ 151 & 1368 & (412.89, 410.90, 396) & 6.01e-01 \\
\hline
\end{tabular}%
\begin{tablenotes}
\item[*] Due to the high uncertainties, the parameter value cannot be determined accurately and is therefore denoted as ``...''.
\item[$\dagger$] Values from the ExoFOP-TESS TOI list on 2022 September 13 (https://exofop.ipac.caltech.edu/tess).
\end{tablenotes}
\end{threeparttable}
\end{minipage}
}
\end{table*}

\begin{table*}
\centering
\rotatebox{90}{%
\begin{minipage}{\textheight} 
\begin{threeparttable}
\caption{Confidence level of ring detection using the transit depth variation caused by the precession of the planetary rotation observed by TESS.
}
\label{tab: precession targets in TESS}
\begin{tabular}{rrrrrrrrr}
  \hline
TOI& $P_{\rm{orb}}$ [day] & $R_{\rm p}$ [$R_{\rm{earth}}$] & $M_{\rm{p}}^*$ [kg] & $P_{\rm{prec}}$ [yr] & $t_{\rm{damp}}$ [Gyr] & Transit Depth Error [ppm] & slope [$10^{-8}$ day$^{-1}$] & $\sigma_{\rm{slope}}$ \\
\hline
173.01 & 29.75361 $\pm$ 3e-05 & 12.15 $\pm$ 0.51 & 2.47e+28  & 74.005 $\pm$ 2e-03 & 2.70 $\pm$ 0.72 & 48.65 & -85 $\pm$ 3 & 26 \\[3pt]
201.01 & 52.97803 $\pm$ 4e-05 & 11.94 $\pm$ 0.61 & 2.47e+28  & 234.626 $\pm$ 5e-03 & 16.08 $\pm$ 4.43 & 48.14 & -29 $\pm$ 2 & 10 \\[3pt]
553.03 & 40.8905 $\pm$ 2e-04 & 2.71 $\pm$ 1.04 & 4.47e+25 $_{-2.67}^{+5.77}$ & 139.77 $\pm$ 2e-02 & 1.14 $\pm$ 1.71 & 109.28 & 21 $\pm$ 3 & 5 \\[3pt]
1456.01 & 18.71167 $\pm$ 1e-05 & 10.17 $\pm$ 0.45 & 2.47e+28  & 29.2691 $\pm$ 5e-04 & 1.15 $\pm$ 0.27 & 40.50 & -181 $\pm$ 10 & 17 \\[3pt]
1670.01 & 40.75014 $\pm$ 3e-05 & 11.10 $\pm$ 0.52 & 2.47e+28  & 138.817 $\pm$ 3e-03 & 9.10 $\pm$ 2.31 & 78.98 & -40 $\pm$ 4 & 9 \\[3pt]
1859.01 & 63.4834 $\pm$ 2e-04 & 11.13 $\pm$ 0.56 & 2.47e+28  & 336.90 $\pm$ 2e-02 & 34.15 $\pm$ 9.00 & 100.86 & -29 $\pm$ 4 & 6 \\[3pt]
1890.01 & 19.96769 $\pm$ 4e-05 & 10.01 $\pm$ 0.52 & 2.47e+28  & 33.330 $\pm$ 2e-03 & 1.46 $\pm$ 0.35 & 245.98 & -117 $\pm$ 17 & 6 \\[3pt]
1899.01 & 29.09023 $\pm$ 7e-05 & 12.18 $\pm$ 0.41 & 2.47e+28  & 70.742 $\pm$ 4e-03 & 2.51 $\pm$ 0.28 & 707.51 & -530 $\pm$ 64 & 8 \\[3pt]
4316.01 & 20.27300 $\pm$ 3e-05 & 11.52 $\pm$ 0.60 & 2.47e+28  & 34.357 $\pm$ 1e-03 & 1.00 $\pm$ 0.23 & 625.37 & -300 $\pm$ 52 & 5 \\[3pt]
4553.01 & 20.313 $\pm$ 1e-03 & 10.80 $\pm$ 0.53 & 2.47e+28  & 34.49 $\pm$ 4e-02 & 1.22 $\pm$ 0.33 & 129.02 & -126 $\pm$ 38 & 3 \\[3pt]

\hline
\end{tabular}
\begin{tablenotes}
\item[*] For $R_{\rm p}$ larger than $10 R_{\rm{earth}}$, $M_{\rm{p}}$ is set to $2.47 \times 10^{28}$ kg, which is an upper bound mass for planets derived from brown dwarfs.
\end{tablenotes}
\end{threeparttable}
\end{minipage}
}
\end{table*}

\begin{table*}
\rotatebox{90}{%
\begin{minipage}{\textheight} 
\begin{threeparttable}
\caption{Confidence level of ring detection using the transit depth variation caused by the precession of the planetary rotation observed by Kepler.}
\label{tab: precession targets in kepler}
\begin{tabular}{rrrrrrrrr}
  \hline
KOI & $P_{\rm{orb}}$ [day] & $R_{\rm p}$ [$R_{\rm{earth}}$] & $M_{\rm{p}}^*$ [kg]  & $P_{\rm{prec}}$ [yr] & $t_{\rm{damp}}$ [Gyr] & Transit Depth Error [ppm] & slope [$10^{-8}$ day$^{-1}$] & $\sigma_{\rm{slope}}$ \\[3pt]
\hline
94.03 & 54.31996 $\pm$ 3e-05 & 6.07 $_{-0.91}^{+1.11}$ & 1.91e+26 $_{-0.93}^{+1.88}$ & 246.662 $\pm$ 4e-03 & 1.02 $\pm$ 0.91 & 21.80 & -7 $\pm$ 1 & 4 \\[3pt]
125.01 & 38.478766 $\pm$ 5e-06 & 13.26 $_{-0.59}^{+1.78}$ & 2.47e+28  & 123.7731 $\pm$ 4e-04 & 4.49 $\pm$ 1.26 & 80.38 & -166 $\pm$ 6 & 25 \\[3pt]
189.01 & 30.360445 $\pm$ 5e-06 & 10.99 $_{-1.11}^{+0.90}$ & 2.47e+28  & 77.0549 $\pm$ 3e-04 & 3.88 $\pm$ 1.19 & 85.74 & -286 $\pm$ 31 & 8 \\[3pt]
209.01 & 50.79035 $\pm$ 2e-05 & 10.60 $_{-1.90}^{+1.91}$ & 2.47e+28  & 215.649 $\pm$ 2e-03 & 20.23 $\pm$ 11.22 & 51.16 & -24 $\pm$ 3 & 6 \\[3pt]
245.01 & 39.79220 $\pm$ 3e-05 & 1.90 $_{-0.07}^{+0.08}$ & 2.83e+25 $_{-1.11}^{+2.02}$ & 132.367 $\pm$ 2e-03 & 1.94 $\pm$ 1.11 & 12.46 & -4 $\pm$ 0 & 5 \\[3pt]
276.01 & 41.74599 $\pm$ 5e-05 & 2.40 $_{-0.13}^{+0.12}$ & 3.86e+25 $_{-1.61}^{+3.04}$ & 145.685 $\pm$ 4e-03 & 1.52 $\pm$ 0.96 & 22.16 & -4 $\pm$ 1 & 3 \\[3pt]
319.01 & 46.15119 $\pm$ 3e-05 & 10.02 $_{-0.40}^{+0.73}$ & 2.47e+28  & 178.053 $\pm$ 3e-03 & 17.97 $\pm$ 4.24 & 29.77 & -10 $\pm$ 2 & 3 \\[3pt]
398.01 & 51.84689 $\pm$ 3e-05 & 8.73 $_{-0.42}^{+0.70}$ & 2.16e+27 $_{-1.91}^{+74.38}$ & 224.714 $\pm$ 3e-03 & 3.37 $\pm$ 59.55 & 160.04 & -34 $\pm$ 10 & 3 \\[3pt]
718.03 & 47.9034 $\pm$ 2e-04 & 3.55 $_{-0.77}^{+0.59}$ & 6.76e+25 $_{-3.35}^{+6.37}$ & 191.83 $\pm$ 2e-02 & 1.24 $\pm$ 1.15 & 52.03 & -12 $\pm$ 2 & 5 \\[3pt]
806.02 & 60.32494 $\pm$ 3e-05 & 12.19 $_{-0.52}^{+1.11}$ & 2.47e+28  & 304.213 $\pm$ 4e-03 & 22.29 $\pm$ 4.85 & 182.32 & -73 $\pm$ 19 & 3 \\[3pt]
918.01 & 39.64316 $\pm$ 1e-05 & 10.98 $_{-0.99}^{+0.72}$ & 2.47e+28  & 131.3774 $\pm$ 8e-04 & 8.65 $\pm$ 2.13 & 111.92 & -121 $\pm$ 14 & 8 \\[3pt]
1070.03 & 92.78 $\pm$ 2e-02 & 1.92 $_{-0.20}^{+0.84}$ & 3.86e+25 $_{-1.79}^{+3.64}$ & 720 $\pm$ 3e+00 & 32.49 $\pm$ 35.64 & 156.22 & -79 $\pm$ 12 & 6 \\[3pt]
1241.01 & 21.4054 $\pm$ 1e-04 & 11.64 $_{-1.71}^{+0.79}$ & 2.47e+28  & 38.303 $\pm$ 5e-03 & 1.14 $\pm$ 0.46 & 49.79 & -16 $\pm$ 3 & 4 \\[3pt]

\hline
\end{tabular}%
\begin{tablenotes}
\item[*] For $R_{\rm p}$ larger than $10 R_{\rm{earth}}$, $M_{\rm{p}}$ is set to $2.47 \times 10^{28}$ kg, which is an upper bound mass for planets derived from brown dwarfs.
\end{tablenotes}
\end{threeparttable}
\end{minipage}
}
\end{table*}


\begin{appendix}
\section{SAP flux analysis} \label{SAP flux analysis}
As pointed out by \cite{2021AJ....162..127W}, when the flux variation due to stellar surface activity is significant, the PDC detrending process may fail to distinguish it from instrumental noise, resulting in time-correlated noise that degrades the quality of the light curve \citep[e.g., ][]{2020AJ....160..155W, 2021AJ....161..131D}.
We applied the same processing described in sections \ref{Data reduction and making phase-folded light curve} and \ref{Comparison of fitting results for models with and without rings} to SAP light curves as we did with PDC-SAP, to search for ringed planet candidates among 308 TOIs.
Since SAP flux includes contamination by light from neighbouring sources, we corrected this using equation (\ref{eq: correct_crowding}), which is the same as that given by equation (2) of \citet{2012PASP..124..985S},

\begin{equation}\label{eq: correct_crowding}
    F_{\rm{corr}}=\frac{F_{\rm{sap}}-F_{\rm{sap, med}}(1-c)}{f}.
\end{equation}

$F_{\rm{corr}}$ is the corrected SAP flux, $F_{\rm{sap}}$ is original SAP flux, $F_{\rm{sap, med}}$ is the median of $F_{\rm{sap}}$, $c$ is crowding metric, i.e. fraction of the flux from the target host star over the total detected flux in the field which contains the contribution from other sources, and $f$ is the fraction of the true target flux detected by TESS which is not always 100 \%. 
Since the objects that satisfy the criterion of $p$-value $< 0.0027$ are the same for both the PDC-SAP and SAP data, we adopt the results from the PDC-SAP light curves for this study.

\section{Exceptional cases} \label{Exceptional cases}
\begin{figure*}
  \begin{tabular}{cc}

  \begin{minipage}[t]{0.50\hsize}
  \centering
  \includegraphics[width=80mm]{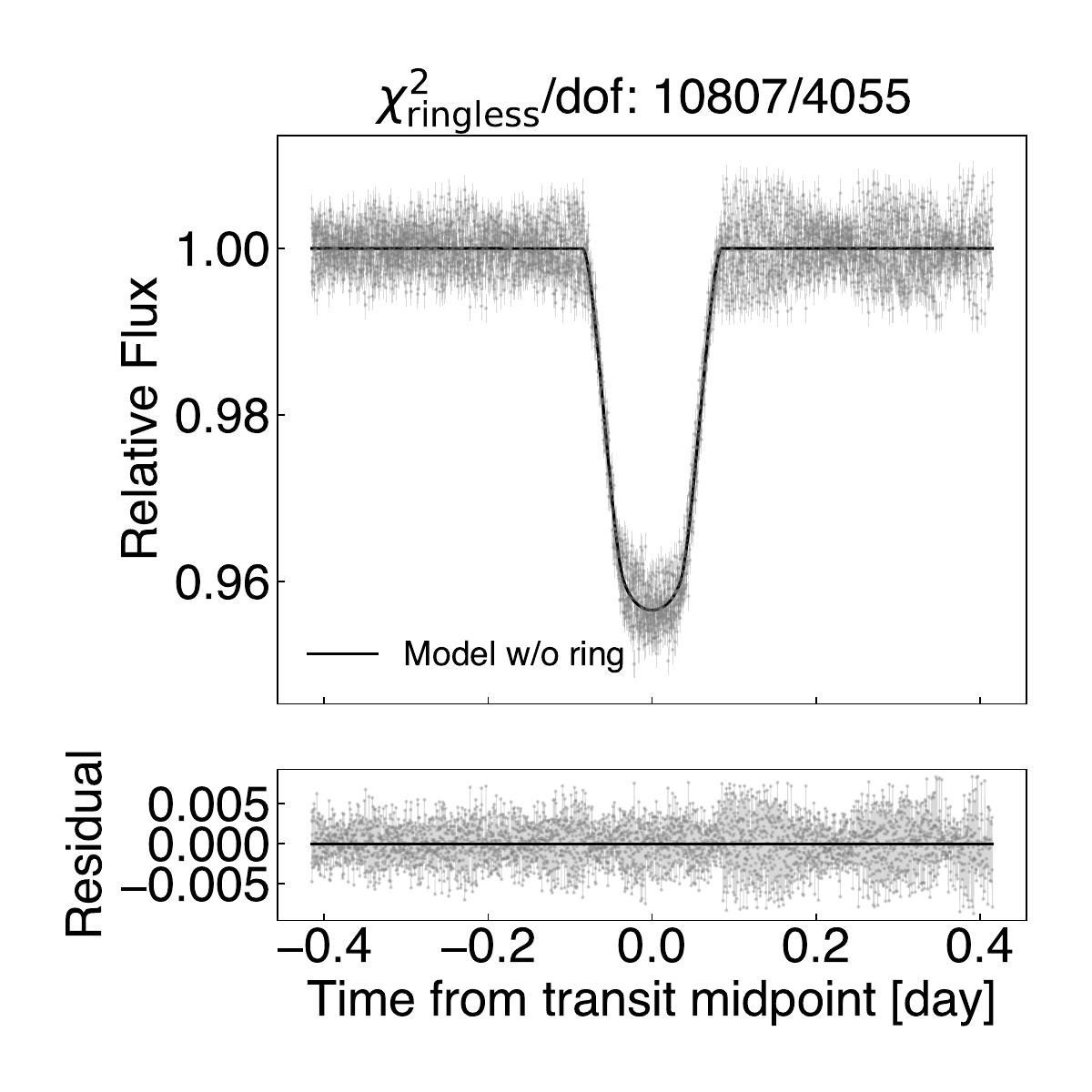}
  \subcaption{Frequent and periodic variations in TOI-890.01.}
  \end{minipage} 
  
  \begin{minipage}[t]{0.50\hsize}
  \centering
  \includegraphics[width=80mm]{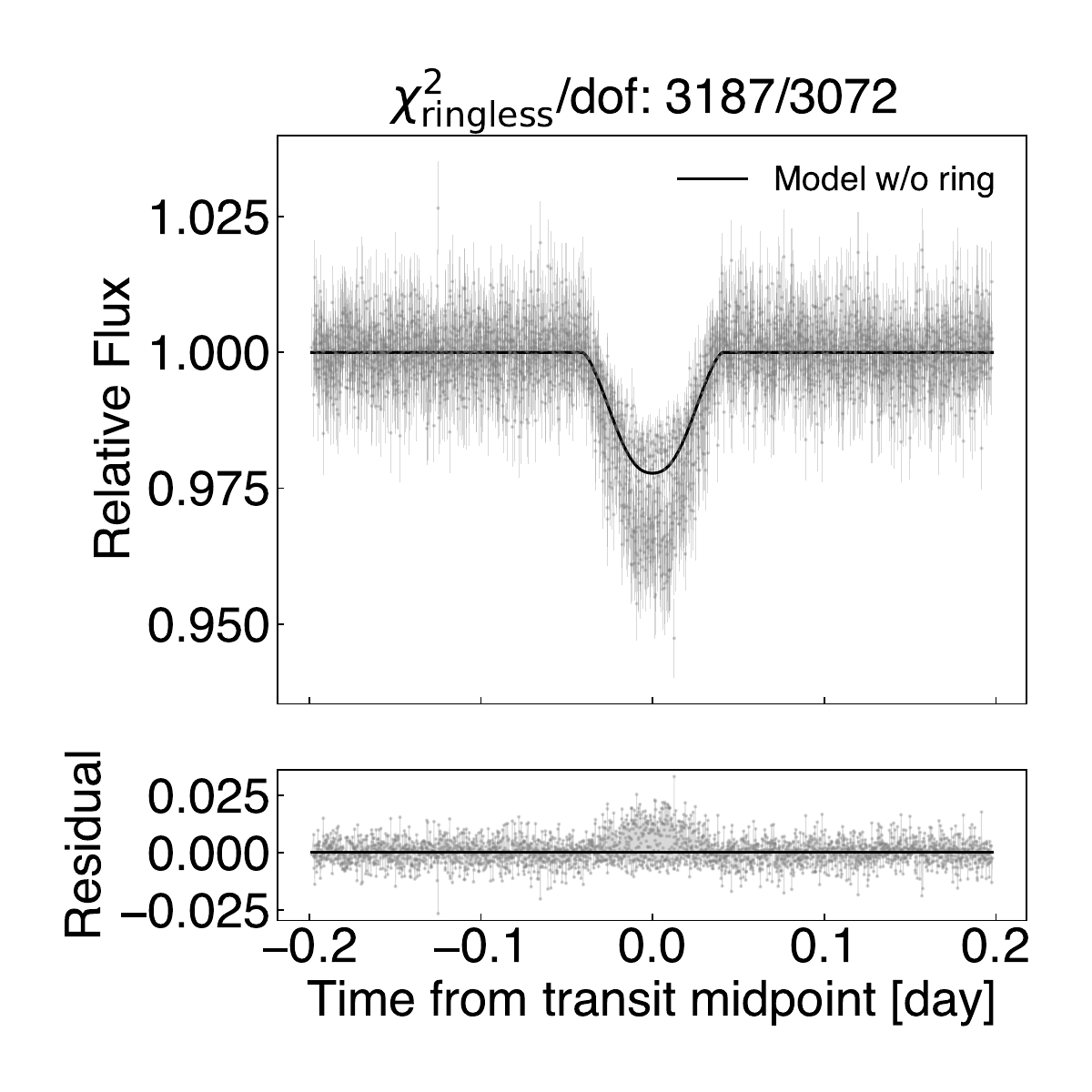}
  \subcaption{Transit depth variations in TOI-1764.01. Light curves with varying transit depths are phase-folded.}
  \end{minipage}
  
  \end{tabular}
  \caption{Examples of data excluded in our preprocessing pipeline.}
  \label{fig: failure pattern}
\end{figure*}

The following exceptions existed in our targets, each of which is addressed here. 
As a result, 14 TOIs are excluded from our analysis. 
Figure \ref{fig: failure pattern} shows some examples of light curves for exceptional cases.
\begin{description}
    \item Failure in determining transit duration: For TOI-182.01, TOI-187.01, TOI-189.01, TOI-275.01, TOI-830.01, TOI-1059.01, TOI-1124.01, TOI-1425.01, and TOI-1834.01, transit duration after folding the light curve cannot be accurately determined through the fitting. 
    This is because the transit shape is close to V-shape, or the planetary $b$ is large, making it difficult to accurately determine transit parameters.
    In such cases, we use the duration value from ETL22 instead of the one from our fitting.
    However, for TOI-189.01, even using the value from ETL22 does not allow for proper extraction of the entire transit. 
    Therefore, we extract the light curve within $\pm\, 0.9\, T_{\rm dur}$ of the transit centre.
    
    \item Significant deviations in transit timing due to other planets: TOI-216 is a multi-planetary system. 
    In the observational data after Sector 27, the actual timing of TOI-216.01 transit events deviated from the predicted timing calculated from the orbital period and mid-transit time, so our pipeline cannot process these light curve data directly. 
    Therefore, we manually set the extraction times to match the actual transit events in preprocessing light curve data after Sector 27.

    \item Anomalous variations in flux: In the case of TOI-1236.01, TOI-1257.01 and TOI-4518.01, sudden peaks, possibly caused by stellar flares, are observed in the light curve. 
    We exclude the events containing flares when folding light curves.

    \item Transit fitting failure due to low $(S/N)$ at each epoch: In the case of TOI-1838.01, the transit depth flux error is too large to find the best fit of the transit model. 
    We apply our pipeline using the fitting parameters from the event where the model successfully fits the transit light curve as the initial values for transit fitting. 
    
    \item The sector with insufficient data quality:
    The light curve of TOI-1838.01 has minus flux values in Sector 50, indicating poor data quality. 
    We exclude the light curve data in this sector from our analysis.

    \item Frequent and periodic variations in flux: In the case of TOI-651.01, TOI-890.01, TOI-1417.01, TOI-1421.01, TOI-1599.01, TOI-2132.01 and TOI-4505.01 variations are too fast to be fitted with the polynomial model and we cannot apply our pipeline process to the data. 
    We exclude these TOIs from our analysis because it would be difficult to detect features caused by planetary rings accurately.

    \item Single transit: We exclude TOI-316.01, TOI-588.01, TOI-850.01, TOI-1356.01 and TOI-2449.01 because only a single transit event is obtained for each. 
    This means that it is impossible to confirm whether detected ring signals are genuinely present or merely a coincidental superposition of other events.

    \item Transit depth variations: We exclude TOI-1764.01 from our analysis because it shows transit depth variations across each transit event, and we cannot fold all transit events accurately.

    \item No usable transit event: We exclude TOI-2180.01 from our analysis because all transit events give fractional useful data below 90 \% of the required length, failing to satisfy prerequisites for an accurate transit fit.
    
\end{description}

\section{Objects with $p$-value $<0.0027$ in excepted TOIs} \label{Objects with p-value<0.0027 in excepted TOIs}
\begin{figure}
  \centering
  \includegraphics[width=80mm]{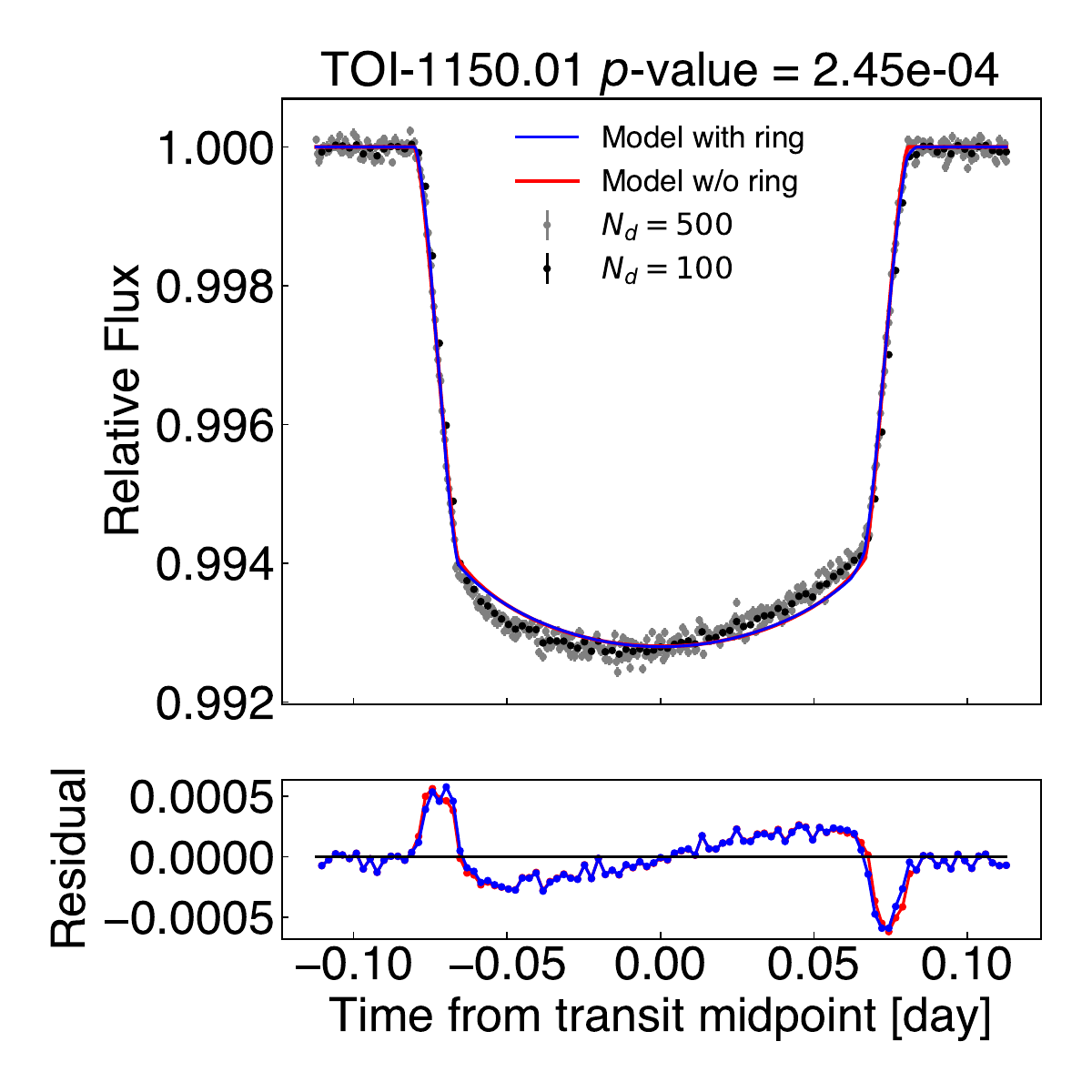}
  \caption{Asymmetric light curve of TOI-1150.01 caused by gravity darkening.}
  \label{fig: TOI1150.01}
\end{figure}

\begin{figure*}[width=160mm]
  \centering
  \begin{minipage}{0.49\textwidth}
    \centering
    \includegraphics[width=\textwidth]{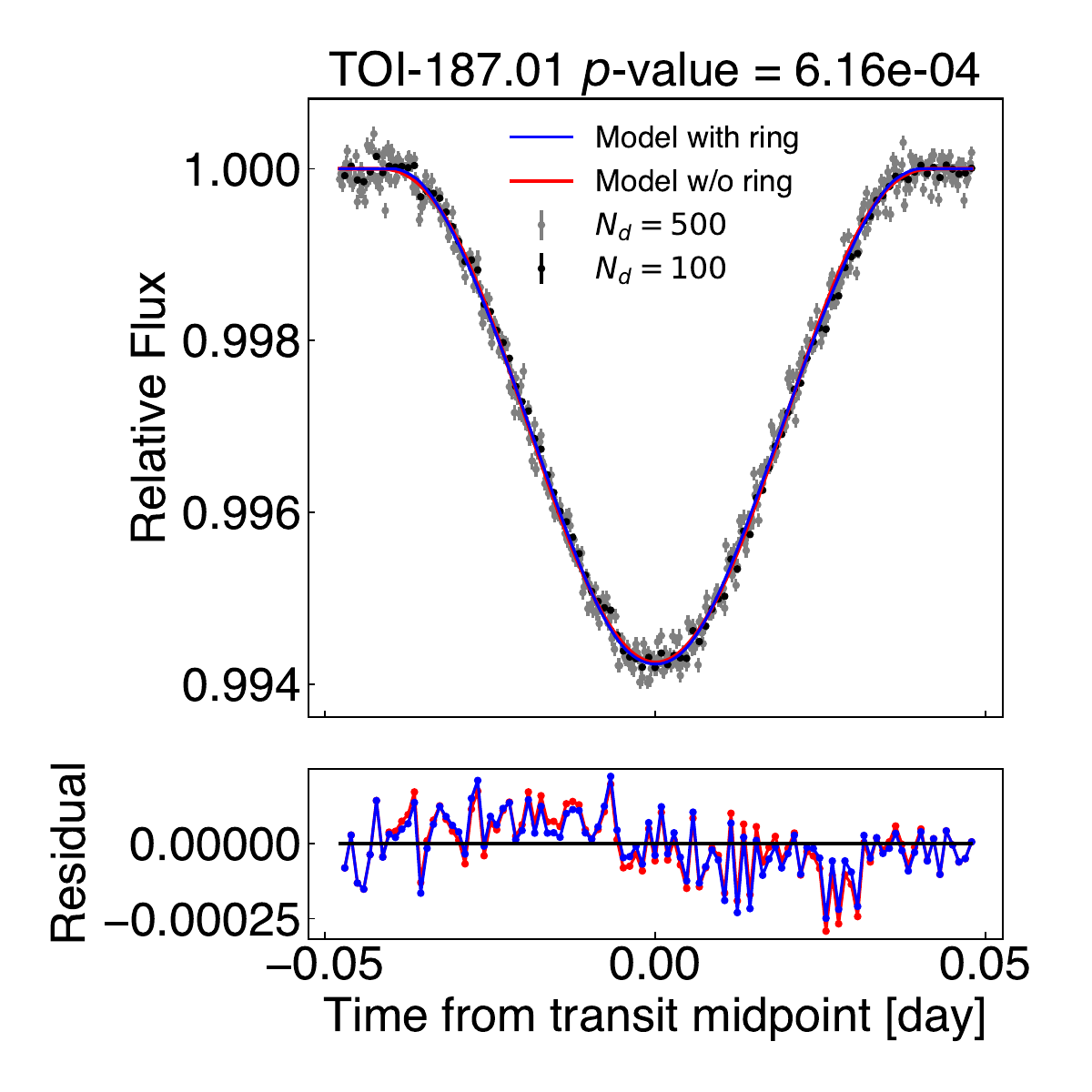}
  \end{minipage}
  \hfill
  \begin{minipage}{0.49\textwidth}
    \centering
    \includegraphics[width=\textwidth]{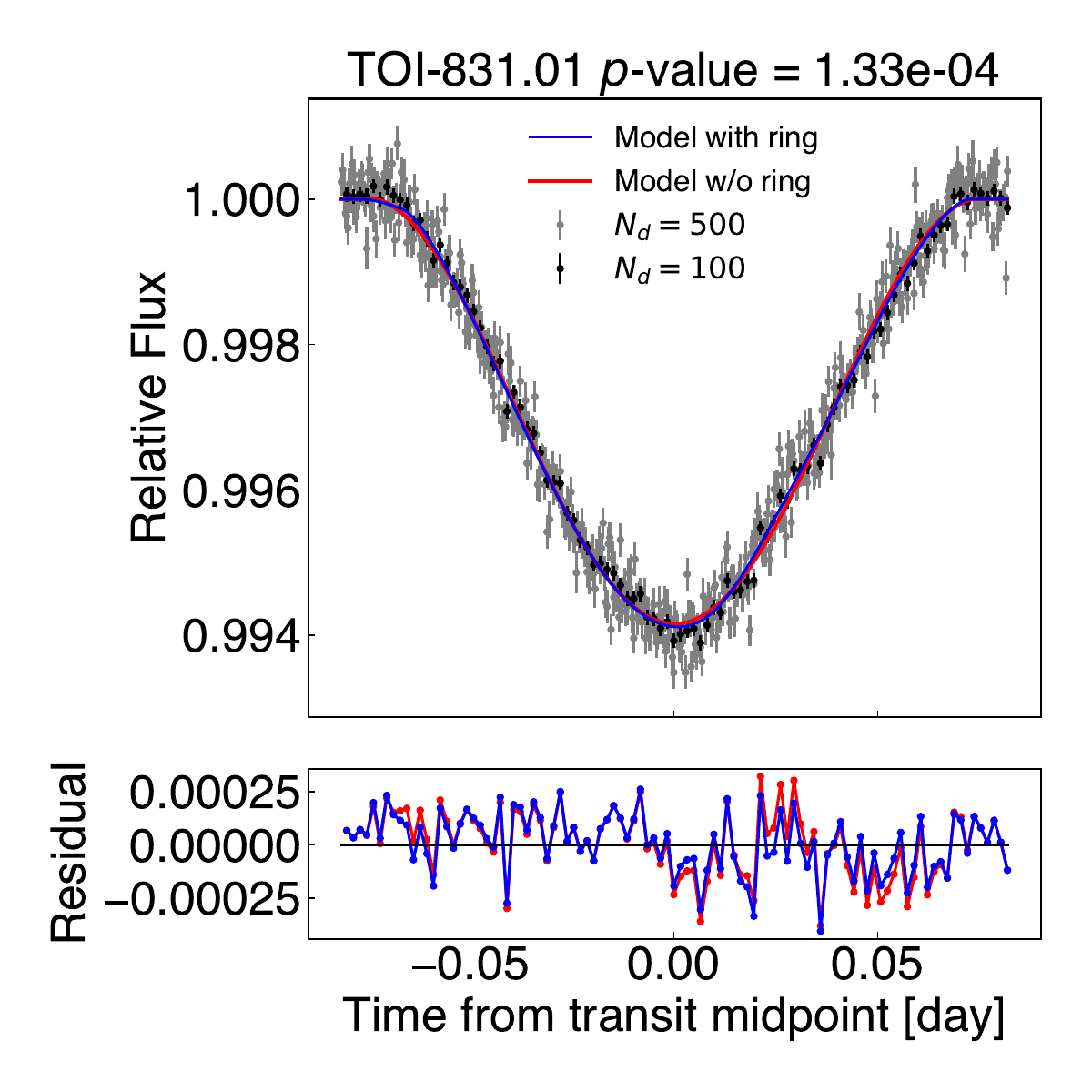}
  \end{minipage}
  \vspace{0.5cm}
  \begin{minipage}{0.49\textwidth}
    \centering
    \includegraphics[width=\textwidth]{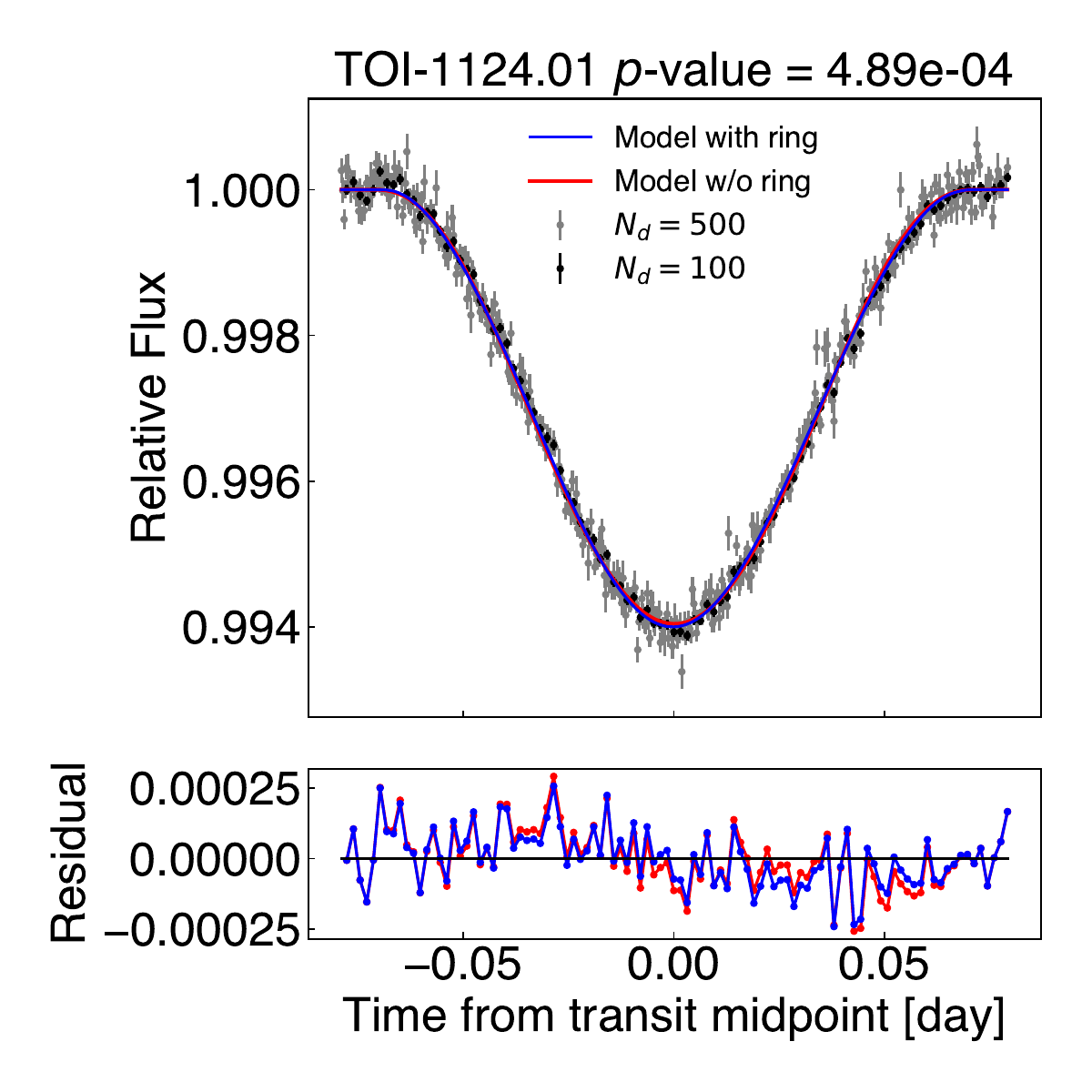}
  \end{minipage}
  
  \caption{Light curves of three systems that satisfy $3\sigma$ detection threshold in excepted TOIs listed in table \ref{ringed model and ringless model fitting result in excepted TOIs}.}
  \label{fig: Light curves of excepted 3 systems}
\end{figure*}

In this section, we describe the interesting objects in 46 TOIs excluded from our main results, as listed in table \ref{ringed model and ringless model fitting result in excepted TOIs}.
Objects characterized by gravity darkening can exhibit a relatively small $p$-value, TOI-1150.01, as shown in figure \ref{fig: TOI1150.01}. 
Gravity darkening is a phenomenon in which the equatorial region becomes darker and cooler than the poles due to the rotation of an oblate host star. 
The transit shape becomes asymmetric when the planet passes in front of the main star with an inclined passage relative to latitudinal lines. 
\citet{2020AJ....160....4A} showed that the asymmetric transit shape in TOI-1150.01 can be well explained by the gravity darkening model. 
TOI-624.01 \citep{2019AJ....158..141Z}, TOI-1924.01 \citep{2020ApJ...888...63A} and TOI-5821.01 \citep{2022A&A...658A..75H} also exhibits distortion in the transit caused by gravity darkening. 
However, the difference in dimming between ingress and egress is minimal, having a negligible impact on the $p$-value.
In the ringless model, no cases required an asymmetric transit shape.
The detection of signals such as gravity darkening indicates that our selection criteria are also effective for identifying targets capable of detecting small and interesting signals unrelated to rings.
The comparison of the best-fit results for the ringless and ringed models in remaining objects with $p$-value $<0.0027$ in excepted TOIs
are shown in figure \ref{fig: Light curves of excepted 3 systems}.
These TOIs are not labelled as ``EB'' but they exhibit V-shaped transits. 

\begin{table*}
\rotatebox{90}{%
\begin{minipage}{\textheight} 
\caption{Fitting results for the excepted TOIs using both ringless and ringed models.}
\label{ringed model and ringless model fitting result in excepted TOIs}
\begin{tabular}{crrrcrrrr}
\hline
\multicolumn{1}{c}{TOI} & \multicolumn{1}{c}{$P_{\rm{orb}}$ [day]} & \multicolumn{1}{c}{$b^{*}$} & \multicolumn{1}{c}{$(R_{\rm p} / R_{\star})_{\rm{ringless}}^{*}$} & \multicolumn{1}{c}{$(r_{\rm out/in })_{\rm{upp}}$} & \multicolumn{1}{c}{$\delta$ [ppm]} & \multicolumn{1}{c}{$\sigma (N_d=500)$ [ppm]} & \multicolumn{1}{c}{$(\chi^{2}_{\rm{ringless,min}},\chi^{2}_{\rm{ring,min}}, N_d)$} & \multicolumn{1}{c}{$p$-value} \\

\hline
106.01 & 2.8493820 $\pm$ 8e-07 & 0.56 $\pm$ 0.01 & 0.0832 $\pm$ 0.0002 & 1.11 & 8050 $\pm$ 15 & 165 & (614.20, 607.59, 500) & 1.51e-01 \\
107.01 & 3.950201 $\pm$ 4e-06 & 0.28 $\pm$ 0.07 & 0.1059 $\pm$ 0.0007 & ... & 12810 $\pm$ 59 & 693 & (716.71, 714.62, 500) & 6.98e-01 \\
135.01 & 4.126906 $\pm$ 2e-06 & 0.27 $\pm$ 0.05 & 0.0952 $\pm$ 0.0003 & 1.69 & 10990 $\pm$ 31 & 360 & (719.82, 712.41, 500) & 1.67e-01 \\
149.01 & 3.339942 $\pm$ 7e-06 & ... & 0.0901 $\pm$ 0.0009 & ... & 8940 $\pm$ 95 & 1078 & (844.83, 841.50, 500) & 5.85e-01 \\
187.01 & 0.5125911 $\pm$ 6e-07 & ... & ... & ... & 5690 $\pm$ 44 & 309 & (1804.44, 1741.94, 500) & 6.16e-04 \\
495.01 & 1.2749250 $\pm$ 3e-07 & ... & 0.1213 $\pm$ 0.0004 & 1.28 & 18700 $\pm$ 42 & 389 & (677.73, 673.14, 500) & 3.42e-01 \\
624.01 & 2.7442 $\pm$ 1e-04 & 0.59 $\pm$ 0.03 & 0.095 $\pm$ 0.001 & ... & 10300 $\pm$ 95 & 968 & (979.96, 977.92, 500) & 7.97e-01 \\
675.01 & 4.178062 $\pm$ 3e-06 & 0.24 $\pm$ 0.09 & 0.1003 $\pm$ 0.0007 & ... & 12540 $\pm$ 48 & 538 & (892.33, 890.13, 500) & 7.50e-01 \\
831.01 & 1.561762 $\pm$ 3e-06 & 1.0 $\pm$ 0.9 & ... & ... & 5700 $\pm$ 41 & 410 & (934.80, 896.51, 500) & 1.33e-04 \\
858.01 & 3.27975 $\pm$ 6e-05 & 0.3 $\pm$ 0.1 & 0.099 $\pm$ 0.001 & ... & 11000 $\pm$ 115 & 1186 & (736.15, 732.94, 500) & 5.44e-01 \\
911.01 & 8.584182 $\pm$ 4e-06 & ... & ... & ... & 8020 $\pm$ 87 & 668 & (711.15, 692.85, 434) & 1.14e-02 \\
1025.01 & 9.683790 $\pm$ 8e-06 & 1.0 $\pm$ 0.4 & ... & ... & 4770 $\pm$ 18 & 153 & (928.73, 919.42, 449) & 2.19e-01 \\
1059.01 & 9.449656 $\pm$ 9e-06 & ... & ... & ... & 19800 $\pm$ 262 & 2040 & (771.80, 758.42, 263) & 2.19e-01 \\
1114.01 & 2.4888 $\pm$ 1e-04 $^{\dagger}$ & 0.85 $\pm$ 0.01 & 0.060 $\pm$ 0.006 & ... & 3470 $\pm$ 36 & 349 & (1632.44, 1614.77, 470) & 1.71e-01 \\
1124.01 & 3.51873 $\pm$ 8e-05 & ... & ... & ... & 5770 $\pm$ 29 & 265 & (668.42, 642.10, 452) & 4.89e-04 \\
1144.01 & 4.887804 $\pm$ 2e-06 & ... & 0.0594 $\pm$ 0.0006 & ... & 4220 $\pm$ 39 & 334 & (675.74, 672.43, 500) & 4.91e-01 \\
1148.01 & 5.551493 $\pm$ 2e-06 & 0.15 $\pm$ 0.05 & 0.0874 $\pm$ 0.0002 & ... & 9410 $\pm$ 14 & 162 & (921.78, 919.69, 500) & 7.73e-01 \\
1149.01 & 2.320165 $\pm$ 4e-06 & 1.0 $\pm$ 0.6 & ... & ... & 5080 $\pm$ 34 & 287 & (887.79, 873.54, 500) & 4.73e-02 \\
1150.01 & 1.4811189 $\pm$ 4e-07 & 0.36 $\pm$ 0.02 & 0.0813 $\pm$ 0.0002 & ... & 6790 $\pm$ 11 & 125 & (3661.57, 3520.79, 500) & 2.45e-04 \\
1151.01 & 3.4740990 $\pm$ 7e-07 & 0.530 $\pm$ 0.006 & 0.1160 $\pm$ 0.0002 & ... & 14230 $\pm$ 16 & 167 & (1247.31, 1240.49, 500) & 4.42e-01 \\
1161.01 & 1.7635880 $\pm$ 7e-07 & ... & 0.0946 $\pm$ 0.0002 & 1.33 & 7800 $\pm$ 35 & 354 & (676.51, 668.20, 500) & 1.09e-01 \\
1163.01 & 3.077155 $\pm$ 2e-06 & 0.87 $\pm$ 0.01 & 0.08 $\pm$ 0.01 & ... & 4700 $\pm$ 58 & 521 & (466.73, 905.03, 500) & 1.00e+00 \\
1236.01 & 3.030069 $\pm$ 1e-06 & ... & 0.1359 $\pm$ 0.0007 & ... & 21800 $\pm$ 101 & 888 & (840.23, 834.81, 500) & 3.65e-01 \\
1271.01 & 6.134985 $\pm$ 8e-06 & ... & 0.0556 $\pm$ 0.0003 & 1.81 & 3860 $\pm$ 14 & 192 & (739.51, 735.63, 500) & 4.61e-01 \\
1274.01 & 19.32034 $\pm$ 2e-05 & 0.45 $\pm$ 0.05 & 0.108 $\pm$ 0.001 & ... & 13910 $\pm$ 89 & 1011 & (343.76, 343.11, 500) & 8.18e-01 \\
1337.01 & 3.3449 $\pm$ 5e-04 & 0.55 $\pm$ 0.03 & 0.108 $\pm$ 0.001 & ... & 13300 $\pm$ 129 & 1297 & (681.45, 674.95, 500) & 1.95e-01 \\
1351.01 & 5.929406 $\pm$ 3e-06 & 0.788 $\pm$ 0.005 & 0.115 $\pm$ 0.003 & 1.10 & 14680 $\pm$ 46 & 440 & (699.84, 697.65, 500) & 6.73e-01 \\
1420.01 & 6.95611 $\pm$ 3e-05 & 0.45 $\pm$ 0.07 & 0.120 $\pm$ 0.002 & ... & 16000 $\pm$ 176 & 1769 & (294.85, 292.84, 453) & 3.86e-01 \\
1465.01 & 1.4200240 $\pm$ 3e-07 & 0.60 $\pm$ 0.02 & 0.144 $\pm$ 0.002 & 1.18 & 23700 $\pm$ 130 & 854 & (324.32, 323.06, 500) & 5.89e-01 \\
1573.01 & 21.21644 $\pm$ 2e-05 & 0.48 $\pm$ 0.04 & 0.0747 $\pm$ 0.0005 & ... & 6010 $\pm$ 42 & 406 & (650.21, 645.83, 476) & 3.68e-01 \\
1629.01 & 4.54217 $\pm$ 1e-05 & 0.47 $\pm$ 0.08 & 0.092 $\pm$ 0.002 & ... & 9760 $\pm$ 92 & 1050 & (333.93, 333.84, 500) & 9.86e-01 \\
1651.01 & 3.764992 $\pm$ 2e-06 & 0.709 $\pm$ 0.009 & 0.115 $\pm$ 0.001 & 1.11 & 14840 $\pm$ 53 & 478 & (693.75, 684.70, 500) & 9.20e-02 \\
1676.01 & 3.1916 $\pm$ 2e-04 & 0.69 $\pm$ 0.03 & 0.088 $\pm$ 0.002 & ... & 8730 $\pm$ 83 & 750 & (245.18, 242.20, 464) & 1.36e-01 \\
1682.01 & 2.7347650 $\pm$ 8e-07 & 0.612 $\pm$ 0.008 & 0.0901 $\pm$ 0.0003 & ... & 9200 $\pm$ 19 & 187 & (802.85, 794.57, 500) & 1.66e-01 \\
1720.01 & 2.615856 $\pm$ 3e-06 & ... & 0.105 $\pm$ 0.002 & ... & 13400 $\pm$ 156 & 1356 & (1029.33, 1024.77, 499) & 5.37e-01 \\
1771.01 & 3.212234 $\pm$ 2e-06 & 0.36 $\pm$ 0.04 & 0.1086 $\pm$ 0.0007 & ... & 14520 $\pm$ 49 & 441 & (834.89, 828.63, 500) & 2.97e-01 \\
1834.01 & 1.216439 $\pm$ 5e-06 & ... & ... & 1.26 & 15400 $\pm$ 128 & 953 & (328.60, 327.21, 500) & 5.58e-01 \\
1924.01 & 2.82408 $\pm$ 1e-05 & 0.39 $\pm$ 0.03 & 0.0876 $\pm$ 0.0003 & ... & 8970 $\pm$ 21 & 232 & (1005.82, 1001.19, 500) & 5.19e-01 \\
2021.01 & 3.941506 $\pm$ 2e-06 & 0.31 $\pm$ 0.06 & 0.132 $\pm$ 0.001 & ... & 21300 $\pm$ 120 & 1115 & (285.22, 283.84, 416) & 5.79e-01 \\
2024.01 & 2.875889 $\pm$ 2e-06 & 0.40 $\pm$ 0.06 & 0.0509 $\pm$ 0.0003 & ... & 2910 $\pm$ 13 & 131 & (686.11, 681.75, 500) & 3.72e-01 \\
2183.01 & 6.50114 $\pm$ 3e-05 & 0.45 $\pm$ 0.06 & 0.0685 $\pm$ 0.0007 & ... & 4370 $\pm$ 19 & 244 & (912.23, 906.99, 497) & 4.22e-01 \\
2403.01 & 1.809881 $\pm$ 1e-06 & 0.19 $\pm$ 0.04 & 0.1072 $\pm$ 0.0003 & 1.33 & 13030 $\pm$ 21 & 230 & (687.19, 675.77, 500) & 4.16e-02 \\
4470.01 & 2.2185750 $\pm$ 5e-07 & 0.654 $\pm$ 0.003 & 0.1551 $\pm$ 0.0005 & ... & 25850 $\pm$ 33 & 244 & (1290.72, 1273.32, 500) & 8.37e-02 \\
\hline
\end{tabular}
\end{minipage}
}
\end{table*}

\begin{table*}
\rotatebox{90}{%
\begin{minipage}{\textheight} 
\begin{threeparttable}
\flushleft{Table \ref{ringed model and ringless model fitting result in excepted TOIs}. (Continued)}
\begin{tabular}{crrrcrrrr}
\hline
\multicolumn{1}{c}{TOI} & \multicolumn{1}{c}{$P_{\rm{orb}}$ [day]} & \multicolumn{1}{c}{$b^{*}$} & \multicolumn{1}{c}{$(R_{\rm p} / R_{\star})_{\rm{ringless}}^{*}$} & \multicolumn{1}{c}{$(r_{\rm out/in })_{\rm{upp}}$} & \multicolumn{1}{c}{$\delta$ [ppm]} & \multicolumn{1}{c}{$\sigma (N_d=500)$ [ppm]} & \multicolumn{1}{c}{$(\chi^{2}_{\rm{ringless,min}},\chi^{2}_{\rm{ring,min}}, N_d)$} & \multicolumn{1}{c}{$p$-value} \\
\hline
4612.01 & 4.11379 $\pm$ 3e-05 & 0.31 $\pm$ 0.05 & 0.0684 $\pm$ 0.0003 & ... & 5650 $\pm$ 16 & 203 & (681.54, 674.83, 500) & 1.83e-01 \\
5074.01 & 3.08017 $\pm$ 2e-05 & 0.58 $\pm$ 0.01 & 0.0919 $\pm$ 0.0004 & 1.28 & 9640 $\pm$ 29 & 290 & (666.23, 658.17, 500) & 1.13e-01 \\
5821.01 & 2.14880 $\pm$ 5e-05 & 0.2 $\pm$ 0.1 & 0.0798 $\pm$ 0.0003 & 1.65 & 7060 $\pm$ 23 & 258 & (715.74, 712.83, 500) & 5.73e-01 \\
\hline
\end{tabular}%
\begin{tablenotes}
\item[*] Due to the high uncertainties, the parameter value cannot be determined accurately and is therefore denoted as ``...''.
\end{tablenotes}
\end{threeparttable}
\end{minipage}
}
\end{table*}

\end{appendix}
\clearpage

\bibliographystyle{aa}
\bibliography{tess_ring}

\end{document}